\documentclass[12pt]{article}
\usepackage{amsmath,amsfonts,amssymb,amsthm,bm,mathtools,graphicx,subfig,color,tcolorbox,algorithm,algorithmic}
\usepackage{caption,fancyhdr,titlesec,indentfirst,booktabs,verbatim,hyperref,cases,titletoc,multirow,wasysym,empheq,setspace,natbib,chngcntr}
\usepackage[font=small,labelfont=bf]{caption}

\newcommand{\rom}[1]{\uppercase\expandafter{\romannumeral #1\relax}}
\usepackage[page,header]{appendix}
\usepackage[english]{babel}
\usepackage[utf8]{inputenc}

\newtheorem{theorem}[]{Theorem}
\newtheorem{lemma}[]{Lemma}

\newtheorem{proposition}[]{Proposition}

\newtheorem{example}[]{Example}
\newtheorem{assumption}[]{Assumption}

\def\cbf{c} 
\def\xbf{x}
\def\ybf{y} 
\def\zbf{z} 
\def\Bbf{B}
\def\epsilonbf{\epsilon}

\def\psibf{\psi} 
\def\thetabf{\theta} 
\def\Sigmabf{\Sigma} 
\def\A{\mathbb A}

\def\E{\mathbb E}

\def\P{\mathbb P}
\def\R{\mathbb R}

\def\({\left(}
\def\){\right)}
\def\[{\left[}
\def\]{\right]}

\def\AA{\mathcal A}
\def\DD{\mathcal D}
\def\FF{\mathcal F}

\def\HH{\mathcal H}

\def\PP{\mathcal P}
\def\TT{\mathcal T}
\def\WW{\mathcal W}
\def\XX{\mathcal X}

\def\Rbf{R} 

\def\eop{\hfill {\large $\Box$}}

\begin{document}

\title{\Large{\textbf{Kernel Ordinary Differential Equations}}} 
\author{
\bigskip
{\sc Xiaowu Dai and Lexin Li} \\
{\it {\normalsize University of California at Berkeley}}
}
\date{}
\maketitle

\begin{abstract}
Ordinary differential equation (ODE) is widely used in modeling biological and physical processes in science. In this article, we propose a new reproducing kernel-based approach for estimation and inference of ODE given noisy observations. We do not assume the functional forms in ODE to be known, or restrict them to be linear or additive, and we allow pairwise interactions. We perform sparse estimation to select individual functionals, and construct confidence intervals for the estimated signal trajectories. We establish the estimation optimality and selection consistency of kernel ODE under both the low-dimensional and high-dimensional settings, where the number of unknown functionals can be smaller or larger than the sample size. Our proposal builds upon the smoothing spline analysis of variance (SS-ANOVA) framework, but tackles several important problems that are not yet fully addressed, and thus extends the scope of existing SS-ANOVA too. We demonstrate the efficacy of our method through numerous ODE examples. 
\end{abstract}
\bigskip

\noindent{\bf Key Words:} 
Component selection and smoothing operator; High dimensionality; Ordinary differential equations; Smoothing spline analysis of variance; Reproducing kernel Hilbert space.

\newpage
\baselineskip=21pt

\section{Introduction}
\label{sec:introduction}

\noindent
Ordinary differential equation (ODE) has been widely used to model dynamic systems and biological and physical processes in a  variety of scientific applications. Examples include infectious disease \citep{Liang2008}, genomics \citep{CaoZhao2008, Chou2009, Ma2009, LuLiang2011, HendersonMichailidis2014, Wu2014}, neuroscience \citep{Izhikevich2007, Zhang2015, Zhang2017, CaoLuo2019}, among many others. A system of ODEs take the form,
\begin{equation} \label{eqn:intermodel}
\frac{d \xbf(t)}{dt} = \left(
\begin{array}{c}
\dfrac{d x_1(t)}{d t} \\
\vdots\\
\dfrac{d x_p(t)}{d t} \\
\end{array}
\right) = \left(
\begin{array}{c}
F_1(\xbf(t))  \\
\vdots\\
F_p(\xbf(t)) \\
\end{array}
\right) 
= F(\xbf(t)), 
\end{equation}
where $\xbf(t) = (x_1(t), \ldots, x_p(t))^\top \in \R^p$ denotes the system of $p$ variables of interest, $F = \{F_1, \ldots,$ $F_p\}$ denotes the set of unknown functionals that characterize the regulatory relations among $\xbf(t)$, and $t$ indexes time in an interval standardized to $\TT = [0,1]$. Typically, the system \eqref{eqn:intermodel} is observed on discrete time points $\{t_1, \ldots, t_n\}$ with measurement errors, 
\begin{equation} \label{eqn:obserdata}
\ybf_i = \xbf(t_i) + \epsilonbf_i, \quad i=1,\ldots,n,
\end{equation}
where $\ybf_i  = (y_{i1},\ldots,y_{ip})^\top \in \R^p$ denotes the observed data, $\epsilonbf_i = (\epsilon_{i1}, \ldots, \epsilon_{ip})^\top \in \R^p$ denotes the vector of measurement errors that are usually assumed to follow independent normal distribution with mean $0$ and variance $\sigma_j^2$,  $j=1,\ldots,p$, and $n$ denotes the number of time points. Besides, an initial condition $\xbf(0) \in \R^p$ is usually given for the system \eqref{eqn:intermodel}. 

In a biological or physical system, a central question of interest is to uncover the structure of the system of ODEs in terms of which variables regulate which other variables, given the observed noisy time-course data $\{ \ybf_i \}_{i=1}^{n}$. Specifically, we say that $x_{k}$ regulates $x_{j}$, if $F_{j}$ is a functional of $x_{k}$. In other words, $x_{k}$ controls the change of $x_{j}$ through the functional $F_{j}$ on the derivative $d x_{j} / d t$. Therefore, the functionals $F = \{F_1, \ldots, F_p\}$ encode the regulatory relations of interest, and are often assumed to take the form, 
\begin{equation} \label{eqn:nonadditivemodel}
F_{j}(\xbf(t)) = \theta_{j0} + \sum_{k=1}^p F_{jk}(x_k(t)) + \sum_{k\neq l, k=1}^p \sum_{l=1}^p F_{jkl}(x_k(t),x_l(t)), \quad j = 1, \ldots, p, 
\end{equation}
where $\theta_{j0} \in \R$ denotes the intercept, and $F_{jk}$ and $F_{jkl}$ represent the main effect and two-way interaction, respectively. Higher-order interactions are possible, but two-way interactions are the most common structure studied in ODE \citep{Ma2009, Zhang2015}.  

There have been numerous pioneering works studying statistical modeling of ODEs. However, nearly all existing solutions constrain the forms of $F$. Broadly speaking, there are three categories of functional forms imposed. The first category considers linear functionals for $F$. For instance, \cite{LuLiang2011} studied a system of linear ODEs to model dynamic gene regulatory networks. \citet{Zhang2015} extended the linear ODE to include the interactions to model brain connectivity networks. The model of \citet{Zhang2015}, other than differentiating between the variables that encode the neuronal activities and the ones that represent the stimulus signals, is in effect of the form,
\begin{equation} \label{eqn:bilinearmodel}
F_{j}(\xbf(t)) = \theta_{j0} + \sum_{k=1}^p \theta_{jk} x_k(t) + \sum_{k\neq l, k=1}^p \sum_{l=1}^p \theta_{jkl} x_k(t) x_l(t), \quad j = 1, \ldots, p, 
\end{equation}
whereas the model of \cite{LuLiang2011} is similar to \eqref{eqn:bilinearmodel} but focuses on the main-effect terms only. In both cases, $F_j$ takes a linear form. \citet{Dattner2015} further extended the functional $F_j$ in \eqref{eqn:bilinearmodel} to a generalized linear form, but without the interactions, i.e.,  
\begin{equation} \label{eqn:genlinear}
F_j(\xbf(t)) = \theta_{j0} + \psibf_j(\xbf(t))^\top \thetabf_j, \quad j=1, \ldots, p,
\end{equation}
where $\theta_{j0} \in \R$, $\thetabf_j \in \R^d$, and $\psibf_j(\xbf) = (\psi_{j1}(\xbf), \ldots, \psi_{jd}(\xbf))^\top \in \R^d$ is a finite set of \emph{known} basis functions. The second category considers additive functionals for $F$. Particularly, \citet{HendersonMichailidis2014, Wu2014, Chen2017} considered the generalized additive model for $F_j$,
\begin{equation} \label{eqn:additivemodel}
F_j(\xbf(t)) = \theta_{j0} + \sum_{k=1}^p F_{jk}(x_k(t)) =  \theta_{j0} + \sum_{k=1}^p \left\{ \psibf(x_k(t))^\top\thetabf_{jk} + \delta_{jk}(x_k(t)) \right\}, \quad j = 1, \ldots, p, 
\end{equation}
where $\theta_{j0} \in \R$, $\thetabf_{jk} \in \R^d$, $\psibf(\xbf) = (\psi_1(\xbf),\ldots,\psi_d(\xbf))^\top \in \R^d$ is a finite set of common basis functions, and $\delta_{jk} \in \R$ is the residual function. Different from \citet{Dattner2015}, the residual $\delta_{jk}$ is unknown. The functional $F_j$ in \eqref{eqn:additivemodel} takes an additive form. Finally, there is a category of ODE solutions focusing on the scenario where the functional forms for $F$ are \emph{known} \citep{gonzalez2014reproducing, zhang2015selection, mikkelsen2017learning}.

These works have laid a solid foundation for statistical modeling of ODE. However, in plenty of scientific applications, the forms of the functionals $F$ are unknown, and the linear or additive forms on $F$ can be restrictive. Besides, it is highly nontrivial to couple the basis function-based solutions with the interactions. We give an example in Section \ref{sec:motivation}, where a commonly used enzyme network ODE system involves both nonlinear functionals and two-way interactions. Such examples are often the rules rather than the exceptions, motivating us to consider a more flexible form of ODE. Moreover, the existing ODE methods have primarily focused on sparse estimation, but few tackled the problem of statistical inference, which is challenging due to the complicated correlation structure of ODE. 

In this article, we propose a novel approach of kernel ordinary differential equation (KODE) for estimation and inference of the ODE system in \eqref{eqn:intermodel} given noisy observations from \eqref{eqn:obserdata}. We adopt the general formulation of \eqref{eqn:nonadditivemodel}, but we do not assume the functional forms of $F$ are known, or restrict them to be linear or additive, and we allow pairwise interactions. As such, we consider a more general ODE system that encompasses \eqref{eqn:bilinearmodel}, \eqref{eqn:genlinear} and \eqref{eqn:additivemodel} as special cases. We further introduce sparsity regularization to achieve selection of individual functionals in \eqref{eqn:nonadditivemodel}, which yields a sparse recovery of the regulatory relations among $F$, and improves the model interpretability. Moreover, we derive the confidence interval for the estimated signal trajectory $x_j(t)$. We establish the estimation optimality and selection consistency of kernel ODE, under both low-dimensional and high-dimensional settings, where the number of unknown functionals $p$ can be smaller or larger than the number of time points $n$, and we study the regime-switching phenomenon. These differences clearly separate our proposal from the existing ODE solutions in the literature. 

Our proposal is built upon the smoothing spline analysis of variance (SS-ANOVA) framework that was first introduced by \citet{Wahba1995}, then further developed in regression and functional data analysis settings by  \cite{huang1998projection, LinZhang2006, Zhu2014}. We adopt a similar component selection and smoothing operator (COSSO) type penalty of \citet{LinZhang2006} for regularization, and conceptually, our work extends COSSO to the ODE setting. However, our proposal considerably differs from COSSO and the existing SS-ANOVA methods, in multiple ways. First, unlike the standard SS-ANOVA models, the regressors of kernel ODE are not directly observed and need to be estimated from the data with error. This extra layer of randomness and estimation error introduces additional difficulty to SS-ANOVA. Second, we employ the integral of the estimated trajectories in the loss function to improve the estimation properties \citep{Dattner2015}. The use of the integral and the inclusion of the interaction terms pose some identifiability question that we tackle explicitly. Third, we establish the estimation optimality and selection consistency in the RKHS framework, which is utterly different from \citet{Zhu2014}, and requires new technical tools. Moreover, our theoretical analysis extends that of \citet{Chen2017} from the finite bases setting of cubic splines to the infinite bases setting of RKHS. Finally, for statistical inference, we derive the confidence bands to provide uncertainty quantification for the penalized estimators of the signal trajectories in the ODE model. Our solution builds on the confidence intervals idea of \citet{wahba1983}. But unlike the classical methods focusing on the fixed dimensionality $p$ \citep{wahba1983, Opsomer1997}, we allow a diverging $p$ that can far exceed the sample size $n$. In summary, our proposal tackles several crucial problems that are not yet fully addressed in the existing SS-ANOVA framework, and it is far from a straightforward extension. We believe the proposed kernel ODE method not only makes a useful addition to the toolbox of ODE modeling, but also extends the scope of SS-ANOVA-based kernel learning. 

The rest of the article is organized as follows. We propose kernel ODE in Section \ref{sec:method}, and develop the estimation algorithm and inference procedure in Section \ref{sec:algorithm}. We derive the consistency and optimality of the proposed method in Section \ref{sec:theory}. We investigate the numerical performance in Section \ref{sec:simulation}, and illustrate with a real data example in Section \ref{sec:application}. We conclude the paper with a discussion in Section \ref{sec:conclusion}, and relegate all proofs and some additional numerical results to the Supplementary Appendix.

\section{Kernel Ordinary Differential Equations}
\label{sec:method}

\subsection{Motivating example}
\label{sec:motivation}

\noindent
We consider an enzymatic regulatory network as an example to demonstrate that nonlinear functionals as well as interactions are common in the system of ODEs. \citet{Ma2009} found that all circuits of three-node enzyme network topologies that perform biochemical adaptation can be well approximated by two architectural classes: a negative feedback loop with a buffering node, and an incoherent feedforward loop with a proportioner node. The mechanism of the first class follows the Michaelis-Menten kinetic equations \citep{Tzafriri2003}, 
\begin{eqnarray} \label{eqn:NFBLB}
\frac{dx_1(t)}{dt} & = & c_1\frac{x_0 \{1-x_1(t)\}}{\{1-x_1(t)\}+C_1} - \tilde{c}_1c_2\frac{x_1(t)}{x_1(t)+C_2}, \nonumber \\
\frac{dx_2(t)}{dt} & = & c_3\frac{\{1-x_2(t)\}  x_3(t)}{\{1-x_2(t)\}+C_3} - \tilde{c}_2c_4\frac{x_2(t)}{x_2(t)+C_4},\\
\frac{dx_3(t)}{dt} & = & c_5\frac{x_1(t) \{1-x_3(t)\}}{\{1-x_3(t)\}+C_5} - c_6\frac{x_2(t) x_3(t)}{x_3(t)+C_6}, \nonumber
\end{eqnarray}
where $x_1(t),x_2(t),x_3(t)$ are three interacting nodes, such that $x_1(t)$ receives the input, $x_2(t)$ plays the diverse regulatory role, and $x_3(t)$ transmits the output, $x_0$ is the initial input stimulus, and $c_1, \ldots, c_6, C_1, \ldots, C_6, \tilde{c}_1,\tilde{c}_2$ denote the catalytic rate parameters, the Michaelis-Menten constants, and the concentration parameters, respectively. See Figure \ref{fig:eg2_1}(a) for a graphical illustration of this ODE system. In this model, the functionals $F_1, F_2, F_3$ are all nonlinear, and both $F_2$ and $F_3$ involve two-way interactions. It is of great interest to estimate $F_j$'s given the observed data, to verify model \eqref{eqn:NFBLB}, and to carry out statistical inference of the unknown parameters. This example, along with many other ODE systems with nonlinear functionals and interaction terms motivate us to consider a general ODE system as in \eqref{eqn:nonadditivemodel}.

\subsection{Two-step collocation estimation}

\noindent
Before presenting our method, we first briefly review the two-step collocation estimation method, which is commonly used for parameter estimation in ODE, and is also useful in our setting. The method was first proposed by \citet{Varah1982}, then extended to various ODE models. In the first step, it fits a smoothing estimate,  
\begin{equation*}
\widehat{x}_j(t) = \underset{z_j\in\mathcal F}{\arg\min} \left\{\frac{1}{n} \sum_{i=1}^n \left\{ y_{ij} - z_{j}(t_i) \right\}^2 + \lambda_{nj}J_1(z_j)\right\}, \quad j = 1, \ldots, p, 
\end{equation*}
where $J_1(\cdot)$ is a smoothness penalty in the function space $\FF$, and $z_j$ is a function in $\mathcal F$ that we minimize over. In the second step, it solves an optimization problem to estimate the model parameters $\theta_{j0} \in \R$ and $\thetabf_j = (\theta_{j1}, \ldots, \theta_{jp})^\top \in \R^p$, for $j = 1, \ldots, p$. Particularly, \citet{Varah1982} considered the derivative $d \widehat{x}_j(t) / d t$ and the following minimization,
\begin{eqnarray*}
\underset{\theta_{j0}, \theta_{j}}{\min} \displaystyle\int_0^1 \left( \dfrac{d \widehat{x}_j(t)}{d t} - \theta_{j0} - \sum_{k=1}^p \theta_{jk} \widehat{x}_k(t)\right)^2 dt, \quad j = 1, \ldots, p.
\end{eqnarray*}
\citet{Wu2014} developed a similar two-step collocation method for their additive ODE model \eqref{eqn:additivemodel}, and estimated the model parameters $\theta_{j0} \in \R$ and $\thetabf_{jk} = (\theta_{jk1}, \ldots, \theta_{jkd})^\top \in \R^d$, for $j, k = 1, \ldots, p$, with a standardized group $\ell_1$-penalty,
\begin{equation*}
\underset{\theta_{j0},\thetabf_{jk}}{\min} \displaystyle\int_0^1 \left\| \dfrac{d \widehat{x}_j(t)}{d t} - \theta_{j0} -\sum_{k=1}^p\thetabf_{jk}^\top\psibf(\widehat{x}_k(t)) \right\|^2_2dt + 
\tau_{nj} \sum_{k=1}^p \left[ \int_0^1\left\{\thetabf_{jk}^\top\psibf(\widehat{x}_k(t))\right\}^2 dt \right]^{1/2}.
\end{equation*}
They further discussed adaptive group $\ell_1$ and regular $\ell_1$-penalties. Meanwhile, \citet{HendersonMichailidis2014} considered an extra $\ell_2$-penalty.

Alternatively, in the second step, \citet{Dattner2015} proposed to focus on the integral $\int_0^t g_j(\widehat{\xbf}(u))du$, rather than the derivative $d \widehat{x}_j(t) / d t$, and they estimated the model parameters $\theta_{j0} \in \R$ and $\thetabf_j = (\theta_{j1}, \ldots, \theta_{jd})^\top \in \R^d$, for $j = 1, \ldots, p$, in \eqref{eqn:genlinear} by,
\begin{equation*}
\underset{\theta_{j0}, \thetabf_j}{\min}\sum_{j=1}^p \displaystyle\int_0^1 \left\{ \widehat{x}_j(t) -\theta_{j0}- \thetabf_j^\top \int_0^t \psi_j(\widehat{\xbf}(u)) du \right\}^2dt.
\end{equation*}
They found that this modification from the derivative to integral leads to a more robust estimate and also an easier derivation of the asymptotic properties. \citet{Chen2017} adopted this idea for their additive ODE model \eqref{eqn:additivemodel}, and estimated the parameters $\theta_{j0} \in \R$, $\tilde\theta_{j} \in \R$, and $\thetabf_{jk} = (\theta_{jk1}, \ldots, \theta_{jkd})^\top \in \R^d$, for $j, k = 1, \ldots, p$, by 
\begin{equation*}
\begin{aligned}
\underset{\theta_{j0}, \tilde\theta_j, \theta_{jk}}{\min}\frac{1}{2n}\sum_{i=1}^n\left\{y_{ij}-\theta_{j0}-b_j t_i-\sum_{k=1}^p\theta_{jk}^\top \int_0^{t_i}\psi(\widehat{x}_k(u))du\right\}^2 \\
+ \tau_{nj}\sum_{k=1}^p\left[\frac{1}{n}\sum_{i=1}^n\left\{\theta_{jk}^\top\int_0^{t_i}\psi(\widehat{x}_k(u)) dt\right\}^2\right]^{1/2}.
\end{aligned}
\end{equation*}

\subsection{Kernel ODE}
\label{subsec:KODE}

\noindent 
We build the proposed kernel ODE within the smoothing spline ANOVA framework; see \citet{Wahba1995} and \citet{gu2013} for more background on SS-ANOVA. Specifically, let $\mathcal H_k$ denote a space of functions of $x_k(t) \in \XX$ with zero marginal integral, where $\XX\subset\R$ is a compact set. Let $\{1\}$ denote the space of constant functions. We construct the tensor product space as
\begin{equation} \label{eqn:spaceH}
\HH = \{1\} \; \oplus \; \sum_{k=1}^p\mathcal H_k \; \oplus \sum_{k=1,k \neq l}^p \sum_{l=1}^p \left( \mathcal H_{k}\otimes\mathcal H_{l} \right).
\end{equation}
We assume the functionals $F_j$, $j=1, \ldots, p$, in the ODE model \eqref{eqn:nonadditivemodel} are located in the space of $\mathcal H$. The identifiability of the terms in \eqref{eqn:nonadditivemodel} is assured by the conditions specified through the averaging operators: $\int_\TT F_{jk}(x_k(t))dt = 0$ for $k=1,\ldots,p$. Let $\|\cdot\|_{\mathcal H}$ denote the norm of $\HH$, and $\PP^{k}F_j$ and $\PP^{kl}F_j$ denote the orthogonal projection of $F_j$ onto $\mathcal H_{k}$ and  $\mathcal H_{k}\otimes \HH_l$, respectively. We consider a two-step collocation estimation method, by first obtaining a smoothing spline estimate $\widehat{\xbf}(t) = (\widehat{x}_1(t), \ldots, \widehat{x}_p(t))^\top$, where 
\begin{equation} \label{eqn:sshatxj}
\widehat{x}_j(t) = \underset{z_j\in\mathcal F}{\arg\min} \left\{\frac{1}{n}\sum_{i=1}^n \left\{ y_{ij} - z_{j}(t_i) \right\}^2 + \lambda_{nj}\|z_j(t)\|_{\mathcal F}^2\right\}, \quad j=1,\ldots,p, 
\end{equation}
then estimating $F_j \in \mathcal H$ and $\theta_{j0} \in \R$ by the following penalized optimization, 
\begin{equation} \label{eqn:kode}
\underset{\theta_{j0}, F_j}{\min} 
\frac{1}{n}\sum_{i=1}^n\left\{y_{ij} - \theta_{j0} - \int_0^{t_i}F_j(\widehat{\xbf}(t))dt\right\}^2+ \tau_{nj} \left( \sum_{k=1}^p\|\PP^k F_j\|_{\HH} + \sum_{k\neq l, k=1}^p \sum_{l=1}^p \|\PP^{kl} F_j\|_\HH \right).
\end{equation}
Our proposal deals with the integral $ \int_0^{t_i}F_j(\widehat{\xbf}(u))du$, rather than the derivative $d \widehat{x}_j(t) / d t$, which is in a similar spirit as \citet{Dattner2015}. Besides, it involves two penalty functions, $J_1 \equiv \| \cdot \|_{\mathcal F}^2$ in \eqref{eqn:sshatxj}, and $J_2(F_j) \equiv \sum_{k=1}^p\|\PP^k F_j\|_{\HH} + \sum_{k=1}^p \sum_{l=1}^p \|\PP^{kl} F_j\|_\HH$ in \eqref{eqn:kode}, with $\lambda_{nj}$ and $\tau_{nj}$ as two tuning parameters. We next make some remarks about this proposal. 

For the functionals, the formulation in \eqref{eqn:kode} is highly flexible, nonlinear, and incorporates two-way interactions. Meanwhile, it naturally covers the linear ODE in \eqref{eqn:bilinearmodel} and \eqref{eqn:genlinear}, and the additive ODE in \eqref{eqn:additivemodel} as special cases. In particular, if $\HH$ is the linear functional space, $\HH = \{1\}\oplus\sum_{k=1}^p\{x_k-1/2\}\oplus\sum_{k\neq l}[\{x_k-1/2\}\otimes \{x_l-1/2\}]$ with the input space $\XX=[0,1]^p$, then any $F$ of the form in \eqref{eqn:bilinearmodel} belongs to $\HH$. If $\HH$ is spanned by some known generalized functions, $\HH = \psi_{j1}(\xbf) \oplus \ldots \oplus \psi_{jp}(\xbf)$, then any $F$ in \eqref{eqn:genlinear} belongs to $\HH$. If $\HH$ is the additive functional space, $\HH=\{1\}\oplus\sum_{k=1}^p\mathcal H_k$ with the $\ell_2$-norm, then for $F_{jk}(x_k(t)) = \psi(x_k(t))^\top\theta_{jk}$, the penalty on the main effects becomes $\sum_{k=1}^p\|\PP^k F_j\|_{\HH} = \sum_{k=1}^p[\int_0^1\{\psi(x_k(t))^\top\theta_{jk}\}^2dt]^{1/2}$, which is exactly the same as the ODE model of \citet{Chen2017}. 

For the penalties, the first penalty function $J_1$ is the squared RKHS norm corresponding to the RKHS $\{ \FF,\|\cdot\|_\FF \}$. It is for estimating $\widehat{\xbf}_j$, and $\FF$ does not have to be the same as $\HH$. The second penalty function $J_2$ is a sum of RKHS norms on the main effects and pairwise interactions. This penalty is similar as the COSSO penalty of \citet{LinZhang2006}. But as we outline in Section \ref{sec:introduction}, our extension is far from trivial. We also note that, we do not impose a hierarchical structure for the main effects and interactions, in that if an interaction term is selected, the corresponding main effect term does not have to be selected \citep{Wang2009}. This is motivated by the observation that, e.g., in the enzymatic regulatory network example in Section \ref{sec:motivation}, the interaction terms $x_1(t) x_3(t)$ and $x_2(t) x_3(t)$ both appear in the ODE regulating $x_3(t)$, but the main effect terms $x_1(t)$ and $x_2(t)$ are not present. 

\begin{theorem}
\label{thm:representer}
Assume that the RKHS $\HH$ can be decomposed as in \eqref{eqn:spaceH}. Then there exists a minimizer of \eqref{eqn:kode} in $\HH$ for any tuning parameter $\tau_{nj}\geq 0$. Moreover, the minimizer is in a finite-dimensional space.
\end{theorem}

\noindent
Theorem \ref{thm:representer} is a generalization of the well-known representer theorem \citep{wahba1990}. The difference is that, unlike the smoothing splines model as studied in \citet{wahba1990}, the minimization of \eqref{eqn:kode} involves an integral in the loss function, and the penalty is not a norm in $\HH$ but a convex pseudo-norm. A direct implication of Theorem \ref{thm:representer} is that, 
although the minimization with respect to $F_j$ is taken over an infinite-dimensional space in \eqref{eqn:kode}, the solution to  \eqref{eqn:kode} can actually be found in a finite-dimensional space. We next develop an estimation algorithm to solve \eqref{eqn:kode}.

\section{Estimation and Inference}
\label{sec:algorithm}

\subsection{Estimation procedure}
\label{sec:functionalestimation}

\noindent
The estimation of the proposed kernel ODE system consists of two major steps. The first step is the smoothing spline estimation in \eqref{eqn:sshatxj}, which is standard and the tuning of the smoothness parameter $\lambda_{nj}$ is often done through generalized cross-validation \citep[see, e.g.,][]{gu2013}. The second step is to solve \eqref{eqn:kode}. Toward that end, we first propose an optimization problem that is equivalent to \eqref{eqn:kode}, but is computationally easier to tackle. We then develop an estimation algorithm to solve this new equivalent problem. 
 
Specifically, we consider the following optimization problem, for $j = 1, \ldots, p$, 
\begin{align} \label{eqn:kodeequiv}
\begin{split}
\underset{\theta_{j0}, \theta_{j}, F_j}{\min} \frac{1}{n}\sum_{i=1}^n & \left\{y_{ij} - \theta_{j0} - \int_0^{t_i}F_j(\widehat{\xbf}(t))dt\right\}^2 
+ \eta_{nj}\left( \sum_{k=1}^p \theta_{jk}^{-1} \|\PP^k F_j\|^2_{\HH} \right. \\
& \left. + \; \theta_{jkl}^{-1} \sum_{k=1,k\neq l}^p \sum_{l=1}^p \|\PP^{kl} F_j\|^2_\HH \right) + \kappa_{nj} \left( \sum_{k=1}^p\theta_{jk} + \sum_{k=1,k\neq l}^p \sum_{l=1}^p \theta_{jkl} \right), 
\end{split}
\end{align}
subject to $\theta_k\geq 0,\theta_{kl} \geq 0, k,l = 1, \ldots, p, k\neq l$, where $\thetabf_j = (\theta_{j1}, \ldots, \theta_{jp}, \theta_{j12}, \ldots, \theta_{j1p}, \ldots, \theta_{jp1},$ $\ldots, \theta_{jp(p-1)})^\top\in\R^{p^2}$ collects the parameters to estimate, and $\eta_{nj},\kappa_{nj\geq 0}$ are the tuning parameters, $j = 1, \ldots, p$. Comparing \eqref{eqn:kodeequiv} to \eqref{eqn:kode}, we introduce the parameters $\theta_{jk}$ and $\theta_{jkl}$ to control the sparsity of the main effect and interaction terms in $F_j$. This is similar to \citet{LinZhang2006}. The two optimization problems \eqref{eqn:kode} and \eqref{eqn:kodeequiv} are equivalent, in the following sense. Let $\kappa_{nj} = \tau_{nj}^2/(4\eta_{nj})$. Then we have, 
\begin{align*}
\eta_{nj}\theta_{jk}^{-1}\|\PP^k F_j\|_\HH^2+\kappa_{nj}\theta_{jk}\geq 2\eta_{nj}^{1/2}\kappa_{nj}^{1/2}\|\PP^kF_j\|_\HH = \tau_{nj}\|\PP^kF_j\|_\HH,
\end{align*}
where the equality holds  if $\theta_{jk}= \eta_{nj}^{1/2}\kappa_{nj}^{-1/2}\|P^kF_j\|_\HH$. A similar result holds for $\theta_{jkl}= \eta_{nj}^{1/2}\kappa_{nj}^{-1/2}\|P^{kl}F_j\|_\HH$. In other words, if $( \widehat\theta_{j0}, \widehat{F}_j )$ minimizes \eqref{eqn:kode}, then $(\widehat\theta_{j0}, \widehat{\thetabf}_j,\widehat{F}_j)$ minimizes \eqref{eqn:kodeequiv}, with $\widehat{\theta}_{jk} = \eta_{nj}^{1/2}\kappa_{nj}^{-1/2}\|\PP^k\widehat{F}_j\|_\HH$, and $\theta_{jkl}= \eta_{nj}^{1/2}\kappa_{nj}^{-1/2}\|\PP^{kl}F_j\|_\HH$, for any $k,l = 1, \ldots, p, k\neq l$. Meanwhile, if $(\widehat\theta_{j0}, \widehat{\thetabf}_j,\widehat{F}_j)$  minimizes \eqref{eqn:kodeequiv}, then $(\widehat\theta_{j0}, \widehat{F}_j)$ minimizes \eqref{eqn:kode}.

Next, we devise an iterative alternating optimization approach to solve \eqref{eqn:kodeequiv}. That is, we first estimate $\theta_{j0}$ given fixed $F_j$ and $\thetabf_j$, then estimate the functional $F_j$ given fixed $\theta_{j0}$ and $\thetabf_j$, and finally estimate $\thetabf_j$ given fixed $\theta_{j0}$ and $F_j$. 

For given $\widehat{F}_j$ and $\widehat{\thetabf}_j$, we have that, 
\begin{equation*}
\widehat{\theta}_{j0} = \bar{y}_j - \int_\TT\bar{T}(t)\widehat{F}_j(\widehat{\xbf}(t))dt,
\end{equation*}
where $T_i(t) = 1\{0\leq t\leq t_i\}$, $\bar{T}(t) = \frac{1}{n}\sum_{i=1}^nT_i(t)$, and $\bar{y}_j = n^{-1} \sum_{i=1}^ny_{ij}$.

For given $\widehat\theta_{j0}$ and $\widehat\thetabf_j$, the optimization problem \eqref{eqn:kodeequiv} becomes, 
\begin{equation} \label{eqn:solveF}
\begin{aligned}
&\underset{F_j}{\min} \left\{\frac{1}{n}\sum_{i=1}^n\left[(y_{ij} - \bar{y}_j)-\int_{\TT}\left\{ T_i(t)-\bar{T}(t) \right\} F_j(\widehat{\xbf}(t))dt\right]^2\right. \\
&\quad\quad\quad\quad\left.+ \eta_{nj} \left( \sum_{k=1}^p \widehat\theta_{jk}^{-1}\|\PP^k F_j\|^2_{\HH} + \widehat\theta_{jkl}^{-1}\sum_{k=1,k\neq l}^p\sum_{l=1}^p\|\PP^{kl} F_j\|^2_\HH \right) \right\}.
\end{aligned}
\end{equation}
Let $K_j(\cdot,\cdot):\XX\times\XX\mapsto\R$ denote the Mercer kernel generating the RKHS $\HH_j$, $j=1,\ldots,p$. Then $K_{kl}\equiv K_kK_l$ is the reproducing kernel of the RKHS $\HH_k\otimes \HH_l$. Let $K_{\thetabf_j} = \sum_{k=1}^p \widehat\theta_{jk}K_k + \sum_{k\neq l} \widehat\theta_{jkl} K_{kl}$. By the representer theorem \citep{wahba1990}, the solution $\widehat{F}_j$ to \eqref{eqn:solveF} is of the form,
\begin{equation} \label{eqn:hatfjxt}
\widehat{F}_j(\widehat{\xbf}(t)) = b_j + \sum_{i=1}^nc_{ij}\int_\TT K_{\thetabf_j}\left( \widehat{\xbf}(t),\widehat{\xbf}(s) \right) \left\{ T_i(s)-\bar{T}(s) \right\} ds
\end{equation}
for some $b_j \in \R$ and $\cbf_j = (c_{1j}, \ldots, c_{nj}) \in \R^n$. Write $\ybf_j = (y_{1j},\ldots,y_{nj})^\top \in \R^n$ and $\bar{\ybf}_j = (\bar{y}_{j},\ldots,\bar{y}_{j})^\top \in \R^n$. Let $\Bbf$ be an $n \times 1$ vector whose $i$th entry is $B_{i} = \int_\TT\{ T_i(t)-\bar{T}(t) \}dt$, $i=1, \ldots, n$. Let $\Sigmabf$ be an $n \times n$ matrix whose $(i,i')$th entry is $\Sigma_{ii'} = \int_\TT\int_\TT \{T_i(s)-\bar{T}(s)\} K_{\thetabf_j}(\widehat{\xbf}(t),\widehat{\xbf}(s))\{T_{i'}(t)-\bar{T}(t)\} ds dt$, $i, i' = 1, \ldots, n$. Plugging \eqref{eqn:hatfjxt} into \eqref{eqn:solveF}, we obtain the following quadratic minimization problem in terms of $\{b_j, \cbf_j\}$, 
\begin{equation*}
\underset{b_j, \cbf_j}{\min}
\frac{1}{n}\|(\ybf_j - \bar\ybf_j) - (\Bbf b_j + \Sigmabf \cbf_j)\|_{2}^2 + \eta_{nj} \cbf_j^\top \Sigmabf \cbf_j,
\end{equation*}
which has a closed-form solution. Consider the QR decomposition $B = [Q_1 \ Q_2][R \ 0]^{\top}$, where $Q_1\in\R^{n\times 1}$, $Q_2\in\R^{n\times(n-1)}$, and $[Q_1\ Q_2]$ is orthogonal such that $B^\top Q_2 = 0_{1\times (n-1)}$.  Write $W_j = \Sigma+n\eta_{nj} I_n$, where $I_n$ is the $n \times n$ identity matrix. Then the minimizers are,
\begin{equation*}
\begin{aligned}
\cbf_j & = Q_2(Q_2^\top W_j Q_2)^{-1}Q_2^\top(\ybf_j-\bar{\ybf}_j),\\
 b_j & = R^{-1}Q_1^\top(\ybf_j-\bar{\ybf}_j - W_j \cbf_j).
\end{aligned}
\end{equation*}
Following the usual smoothing splines literature, we tune the parameter $\eta_{nj}$ in \eqref{eqn:solveF} by minimizing the generalized cross-validation criterion \citep[GCV,][]{Wahba1995}, 
\begin{equation*}
\text{GCV} =  \frac{\|A_j(\eta_{nj})(\ybf_j-\bar{\ybf}_j) - (\ybf_j-\bar{\ybf}_j)\|^2}{[n^{-1}\text{tr}\{I_n - A_j(\eta_{nj})\}]^2}, 
\end{equation*}
where the smoothing matrix $A_j(\eta_{nj}) \in \R^{n \times n}$ is of the form, 
\begin{equation} \label{eqn:smoothmatrix}
A_j(\eta_{nj}) = I_n - n \eta_{nj} Q_2(Q_2^\top W_jQ_2)^{-1}Q_2^\top.
\end{equation} 

For given $\widehat\theta_{j0}$ and $\widehat{F}_j$, $\thetabf_j$ is the solution to a usual $\ell_1$-penalized regression problem,
\begin{equation} \label{eqn:regfortheta}
\begin{aligned}
\min_{\thetabf_j}\left\{(\zbf_j - G\thetabf_j)^\top (\zbf_j - G\thetabf_j) + n \kappa_{nj} \left( \sum_{k=1}^p\theta_{jk}+\sum_{k\neq l, k=1}^p\sum_{l=1}^{p}\theta_{jkl} \right)\right\},
\end{aligned}
\end{equation}
subject to $\theta_k\geq 0,\theta_{kl} \geq 0, k,l = 1, \ldots, p, k\neq l$, where the ``response" is $\zbf_j = (\ybf_j - \bar{\ybf}_j) - (1/2)n\eta_{nj} \cbf_j - Bb_j$, the ``predictor" is $G \in \R^{n \times p^2}$, whose first $p$ columns are $\Sigmabf^k \cbf_j$ with $k=1,\ldots,p$, and the last $p(p-1)$ columns are $\Sigma^{kl} \cbf_j$ with $k,l=1,\ldots,p, k\neq l$, and $\Sigma^k = (\Sigma^k_{ii'}), \Sigma^{kl} = (\Sigma^{kl}_{ii'})$ are both $n\times n$ matrices whose $(i,i')$th entries are $\Sigma^k_{ii'} = \int_\TT\int_\TT\{ T_i(s)-\bar{T}(s) \} K_{k}(\widehat{\xbf}(t),\widehat{\xbf}(s)) \{ T_{i'}(t)-\bar{T}(t) \} dsdt$, and $\Sigma^{kl}_{ii'} = \int_\TT\int_\TT\{ T_i(s)-\bar{T}(s) \} K_{kl}(\widehat{\xbf}(t),\widehat{\xbf}(s)) \{ T_{i'}(t)-\bar{T}(t) \} dsdt$, respectively, where $i, i'=1,\ldots,n, j=1,\ldots,p$. We employ Lasso for \eqref{eqn:regfortheta} in our implementation, and tune the parameter $\kappa_{nj}$ using tenfold cross-validation, following the usual Lasso literature. 

We repeat the above optimization steps iteratively until some stopping criterion is met; i.e., when the estimates in two consecutive iterations are close enough, or when the number of iterations reaches some maximum number. In our simulations, we have found that the algorithm converges quickly, usually within 10 iterations. Another issue is the identifiability of $\PP^k F_j$'s and $\PP^{kl} F_j$'s in \eqref{eqn:kodeequiv} in the sense of unique solutions. We introduce the collinearity indices $\mathcal C_{jk}$ and $\mathcal C_{jkl}$ to reflect the identifiability. Specifically, let $\WW$ denote a $p^2 \times p^2$ matrix, whose entries are $\cos(\PP^{k} F_j, \PP^{k'} F_j), \cos(\PP^{k} F_j, \PP^{k'l'} F_j), \cos(\PP^{kl} F_j, \PP^{k'} F_j), \cos(\PP^{kl} F_j, \PP^{k'l'} F_j)$, $j,k,l=1,\ldots,p$. Then $\mathcal C_{jk}^2$ and $\mathcal C_{jkl}^2$ are defined by the diagonals of $\WW^{-1}$. When some $\mathcal C_{jk}$ and $\mathcal C_{jkl}$ are much larger than one, then the  identifiability issue occurs \citep{gu2013}. This is often due to insufficient amount of data relative to the complexity of the model we fit. In this case, we find that increasing $\eta_{nj}$ and $\kappa_{nj}$ in \eqref{eqn:kodeequiv} often helps with the identifiability issue, as it helps reduce the model complexity. 

We summarize the above estimation procedure in Algorithm \ref{alg:trainofkode}. 

\begin{algorithm}[t!]
\caption{Iterative optimization algorithm for kernel ODE.} 
\begin{algorithmic}[1]
\STATE Initialization: the initial values for $\theta_{jk}=\theta_{jkl}=1$, $j,k,l=1,\ldots,p, k\neq l$, and the tuning parameters: $(\eta_{nj},\kappa_{nj})$.
\STATE Fit smoothing spline model \eqref{eqn:sshatxj}, and obtain $\widehat{x}_j(t)$, $j=1,\ldots,p$.
\REPEAT  
\STATE Solve $\widehat{\theta}_{j0}$ given $\widehat{F}_j$ and $\widehat{\thetabf}_j$, $j=1,\ldots,p$.
\STATE Solve $\widehat{F}_j$ in \eqref{eqn:solveF} given $\widehat\theta_{j0}$ and $\widehat\thetabf_j$, $j=1,\ldots,p$.
\STATE Solve $\widehat{\thetabf}_j$ in \eqref{eqn:regfortheta} given $\widehat\theta_{j0}$ and $\widehat{F}_j$, $j=1,\ldots,p$. 
\UNTIL{the stopping criterion is met.}
\end{algorithmic} 
\label{alg:trainofkode}
\end{algorithm}

\subsection{Confidence intervals}
\label{sec:bayesianci}

\noindent 
Next, we derive the confidence intervals for the estimated trajectory $\widehat{x}_j(t_i)$. This is related to post-selection inference, as the actual coverage probability of the confidence interval ignoring the preceding sparse estimation uncertainty can be dramatically smaller than the nominal level. Our result extends the recent work of \citet{Berk2013, Bachoc2019} from linear regression models to nonparametric ODE models, while our setting is more challenging, as it involves infinite-dimensional functional objects. 

Let $\widehat\thetabf_j$ denote the estimator of $\thetabf_j$ obtained from Algorithm \ref{alg:trainofkode}. Denote $\mathcal M \equiv \{1, \ldots, p,$ $(1,2), \ldots, (1,p), \ldots, (p,1), \ldots, (p,p-1)\}$, and denote $M_j \subseteq \mathcal M$ as the index set of the nonzero entries of the sparse estimator $\widehat\thetabf_j$. Note that $M_j$ is allowed to be an empty set. Let $\widehat{\thetabf}_{M_j}$ be the least squares estimate with $M_j$ as the support that minimizes the unpenalized objective function in \eqref{eqn:regfortheta}, i.e., $(\zbf_j - G \thetabf_j)^\top (\zbf_j - G \thetabf_j)$. Plugging this estimate $\widehat{\thetabf}_{M_j}$ into \eqref{eqn:hatfjxt} gets the corresponding estimate of the functional $F_j$ as, 
\begin{equation*}
\widehat{F}_{j,\widehat{\thetabf}_{M_j}}(\widehat{\xbf}(t)) = b_j + \sum_{i=1}^nc_{ij}\int_\TT K_{\widehat{\thetabf}_{M_j}}(\widehat{\xbf},\widehat{\xbf}(s)) \left\{ T_i(s)-\bar{T}(s) \right\}ds.
\end{equation*}
For a nominal level $\alpha \in (0,1)$ and $i=1,\ldots,n$, define $c_0(\widehat{\xbf}_j(t_i))$ as the smallest constant satisfying that, 
\begin{equation} \label{eqn:defofc0}
\P_{n,F_j,\sigma_j}\left[ \max_{M_j\subseteq\mathcal M} \sigma_j^{-1} \left| \{\widetilde{A}_{M_j}\}_{i\cdot} ( \ybf_j -\bar{y}_j ) \right| \leq c_0(\widehat{\xbf}_j(t_i)) \right] \geq 1-\alpha,
\end{equation}
where $\{\widetilde{A}_{M_j}\}_{i\cdot} = \{A_{M_j}\}_{i\cdot}/\|\{A_{M_j}\}_{i\cdot}\|_{l_2}$, $\{A_{M_j}\}_{i\cdot}$ is the $i$th row of $A_{M_j}$, $A_{M_j}$ is the smoothing matrix as defined in \eqref{eqn:smoothmatrix} with the corresponding $\widehat{\thetabf}_{M_j}$, and $\sigma_j^2$ is the variance of the error term $\epsilon_{ij}$ in \eqref{eqn:obserdata}. We then construct the confidence interval CI$(\widehat{\xbf}_j(t))$ for the prediction of true trajectory $x_j(t)$ following model selection as, 
\begin{equation} \label{eqn:funcci}
\text{CI}(\widehat{\xbf}_j(t_i)) = \int_\TT \left\{ T_i(t)-\bar{T}(t) \right\} \widehat{F}_{j,\widehat{\thetabf}_{M_j}}(\widehat{\xbf}(t))dt \; \pm \; c_0(\widehat{\xbf}_j(t_i))\sigma_j\|\{A_{M_j}\}_{i\cdot}\|,
\end{equation}
for any $i=1,\ldots,n$ and $j=1,\ldots,p$.

Next, we show that the confidence interval in \eqref{eqn:funcci} has the desired coverage probability. Later we develop a procedure to estimate the cutoff value $c_0(\widehat{\xbf}_j)$ in \eqref{eqn:defofc0} given the data. 

\begin{theorem} \label{thm:postselection}
Let $M_j \subseteq \mathcal M$ be the index set of the nonzero entries of the sparse estimator $\widehat\thetabf_j$. Then the choice of $c_0(\widehat{\xbf}_j(t_i))$ in \eqref{eqn:defofc0} does not depend on $F_j$, and CI$(\widehat{\xbf}_j(t_i))$ in \eqref{eqn:funcci} satisfies the coverage property, for any $i=1,\ldots,n$ and $j=1,\ldots,p$, in that, 
\begin{equation*}
\inf_{F_j\in\HH,\sigma_j>0} \P \left\{ \int_\TT \left\{ T_i(t)-\bar{T}(t) \right\} \E\left[\widehat{F}_{j,\widehat{\thetabf}_{M_j}}(\widehat{\xbf}(t))\right]dt\in \text{CI}(\widehat{\xbf}_j(t_i)) \right\} \geq 1- \alpha.
\end{equation*}
\end{theorem}

\noindent
A few remarks are in order. First, the coverage in Theorem \ref{thm:postselection} is guaranteed for all sparse estimation and selection procedures. As such, CI$(\widehat{\xbf}_j)$ in \eqref{eqn:funcci}, following the terminology of \citet{Berk2013}, is a universally valid post-selection confidence interval. Second, if we replace $c_0(\widehat{\xbf}_j(t_i))$ in \eqref{eqn:funcci} by $z_{\alpha/2}$, i.e., the $\alpha/2$ cutoff value of a standard normal distribution, then CI$(\widehat{\xbf}_j(t_i))$ reduces to the ``naive" confidence interval. It is constructed as if $M_j$ were fixed a priori, and it ignores any uncertainty or error of the sparse estimation step.  This naive confidence interval, however, does not have the coverage property as in Theorem \ref{thm:postselection}, and thus is not a truly valid confidence interval. Finally, data splitting is a commonly used alternative strategy for post-selection inference. But it is not directly applicable in our ODE setting, because it is difficult to split the time series data into independent parts.

Next, we devise a procedure to compute the cutoff value $c_0(\widehat{\xbf}_j(t_i))$. 
\begin{proposition}
\label{lem:findingc0}
The value $c_0(\widehat{\xbf}_j(t_i))$ in (\ref{eqn:defofc0}) is the same as the solution of $t\geq 0$ satisfying, 
\begin{equation*}
\E_U \P \left.\left( \max_{M_j\subseteq\mathcal M} \left| \{\widetilde{A}_{M_j}\}_{i\cdot}V \right| \leq t / U \right | U \right) = 1-\alpha,
\end{equation*}
where $V$ is uniformly distributed on the unit sphere in $\mathbb R^n$, and $U$ is a nonnegative random variable such that $U^2$ follows a chi-squared distribution $\chi^2(n)$.
\end{proposition}

\noindent
Following Proposition \ref{lem:findingc0}, we compute $c_0(\widehat{\xbf}_j(t_i))$ as follows. We first generate $N$ i.i.d.\ copies of random vectors $V_1,\ldots,V_N$ uniformly distributed on the unit sphere in $\mathbb R^n$. We then calculate the quantity, $c_\nu = \max_{M_j\subseteq\mathcal M}|\{\widetilde{A}_{M_j}\}_{i\cdot}V_\nu|$ for $\nu=1,\ldots,N$. Let $D_U$ denote the cumulative distribution function of $U$, and $D_{\chi^2}$ denote the cumulative distribution function of a $\chi^2(n)$ distribution. Then $D_U(t) = D_{\chi^2}(t^2)$. We next obtain $c_0(\widehat{\xbf}_j(t_i))$ by searching for $c$ that solves $N^{-1} \sum_{i=1}^N D_{U}(c/c_i) = 1-\alpha$, using, for example, a bisection searching method. 

Finally, we estimate the error variance $\sigma^2_j$ in \eqref{eqn:funcci} using the usual noise estimator in the context of RKHS \citep{wahba1990}; i.e., $\widehat{\sigma}_j^2 = \| A_{M_j}(\ybf_j - \bar\ybf_j) - (\ybf_j - \bar\ybf_j) \|^2 / \textrm{tr}(I-A_{M_j})$. 

We also remark that, the inference on the prediction of the trajectory $x_j(t)$ following model selection as described in Theorem 2 amounts to the inference on the estimation of the integration $\int_0^t F_j(x(s))ds$. This type of inference is of great importance in dynamic systems \citep{Izhikevich2007, Chou2009, Ma2009}. Our solution takes the selected model as an approximation to the truth, but does not require that  the true data generation model has to be among the candidates of model selection. We note that, it is also possible to do inference on the individual components of $F_j$ directly; e.g., one could construct the confidence interval for $F_{jk}$ in \eqref{eqn:nonadditivemodel}. But this is achieved at the cost of imposing additional assumptions, including the requirement that the true data generation model is among the class of pairwise interaction model as in \eqref{eqn:nonadditivemodel}, and the orthogonality property as in \citet{Chernozhukov2015}, or its equivalent characterization as in \citet{Zhang2014CI, Javanmard2014}. For nonparametric kernel estimators, the orthogonality property is shown to hold if the covariates $x_j$'s are assumed to be weakly dependent \citep{Lu2020}. It is interesting to further investigate if such a property holds in the context of kernel ODE model under a similar condition of weakly dependent covariates. We leave this as our future research.

\section{Theoretical Properties}
\label{sec:theory}

\noindent
We next establish the estimation optimality and selection consistency of kernel ODE. These theoretical results hold for both the low-dimensional and high-dimensional settings, where the number of functionals $p$ can be smaller or larger than the sample size $n$. We first introduce two assumptions. 

\begin{assumption}
\label{assump:complexity}
The number of nonzero functional components is bounded, i.e., $\text{card}\big( \{k: F_{jk}\neq 0\}\cup\{1\le l\neq k\le p: F_{jkl}\neq 0\} \big)$ is bounded for any $j=1,\ldots,p$.
\end{assumption}

\begin{assumption}
\label{assump:fluctuation}
For any $F_j\in\HH$, there exists a random variable $B$, with $\mathbb E(B)<\infty$, and 
\begin{equation*}
\left|\frac{\partial F_j(x)}{\partial x_k}\right| \leq B \|F_j\|_{L_2}, \; \textrm{almost surely}. 
\end{equation*}
\end{assumption}

\noindent 
Assumption \ref{assump:complexity} concerns the complexity of the functionals. Similar assumptions have been adopted in the sparse additive model over RKHS when $F_{jkl} = 0$ \citep[see, e.g.,][]{Koltchinskii2010, Raskutti2011}. Assumption \ref{assump:fluctuation} is an inverse Poincar\'e inequality type condition, which places regularization on the fluctuation in $F_j$ relative to the $\ell_2$-norm. The same assumption was also used in additive models in RKHS \citep{Zhu2014}.

We begin with the error bound for the estimated trajectory $\widehat{\xbf}(t)$ uniformly for $j=1,\ldots,p$. This is a relatively standard result, which is needed for both analyzing the error of the functional estimators in kernel ODE, and establishing the selection consistency later.

\begin{theorem}[Optimal estimation of the trajectory]
\label{thm:optimalestofpredictor}
Suppose that $x_j(t) \in \mathcal F$, $j=1,\ldots,p$, and the RKHS $\mathcal F$ is embedded to a $\beta_1$th-order Sobolev space, $\beta_1>1/2$. Then the smoothing spline estimate from \eqref{eqn:sshatxj} satisfies that, for any $j=1, \ldots, p$,
\begin{equation*}
\min_{\lambda_{nj}\geq 0}\int_{\TT} \left\{ \widehat{x}_j(t)-x_j(t) \right\}^2dt = O_p\left( n^{-\frac{2\beta_1}{2\beta_1+1}} \right), 
\end{equation*}
which achieves the minimax optimal rate.
\end{theorem}

Next, we derive the convergence rate for the estimated functional $F_j$. Because the trajectory $\widehat{\xbf}$ is estimated, to establish the optimal rate of convergence, it requires extra theoretical attention, which is related to recent work on errors in variables for lasso-type regressions \citep{loh2012, Zhu2014}. The proof involves several tools for the Rademacher processes \citep{Vandevaart1996}, and the concentration inequalities for empirical processes \citep{Talagrand1996, Yuan2016}. 

\begin{theorem}[Optimal estimation of the functional]
\label{thm:optimalestoffunctional}
Suppose that $F_j \in \mathcal H$, $j=1,\ldots,p$, where $\HH$ satisfies \eqref{eqn:spaceH}, and the RKHS $\HH_j$ is embedded to a $\beta_2$th-order Sobolev space, $\beta_2>1$. Suppose Assumptions \ref{assump:complexity} and \ref{assump:fluctuation} hold.
Then, as long as $F_j$ is not a constant function, the KODE estimate $\widehat{F}_j$ from \eqref{eqn:kode} satisfies that, for any $j=1, \ldots, p$,
\begin{equation*}
\min_{\tau_{nj}\geq 0}\int_\TT \left\{ \widehat{F}_j(x(t))-F_j(x(t)) \right\}^2dt= O_p\left( \left(\frac{n}{\log n}\right)^{-\frac{2\beta_2}{2\beta_2+1}} + \frac{\log p}{n} + n^{-\frac{2\beta_1}{2\beta_1+1}} \right),  
\end{equation*}
which achieves the minimax optimal rate. 
\end{theorem}

\noindent
This theorem is one of our key results, and we make a few remarks. First, there are three error terms in Theorem \ref{thm:optimalestoffunctional}, which are attributed to the estimation of the interactions, the Lasso estimation, and the measurement errors in variables, respectively. Particularly, the error term $O_p\left( n^{-2\beta_1/(2\beta_1+1)} \right)$ arises due to the unobserved $x(t)$, which is instead measured at discrete time points and is subject to measurement errors. Since this error term achieves the optimal rate, it fully characterizes the influence of the estimated $\widehat{x}(t)$ on the resulting estimator $\widehat{F}_j$. Moreover, $\beta_1$ and $\beta_2$ measure the orders of smoothness for estimating $x_j$ and $F_j$, respectively. They can be different, which makes it flexible when choosing kernels for the estimation procedure. For instance, if there is prior knowledge that  $x(t)$ is smooth, we may then choose $\beta_1>\beta_2$, and the resulting estimator $\widehat{F}_j$ achieves a convergence rate of $O_p\left( (n/\log n)^{-2\beta_2/(2\beta_2+1)} + \log p / n \right)$. It is interesting to note that this rate is the same as the rate as if $x(t)$ were directly observed and there were no integral involved in the loss function, for example, in the setting of \citet{LinZhang2006}.

Second, there exists a regime-switching phenomenon, depending on the dimensionality $p$ with respect to the sample size $n$. On one hand, if it is an ultrahigh-dimensional setting, i.e., $p > \exp\left[ \left\{ n(\log n)^{2\beta_2} \right\}^{\frac{1}{2\beta_2+1}} \right]$, then 
the minimax optimal rate in Theorem \ref{thm:optimalestoffunctional} becomes $O_p\left( \log p / n + n^{-2\beta_1/(2\beta_1+1)} \right)$. Here, the first rate $O_p(\log p / n)$ matches with the minimax optimal rate for estimating a $p$-dimensional linear regression when the vector of regression coefficients has a bounded number of nonzero entries \citep{Raskutti2011}. Hence, we pay no extra price in terms of the rate of convergence for adopting a nonparametric modeling of  $F_j$ in \eqref{eqn:nonadditivemodel}, when compared with the more restrictive linear ODE model in \eqref{eqn:bilinearmodel} \citep{Zhang2015}. On the other hand, if it is a low-dimensional setting, i.e., $p \leq \exp\left[ \left\{ n(\log n)^{2\beta_2} \right\}^{\frac{1}{2\beta_2+1}} \right]$, then the optimal rate becomes $O_p\left( (n/\log n)^{-2\beta_2/(2\beta_2+1)} + \right.$ $\left. n^{-2\beta_1/ (2\beta_1+1)} \right)$. Here, the first rate $O_p\left( (n/\log n)^{-2\beta_2/(2\beta_2+1)} \right)$ is the same as the optimal rate of estimating $F_j$ as if we knew a priori that $F_j$ comes from a two-dimensional tensor product functional space, rather than the $p$-variate functional space $\HH$ in (\ref{eqn:spaceH}); see also \citet{Lin2000} for a similar observation. 

Third, the optimal rate in Theorem \ref{thm:optimalestoffunctional} is immune to the ``curse of dimensionality", in the following sense. We introduce $p(p-1)$ pairwise interaction components to $\HH$ in \eqref{eqn:spaceH}, and henceforth, for each $x_j(t)$, $j=1,\ldots,p$, it requires to estimate a total of $p^2$ functions. A direct application of an existing basis expansion approach, for instance, \citet{Brunton2016}, leads to a rate of $O_p\left( n^{-O(1/p^2)} \right)$. This rate degrades fast when $p$ increases. By contrast, we proceed in a different way, where we simultaneously aim for the flexibility of a nonparametric ODE model by letting $\HH$ obey a tensor product structure as in \eqref{eqn:spaceH}, while exploiting the interaction structure of the system. As a result, our optimal error bound $O_p\left( (n/\log n)^{-2\beta_2/(2\beta_2+1)} \right)$ does not depend on the dimensionality $p$.

Lastly, the incorporation of the integral, $\int_0^{t_i}F_j(\widehat{\xbf}(t))dt$, in the loss function in \eqref{eqn:kode} makes the estimation error of $\widehat{F}_j$ depend on the convergence of $\E\int_\TT\{ \widehat{x}_j(t)-x_j(t) \}^2dt$. As a comparison, if we use the derivative instead of the integration, then the estimation error would depend on the convergence of the derivative, $\E\int_\TT\{ d\widehat{x}_j(t)/dt-dx_j(t)/dt \}^2dt$ \citep{Wu2014}. However, it is known that the derivative estimation in the reproducing kernel Hilbert space has a slower convergence rate than the function estimation \citep{Cox1983}. That is, $\E\int_\TT\{ d\widehat{x}_j(t)/dt-dx_j(t)/dt \}^2dt$ converges at a slower rate than $\E\int_\TT\{ \widehat{x}_j(t)-x_j(t) \}^2dt$. This demonstrates the advantage of working with the integral in our KODE formulation, and our result echos the observation for the additive ODE model \citep{Chen2017}. 

Next, we establish the selection consistency of KODE. Putting all the functionals $\{F_1, \ldots,$ $F_p\}$ together forms a network of regulatory relations among the $p$ variables $\{x_1(t),$ $\ldots, x_p(t)\}$. Recall that, we say $x_k$ is a regulator of $x_j$, if in \eqref{eqn:nonadditivemodel} $F_{jk}$ is nonzero, or if $F_{jkl}$ is nonzero for some $l \neq k$. Denote the set of the true regulators and the estimated regulators of $x_j(t)$ by 
\begin{align*}
S^0_j & = \big\{1 \leq k \leq p : F_{jk}\neq 0, \textrm{ or } F_{jkl}\neq 0 \textrm{ for some } 1 \le l \neq k \le p \big\}, \\
\widehat{S}_j & = \big\{ 1 \leq k \leq p: \|\widehat{F}_{jk}\|_{\HH}\neq 0, \text{ or } \|\widehat{F}_{jkl}\|_\HH \neq 0 \textrm{ for some } 1 \le l \neq k \le p \big\},
\end{align*}
respectively, $j = 1, \ldots, p$. We need some extra regularity conditions on the minimum regulatory effect and the design matrix, which are commonly adopted in the literature of Lasso regression \citep{Zhao2006, Ravikumar2010}. In the interest of space, we defer those conditions to Section \ref{sec:condadd} of the Appendix. The next theorem establishes that KODE is able to recover the true regulatory network asymptotically.

\begin{theorem}[Recovery of the regulatory network]
\label{thm:optimalrecovery}
Suppose that $F_j \in \HH$, $j=1,\ldots,p$, where $\HH$ satisfies \eqref{eqn:spaceH}, and the RKHS $\HH_j$ is embedded to a $\beta_2$th-order Sobolev space, with $\beta_2>1$. Suppose Assumption \ref{assump:complexity}, and Assumptions \ref{network:dependency}, \ref{network:incoherence}, \ref{network:minregeffect} in the Appendix hold. Then, the KODE correctly recovers the true regulatory network, in that, for all $j = 1, \ldots, p$, 
\begin{eqnarray*}
\P\left( \widehat{S}_j = S^0_j \right) \to 1, \;\; \textrm{ as } \; n \to \infty.
\end{eqnarray*}
\end{theorem}

\section{Simulation Studies}
\label{sec:simulation}

\subsection{Setup}

\noindent 
We study the empirical performance of the proposed KODE using two ODE examples, the enzyme regulatory network in Section \ref{sec:enzymatic}, and the Lotka-Volterra equations in Section \ref{sec:lotkavolterra}. For a given system of ODEs and the initial condition, we obtain the numerical solutions of the ODEs using the Euler method with step size $0.01$. The data observations are drawn from the solutions at an evenly spaced time grid, with measurement errors. To implement KODE, we fit the smoothing spline to estimate $x_j(t)$ in \eqref{eqn:sshatxj} using a Mat\'{e}rn kernel, $K_\FF(x,x') = (1+\sqrt{3}\|x-x'\|/\nu)\exp(-\sqrt{3}\|x-x'\|/\nu)$, where the smoothing parameter $\lambda_{nj}$ is chosen by GCV, and the bandwidth $\nu$ is chosen by tenfold cross-validation. We compute the integral $\int_0^{t_i}F_j(\widehat{x}(t))dt$ in \eqref{eqn:kode} numerically with independent sets of $1000$ Monte Carlo points. We compare KODE with linear ODE with interactions in \eqref{eqn:bilinearmodel} \citep{Zhang2015}, and additive ODE in \eqref{eqn:additivemodel} \citep{Chen2017}. Due to the lack of available code online, we implement the two competing methods in the framework of Algorithm  \ref{alg:trainofkode}, using a linear kernel for \eqref{eqn:additivemodel}, and using an additive Mat\'{e}rn kernel for \eqref{eqn:additivemodel}. We evaluate the performance using the prediction error, plus the false discovery proportion and power for edge selection of the corresponding regulatory network. Furthermore, we compare with the family of ODE solutions assuming known $F$ \citep{zhang2015selection, mikkelsen2017learning} in Section \ref{asec:morecomparison} of the Appendix. We also carry out a sensitivity analysis in Section \ref{asec:sensitivity} of the Appendix to study the robustness of the choice of kernel function and initial parameters.

\subsection{Enzymatic regulatory network}
\label{sec:enzymatic}

\begin{figure}[b!]
\centering
\includegraphics[width=\textwidth]{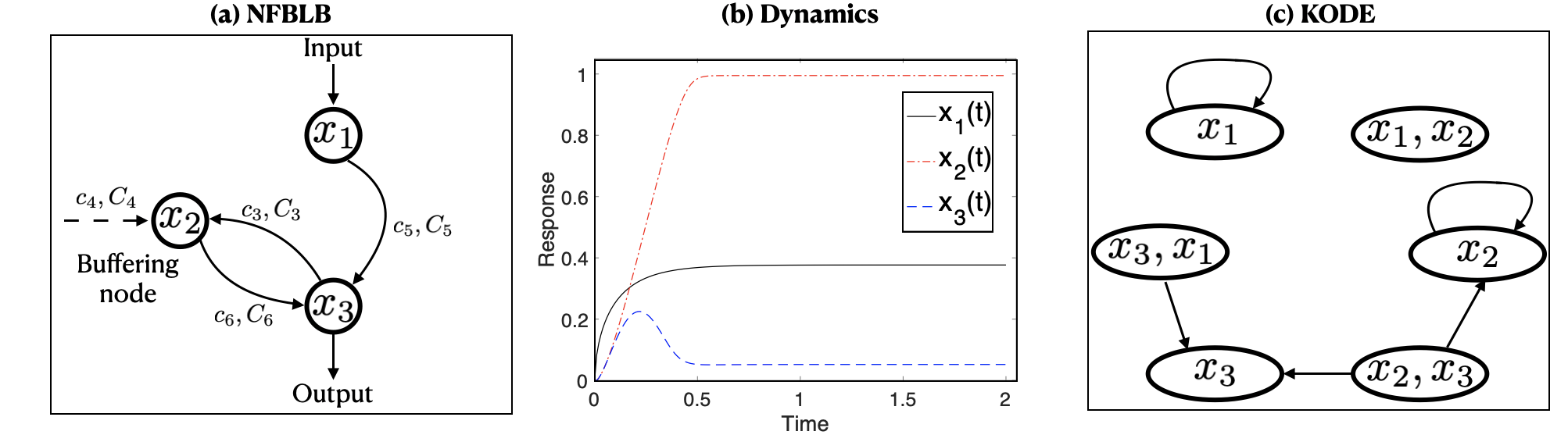}
\caption{(a) Diagram of the NFBLB regulatory network following \eqref{eqn:NFBLB}. (b) Phase dynamics for the three nodes $x_1,x_2,x_3$ over time $[0,1]$, with a random input $x_0$ uniformly drawn from $[0.5,1.5]$. (c) Illustration of  the NFBLB network in terms of the interactions in KODE.}
\label{fig:eg2_1}
\end{figure}

\noindent
The first example is a three-node enzyme regulatory network of a negative feedback loop with a buffering node \citep[NFBLB]{Ma2009}. The ODE system is given in \eqref{eqn:NFBLB} in Section \ref{sec:motivation}. Figure \ref{fig:eg2_1}(a) shows the NFBLB network diagram consisting of the three interacting nodes: $x_1$ receives the input, $x_3$ transmits the output, and $x_2$ plays a regulatory role, leading a negative regulatory link to $x_3$. We note that, although biological circuits can have more than three nodes, many of those circuits can be reduced to a three-node framework, given that multiple molecules often function as a single virtual node. Moreover, despite the diversity of possible network topologies, NFBLB is one of the two core three-node topologies that could perform adaption in the sense that the system resets itself after responding to a stimulus; see \citet{Ma2009} for more discussion of NFBLB.  For the ODE system in \eqref{eqn:NFBLB}, we set the catalytic rate parameters of the enzymes as $c_1=c_2=c_3=c_5=c_6=10, c_4=1$, the Michaelis-Menten constants as $C_1=\cdots=C_6=0.1$, and the concentration parameters of enzymes as $\tilde{c}_1 = 1, \tilde{c}_2 = 0.2$. These parameters achieve the adaption as shown in Figure \ref{fig:eg2_1}(b). The output node $x_3$ shows a strong initial response to the stimulus, and also exhibits strong adaption, since its post-stimulus steady state $x_3=0.052$ is close to the pre-stimulus state $x_3=0$. The input $x_0 \in \R^3$ is drawn uniformly from $[0.5,1.5]$, with the initial value $\xbf(0) = 0$, and the measurement errors are i.i.d.\ normal with mean zero and variance $\sigma^2_j$. The time points are evenly distributed,  $t_i=(i-1)/20, i=1,\ldots,n$. In this example, $p=3$, and for each function $x_j(t)$, $j=1,2,3$, there are $p^2 = 9$ functions to estimate, and in total there are $27$ functions to estimate under the sample size $n=40$.

\begin{figure}[t!]
\centering
\includegraphics[width=\textwidth, height=2.25in]{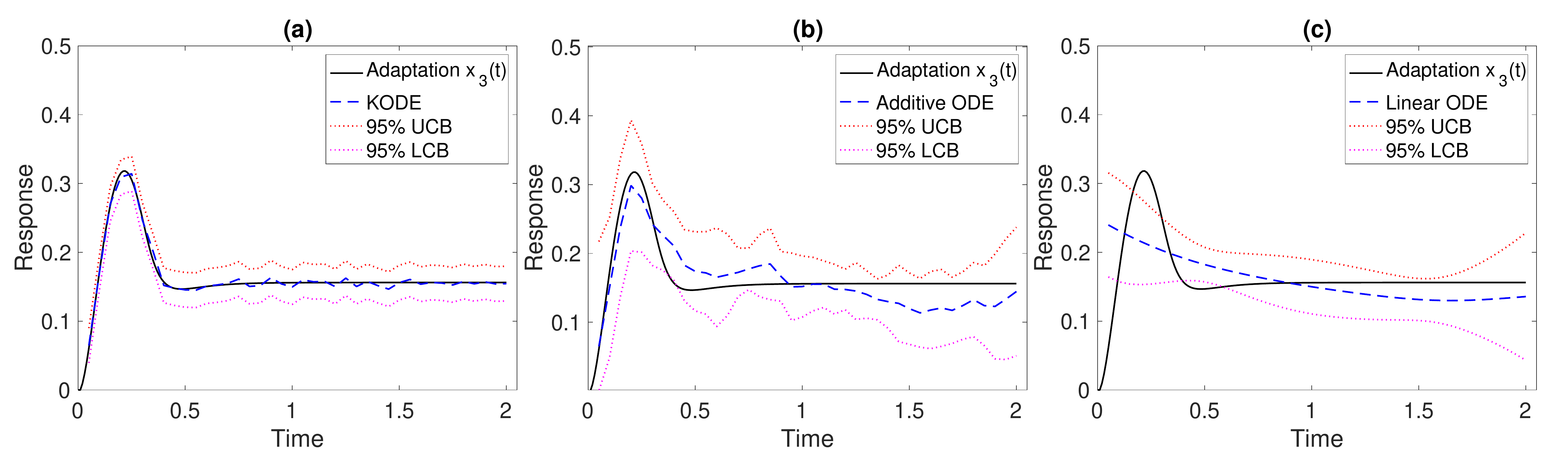}
\caption{The true (black solid line) and the estimated (blue dashed line) trajectory of $x_3(t)$, with the $95\%$ upper and lower confidence bounds (red dotted lines). The results are averaged over 500 data replications. (a) KODE; (b) Additive ODE; (c) Linear ODE.}
\label{fig:eg2_3}
\end{figure}

\begin{figure}[t!]
\centering
\includegraphics[width=\textwidth, height=2.25in]{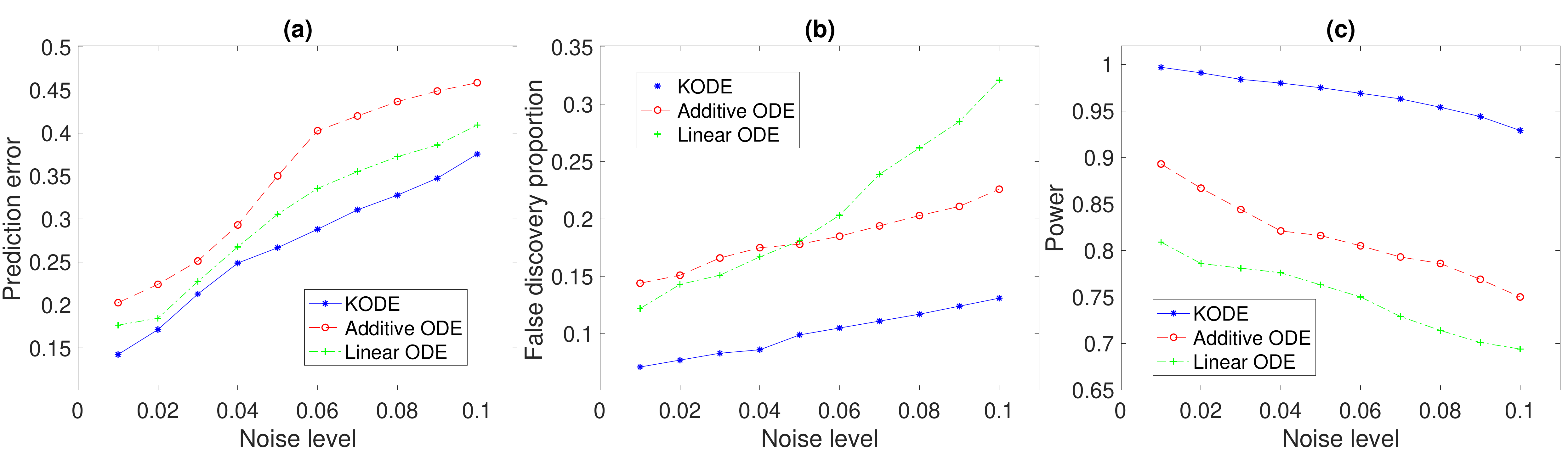}
\caption{The prediction and selection performance of three ODE methods with varying noise level. The results are averaged over 500 data replications. (a) Prediction error; (b) False discovery proportion; (c) Empirical power.}
\label{fig:eg2_2}
\end{figure}

Figure \ref{fig:eg2_3} reports the true and estimated trajectory of $x_3(t)$, with $95\%$ upper and lower confidence bounds, of the three ODE methods, where we use the tensor product Mat\'{e}rn kernel for KODE in \eqref{eqn:kode}. The noise level is set as $\sigma_j=0.1, j=1,2,3$, and the results are averaged over 500 data replications. It is seen that the KODE estimate has a smaller variance than the additive and  linear ODE estimates. Moreover, the confidence interval of KODE achieves the desired coverage for the true trajectory. In contrast, the confidence intervals of additive and linear ODE models mostly fail to include the truth. This is because there is a discrepancy between the additive and linear ODE model specifications and the true ODE model in \eqref{eqn:NFBLB}, and this discrepancy accumulates as the course of the ODE evolves. 

Figure \ref{fig:eg2_2} reports the prediction and selection performance of the three ODE methods, with varying noise level $\sigma_j \in \{0.01,0.02,\ldots,0.1\}, j=1,2,3$. The results are averaged over $500$ data replications. The prediction error is defined as the squared root of the sum of predictive mean squared errors for $x_1(t),x_2(t),x_{3}(t)$ at the unseen ``future" time point $t=2$. The false discovery proportion is defined as the proportion of falsely selected edges in the regulatory network out of the total number of edges. The empirical power is defined as the proportion of selected true edges in the network. It is seen that KODE clearly outperforms the two alternative solutions in both prediction and selection accuracy. Moreover, we report graphically the sparse recovery of this regulatory network in Section \ref{asec:enzymatic} of the Appendix.

\subsection{Lotka-Volterra equations}
\label{sec:lotkavolterra}

\begin{figure}[t!]
\centering
\includegraphics[width=\textwidth, height=2.25in]{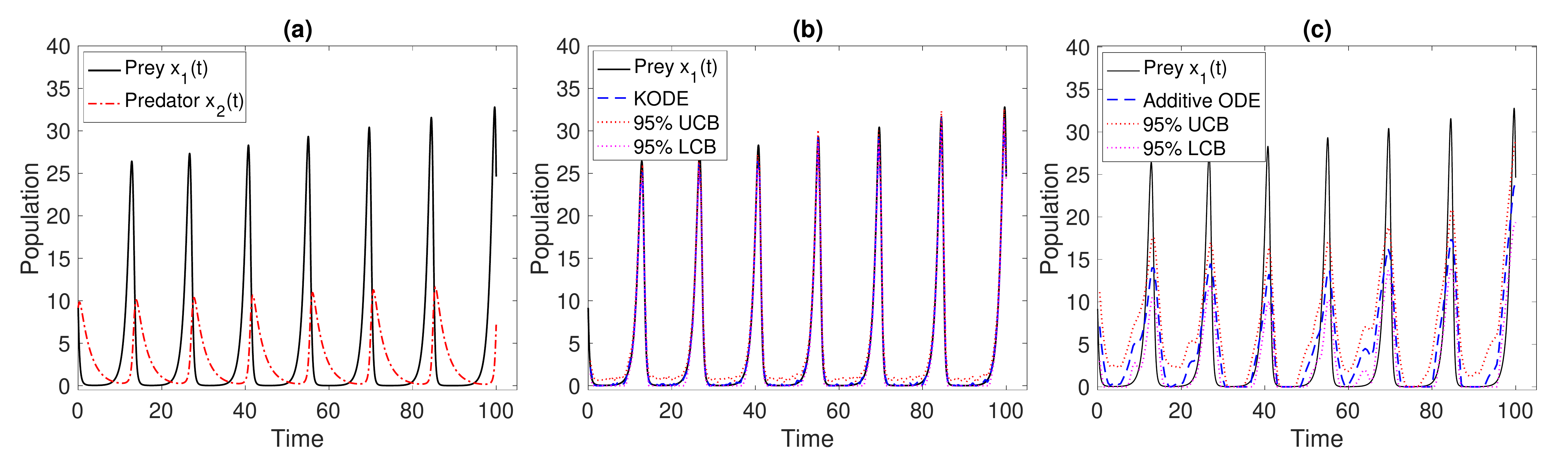}
\caption{(a) The true trajectories of the prey $x_1(t)$ and the predator $x_2(t)$. (b) The estimated trajectory $\widehat{x}_1(t)$ (blue dashed line), with the $95\%$ upper and lower confidence bounds (red dotted lines), by KODE. (c) By additive ODE. The results are averaged over 500 data replications.}
\label{fig:eg1_1}
\end{figure}

\noindent
The second example is the high-dimensional Lotka-Volterra equations, which are pairs of first-order nonlinear differential equations describing the dynamics of biological systems in which predators and prey interact \citep{Volterra1928}. We consider a ten-node system, 
\begin{equation}
\label{eqn:lotkavolterra}
\begin{aligned}
\frac{dx_{2j-1}(t)}{dt} & = &  \alpha_{1,j}x_{2j-1}(t)-\alpha_{2,j} x_{2j-1}(t)x_{2j}(t),  \\
\frac{dx_{2j}(t)}{dt} & = &  \alpha_{3,j} x_{2j-1}(t)x_{2j}(t)-\alpha_{4,j}x_{2j}(t),
\end{aligned}
\end{equation}
where $\alpha_{1,j} = 1.1+0.2(j-1), \alpha_{2,j} = 0.4+0.2(j-1), \alpha_{3,j} = 0.1+0.2(j-1)$, and $\alpha_{4,j} = 0.4+0.2(j-1)$, $j=1,\ldots,5$. The parameters $\alpha_{2,j}$ and $\alpha_{3,j}$ define the interaction between the two populations such that $dx_{2j-1}(t)/dt$ and $dx_{2j}(t)/dt$ are nonadditive functions of $x_{2j-1}$ and $x_{2j}$, where $x_{2j-1}$ is the prey and $x_{2j}$ is the predator. Figure \ref{fig:eg1_1}(a) shows an illustration of the interaction between $x_1(t)$ and $x_2(t)$. The input $x_0 \in \R^{10}$ is drawn uniformly from $[5,15]^{10}$, with the initial value $x_{2j-1}(0) = x_{2j}(0)$, and the measurement errors are i.i.d.\ normal $N(0,\sigma^2_j)$, where $\sigma_j$ again reflects the noise level. The time points are evenly distributed in $[0,100]$ with $n=200$. In this example, $p=10$, and for each function $x_j(t)$, $j=1,\ldots,10$, there are $p^2 = 100$ functions to estimate, and in total there are $1,000$ functions to estimate under the sample size $n=200$.

\begin{figure}[t!]
\centering
\includegraphics[width=\textwidth, height=2.25in]{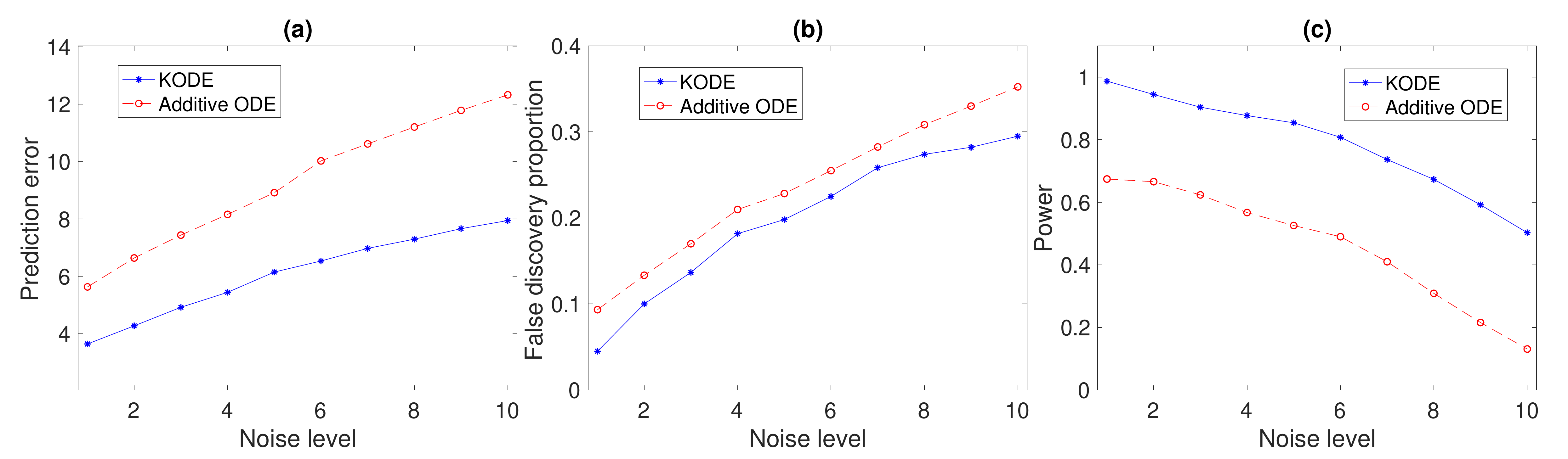}
\caption{The prediction and selection performance of two ODE methods with varying noise level. The results are averaged over 500 data replications. (a) Prediction error; (b) False discovery proportion; (c) Empirical power.}
\label{fig:eg1_2}
\end{figure}

Figure \ref{fig:eg1_1}(b) and (c) report the estimated trajectory of $x_1(t)$, with $95\%$ upper and lower confidence bounds, of KODE and additive ODE, where the noise level is set as $\sigma_j = 1, j=1,\ldots,10$. The confidence interval of KODE achieves a better empirical coverage for the true trajectory compared to that of additive ODE. For this example, we use the linear kernel for KODE in \eqref{eqn:kode}, since the functional forms in \eqref{eqn:lotkavolterra} are known to be linear. For this reason, we only compare KODE with the additive ODE method. Moreover, in the implementation, the estimates $\widehat{F}_j(\widehat{x}(t))$ are thresholded to be nonnegative to ensure the physical constraint that the number of population cannot be negative. Figure \ref{fig:eg1_2} reports the prediction and selection performance of the two ODE methods, with varying noise level $\sigma_j \in \{1,2,\ldots,10\}, j=1,\ldots,10$. All the results are averaged over 500 data replications. It is seen that the KODE estimate achieves a smaller prediction error, and a higher selection accuracy, since KODE allows flexible non-additive structures, which results in significantly smaller bias and variance in functional estimation as compared to the additive modeling.

\section{Application to Gene Regulatory Network}
\label{sec:application}

\begin{figure}[t!]
\centering
\begin{tabular}{c}
\includegraphics[width=\textwidth, height=2.25in]{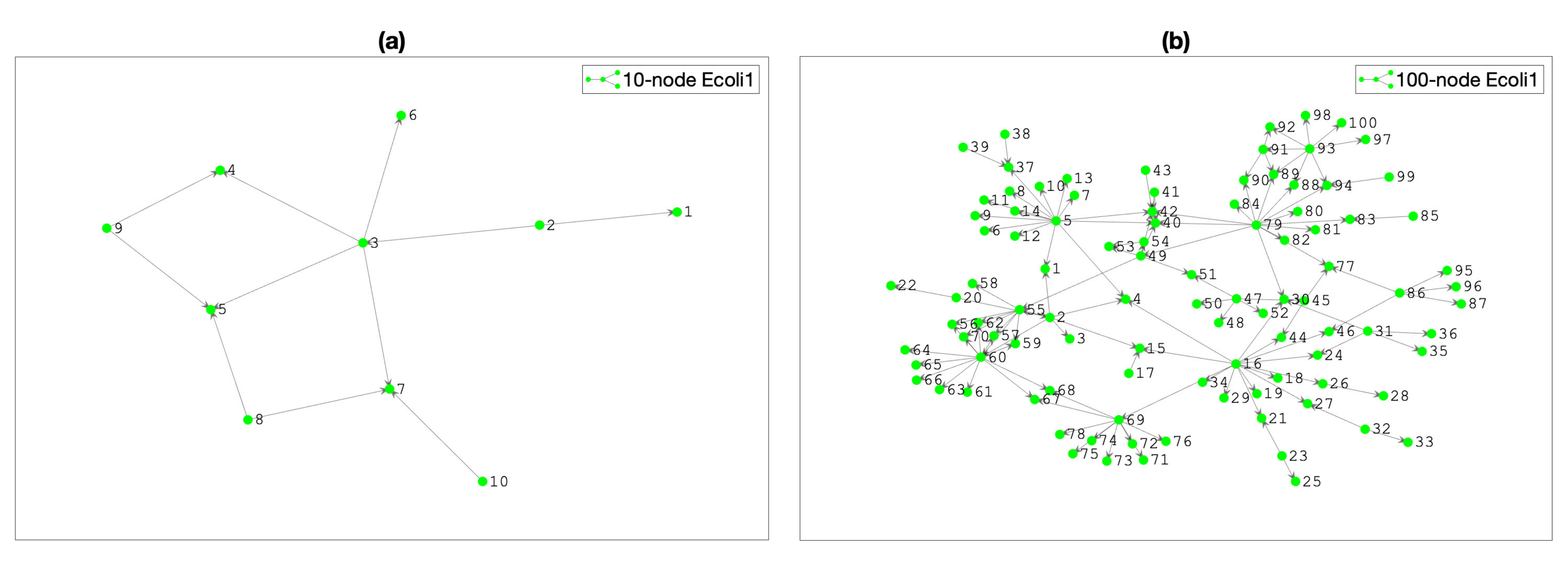} \\
\includegraphics[width=\textwidth]{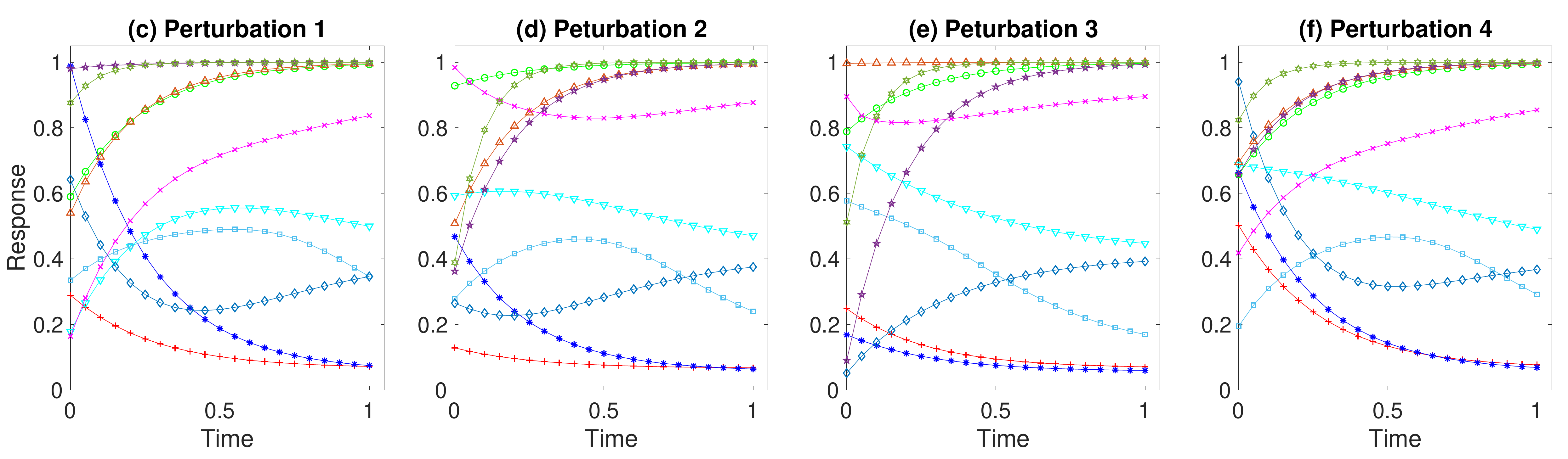}
\end{tabular}
\caption{(a) The 10-node \emph{E.coli1} network.  (b) The 100-node \emph{E.coli1} network. (c)-(f) Four perturbation experiments for the $10$-node \emph{E.coli1} network, where each experiment corresponds to a different initial condition of the ODE system.}
\label{fig:ecoli1}
\end{figure}

\noindent 
We illustrate KODE with a gene regulatory network application. \citet{Schaffter2011} developed an open-source platform called GeneNetWeaver (GNW) that generates in silico benchmark gene expression data using dynamical models of gene regulations and nonlinear ODEs. The generated data have been used for evaluating the performance of network inference methods in the DREAM3 competition \citep{Marbach2009}, and were also analyzed by \citet{HendersonMichailidis2014, Chen2017} in additive ODE modeling. GNW extracts two regulatory networks of \emph{E.coli} (\emph{E.coli1}, \emph{E.coli2}), and three regulatory networks of yeast (yeast1, yeast2, yeast 3), each of which has two dimensions, $p=10$ nodes and $p=100$ nodes. This yields totally 10 combinations of network structures. Figure \ref{fig:ecoli1}(a)-(b) show an example of the $10$-node and the $100$-node \emph{E.coli1} networks, respectively. The systems of ODEs for each extracted network are based on a thermodynamic approach, which leads to a non-additive and nonlinear ODE structure \citep{Marbach2010}. Besides, the network structures are sparse; e.g., for the $10$-node \emph{E.coli1} network, there are $11$ edges out of $90$ possible pairwise edges, and for the $100$-node \emph{E.coli1} network, there are $125$ edges out of $9,900$ possible pairwise edges. Moreover, for the $10$-node network, GNW provides $R=4$ perturbation experiments, and for the 100-node network, GNW provides $R=46$ experiments. In each experiment, GNW generates the time-course data with different initial conditions of the ODE system to emulate the diversity of gene expression trajectories \citep{Marbach2009}. Figure \ref{fig:ecoli1}(c)-(f) show the time-course data under $R=4$ experiments for the $10$-node \emph{E.coli1} network. All the trajectories are measured at $n=21$ evenly spaced time points in $[0,1]$. We add independent measurement errors from a normal distribution with mean zero and standard deviation $0.025$, which is the same as the DREAM3 competition and the data analysis done in \citet{HendersonMichailidis2014, Chen2017}.

The kernel ODE model we have developed focuses on a single experiment data, but it can be easily generalized to incorporate multiple experiments. Specifically, let $\big\{ y_{ij}^{(r)};i=1,\ldots,n, j=1,\ldots,p, r=1,\ldots,R \big\}$ denote the observed data from $n$ subjects for $p$ variables under $R$ experiments, with unknown initial conditions $\xbf^{(r)}(0) \in \R^p, r=1,\ldots,R$. Then we modify the KODE method in \eqref{eqn:sshatxj} and \eqref{eqn:kode}, by seeking $F_j\in\mathcal H$ and $\theta_{j0}\in\R$ that minimize 
\begin{equation}
\label{eqn:heterogeneousODEs}
\frac{1}{Rn}\sum_{r=1}^R\sum_{i=1}^n\left\{y^{(r)}_{ij} - \theta_{j0} - \int_0^{t_i}F_j(\widehat{\xbf}^{(r)}(t))dt\right\}^2+ \tau_{nj} \left( \sum_{k=1}^p\|\PP^k F_j\|_{\HH} + \sum_{k\neq l, k=1}^p \sum_{l=1}^p \|\PP^{kl} F_j\|_\HH \right),
\end{equation}
where $\widehat{\xbf}^{(r)}(t) = (\widehat{x}^{(r)}_1(t), \ldots, \widehat{x}^{(r)}_p(t))^\top$ is the smoothing spline estimate obtained by, 
\begin{equation*}
\widehat{x}^{(r)}_j(t) = \underset{z_j\in\mathcal F}{\arg\min} \left\{\frac{1}{n}\sum_{i=1}^n(y^{(r)}_{ij} - z_{j}(t_i))^2 + \lambda_{nj}\|z_j(t)\|_{\mathcal F}^2\right\}, \quad j=1,\ldots,p,\ r=1,\ldots,R.
\end{equation*}
Algorithm  \ref{alg:trainofkode} can be modified accordingly to work with multiple experiments.

\begin{table}[t!]
\centering
\caption{The area under the ROC curve, and the $95\%$ confidence interval, for 10 combinations of network structures from GNW. The results are averaged over $100$ data replications.}
\resizebox{\textwidth}{!}{
\begin{tabular}{ccccccc}
\toprule
& \multicolumn{3}{c}{$p=10$} & \multicolumn{3}{c}{$p=100$} \\ \cline{2-4} \cline{5-7}
& KODE & Additive ODE & Linear ODE & KODE & Additive ODE & Linear ODE \\ [0.2ex]
 \midrule
\emph{E.coli1} & $\textbf{0.582}$ & $0.541$ &   $0.460$ & $\textbf{0.711}$ & $0.677$ & $0.640$ \\
 & $(0.577, 0.587)$ & $(0.535, 0.547)$ & $(0.453, 0.467)$ & $(0.708, 0.714)$ & $(0.672, 0.682)$ & $(0.637, 0.643)$ \\ [0.2ex]
\emph{E.coli2} &  $\textbf{0.662}$ & $0.632$ &   $0.562$ & $\textbf{0.685}$ & $0.659$ &   $0.533$ \\
 & $(0.658, 0.666)$ & $(0.625, 0.639)$ & $(0.555, 0.569)$ & $(0.681, 0.689)$ & $(0.652, 0.666)$ & $(0.527, 0.539)$ \\ [0.2ex]
Yeast1 & $\textbf{0.603}$ & $0.541$ &   $0.436$ & $\textbf{0.619}$ & $0.589$ & $0.569$ \\
 & $(0.599, 0.607)$ & $(0.536, 0.546)$ & $(0.430, 0.442)$ & $(0.616, 0.622)$ & $(0.581, 0.597)$ & $(0.562, 0.576)$ \\ [0.2ex]
Yeast2 & $\textbf{0.599}$ & $0.562$ &   $0.536$ & $\textbf{0.606}$ & $0.588$ & $0.541$ \\
 & $(0.595, 0.603)$ & $(0.555, 0.570)$ & $(0.530, 0.542)$ & $(0.603, 0.609)$ & $(0.582, 0.594)$ & $(0.536, 0.546)$ \\ [0.2ex]
Yeast3 & $\textbf{0.612}$ & $0.569$ &   $0.487$ & $\textbf{0.621}$ & $0.613$ & $0.609$ \\
 & $(0.608, 0.616)$ & $(0.564, 0.573)$ & $(0.481, 0.493)$ & $(0.617, 0.625)$ & $(0.607, 0.619)$ & $(0.605, 0.613)$ \\ [0.2ex]
\bottomrule
\end{tabular}
}
\label{table:gnw}
\end{table}

We again compare KODE with the additive ODE \citep{Chen2017} and the linear ODE \citep{Zhang2015}, adopting the same implementation as in the simulations. Since we know the true edges of the generated gene regulatory networks, we use the area under the ROC curve (AUC) as the evaluation criterion. Table \ref{table:gnw} reports the results averaged over 100 data realizations for all ten combinations of network structures. It is clearly seen that KODE outperforms both alternative methods in all cases. We further report graphically the sparse recovery of the 10-node \emph{E.coli1} network in Section \ref{asec:genenet} of the Appendix.  This example shows that our proposed KODE is a competitive and useful tool for ODE modeling. In addition, it also shows that the proposed method can scale up and work with reasonably large networks. For instance, for the network with $p=100$ nodes, there are $p^2 = 10,000$ functions to estimate, and the sample size is $n=21$ with $R=46$ perturbations.

\section{Conclusion and Discussion}
\label{sec:conclusion}

\noindent
In this article, we have developed a new reproducing kernel-based approach for a general family of ODE models to learn a dynamic system from noisy time-course data. We employ sparsity regularization to select individual functionals and recover the underlying regulatory network, and we derive the post-selection confidence interval for the estimated signal trajectory. Our proposal is built upon but also extends the smoothing spline analysis of variance framework. We establish the theoretical properties of the method, while allowing the number of functionals to be either smaller or larger than the number of time points. 

In numerous scientific applications, ODE is often employed to understand the regulatory effects and causal mechanisms within a dynamic system under interventions. Our proposed KODE method can be applied for this very purpose. There are different formulations of causal modeling for dynamic systems in the literature. We next consider and illustrate with two relatively common scenarios, one regarding dynamic causal modeling under experimental stimuli \citep{friston2003dynamic}, and the other about kinetic modeling that is invariant across heterogeneous experiments \citep{pfister2019learning}. 

The first scenario concerns dynamic causal modeling (DCM) that infers the regulatory effects within a dynamic system under experimental stimuli \citep{friston2003dynamic}. Specifically, the DCM characterizes the variations of the state variables $x(t) = (x_1(t),\ldots,x_p(t))^\top\in\R^p$ under the stimulus inputs  $u(t)= (u_1(t), \ldots, u_q(t))^\top\in\R^q$ via a set of ODEs, $dx(t) / dt = F(x(t),u(t))$, where the functional $F$ is modeled by a bilinear form, 
\begin{equation} \label{eqn:DCM}
F_j(x(t),u(t)) = \theta_{j0} + \sum_{k=1}^p \theta^{(1)}_{jk} x_k(t) + \sum_{l=1}^q \theta^{(2)}_{jl} u_l(t)  + \sum_{k=1}^p \sum_{l=1}^q \theta^{(1,2)}_{jkl} x_k(t) u_l(t), \;\; j=1,\ldots,p.
\end{equation}
In this model, $\theta^{(1)}_{jk} \in \R$ reflects the strength of intrinsic connection from $x_k(t)$ to $x_j(t)$, $\theta^{(2)}_{jl} \in \R$ reflects the effect of the $l$th input stimulus $u_l(t)$ on $x_j(t)$, and $\theta^{(1,2)}_{jkl} \in \R$ reflects the influence of $u_l(t)$ on the directional connection between $x_k(t)$ and $x_j(t)$, $j, k = 1, \ldots, p, l = 1, \ldots, q$. Note that $\theta^{(1)}_{jk}$ and $\theta^{(1)}_{kj}$ can be different, and thus the effect from $x_k(t)$ to $x_j(t)$ and that from $x_j(t)$ to $x_k(t)$ can be different. Similarly, $\theta^{(1,2)}_{jkl}$ and $\theta^{(1,2)}_{kjl}$ can be different. As such, model \eqref{eqn:DCM} encodes a directional network, and under certain conditions, a causal network. DCM has been widely used in biology and neuroscience \citep[see, e.g.,][]{friston2003dynamic, Zhang2015, Zhang2017, CaoLuo2019}.

We can combine the proposed KODE with the DCM model \eqref{eqn:DCM} straightforwardly. Such a combination allows us to estimate and infer the causal regulatory effects under experimental stimuli without specifying the forms of the functionals $F$. This is appealing, as there have been evidences suggesting that the regulatory effects can be nonlinear \citep{buxton2004modeling, friston2019dynamic}. More specifically, we model $F$ such that, 
\begin{equation}
\label{eqn:kodedcm}
F_{j}(\xbf(t),u(t)) = \theta_{j0} + \sum_{k=1}^p F^{(1)}_{jk}(x_k(t)) + \sum_{l=1}^q F^{(2)}_{jl}(u_l(t)) + \sum_{k=1}^p \sum_{l=1}^q F^{(1,2)}_{jkl}(x_k(t),u_l(t)), \ j=1,\ldots,p.
\end{equation}
Similar as the tensor product space defined in \eqref{eqn:spaceH}, let $\mathcal H_{k}^{(1)}$ and $\mathcal H_{l}^{(2)}$ denote the space of functions of $x_k(t)$ and $u_l(t)$ with zero marginal integral, respectively. We impose that the functionals $F_j,j=1,\ldots,p$ in \eqref{eqn:kodedcm} are located in the following space, 
\begin{equation*}
\HH = \{1\} \; \oplus \; \sum_{k=1}^p\mathcal H^{(1)}_k \; \oplus \; \sum_{l=1}^q\mathcal H_l^{(2)} \; \oplus \; \sum_{k=1}^p \sum_{l=1}^q \left( \mathcal H^{(1)}_{k}\otimes\mathcal H^{(2)}_{l} \right).
\end{equation*}
Parallel to \eqref{eqn:DCM}, the functions $F_{jk}^{(1)}, F_{jl}^{(2)}$ and $F_{jkl}^{(1,2)}$ in \eqref{eqn:kodedcm} capture the causal regulatory effects, and together, they encode a directional network. Moreover, Algorithm \ref{alg:trainofkode} of KODE is directly applicable to estimate $F_{jk}^{(1)}, F_{jl}^{(2)}$ and $F_{jkl}^{(1,2)}$. As we have shown in our simulations, the DCM model \eqref{eqn:kodedcm} based on KODE is to outperform \eqref{eqn:DCM} that is based on linear ODE. 

The second scenario concerns learning the causal structure of kinetic systems by identifying a stable model from noisy observations generated from heterogeneous experiments. \citet{pfister2019learning} proposed the CausalKinetiX method, where the main idea is to optimize a noninvariance score to identify a causal ODE model that is invariant across heterogeneous experiments. Again, we can combine the proposed KODE with CausalKinetiX to learn the causal structure, while balancing between predictability and causality of the ODE model, and extending from a linear ODE model to a more flexible ODE model. We refer to this integrated method as KODE-CKX. 

More specifically, consider $R$ heterogenous experiments, which stem from interventions such as manipulations of initial or environmental conditions. Following Algorithm  \ref{alg:trainofkode} of KODE, we obtain $\widehat{\theta}_j^{(r)}$ for each experiment $r\in\{1,\ldots,R\}$, and $j=1,\ldots,p$. Let $M_j^{(r)}\subseteq\mathcal M$ denote the index set of the nonzero entries of the sparse estimator $\widehat{\theta}_j^{(r)}$. We propose the following four-step procedure to score each model $M_j^{(r)}$. In the first step, we obtain the smoothing spline estimate $\widehat{x}_j^{(r)}(t)$ by \eqref{eqn:sshatxj} using the data from the $r$th experiment. In the second step, we apply Algorithm  \ref{alg:trainofkode} to compute $\widehat{F}_j^{(r)}$, by setting $\kappa_{nj}=0$, restricting $\theta_j \in M_j^{(r)}$, and using the data from all other experiments except for  the $r$th experiment. Here leaving out the $r$th experiment is to ensure a good generalization capability. In the third step, we estimate the signal trajectory under the  derivative constraint, 
\begin{equation} \label{eqn:dericonstraints}
\begin{aligned}
\widetilde{x}^{(r)}_j(t) & = \underset{z_j\in\mathcal F}{\arg\min} \left\{\frac{1}{n}\sum_{i=1}^n \left\{ y_{ij} - z_{j}(t_i) \right\}^2 + \lambda_{nj}\|z_j(t)\|_{\mathcal F}^2\right\},  \text{ such that } \widetilde{x}^{(r)}_j(t_i) = \widehat{F}_j^{(r)}(\widehat{x}_j^{(r)}(t_i)),
\end{aligned}
\end{equation}
for $i=1, \ldots, n, j = 1, \ldots, p$. In the last step, similar as CausalKinetiX, we obtain for each model $M_j^{(r)}\subseteq \mathcal M$ the noninvariance score, 
\begin{equation*}
\text{NS}\left( M_j^{(r)} \right) \equiv \frac{1}{R} \sum_{r=1}^R \frac{ \text{RSS}_B^{(r)}-\text{RSS}_A^{(r)} }{ \text{RSS}_A^{(r)} },
\end{equation*}
where $\text{RSS}_A^{(r)} = n^{-1} \sum_{i=1}^n \left\{ y^{(r)}_{ij}-\widehat{x}^{(r)}_j(t_{ij}) \right\}^2$, and $\text{RSS}_B^{(r)} = n^{-1} \sum_{i=1}^n \left\{ y^{(r)}_{ij}-\widetilde{x}^{(r)}_j(t_{ij}) \right\}^2$ are the residual sums of squares based on $\widehat{x}_j^{(r)}(t)$ and $\widetilde{x}_j^{(r)}(t)$, respectively. Due to the additional constraint in \eqref{eqn:dericonstraints}, $\text{RSS}_B^{(r)} $ is always larger than $\text{RSS}_A^{(r)}$. Following \citet{pfister2019learning}, the model $M_j^{(r)}\subseteq \mathcal M$ with a small score $\text{NS}(M_j^{(r)})$ is predictive and invariant. Such an invariant ODE model allows researchers to predict the behavior of the dynamic system under interventions, and it is closely related to the causal mechanism of the underlying dynamic system from the structural casual model and modularity perspective \citep{pfister2019learning, rubenstein2018deterministic}. Compared to CausalKinetiX, our proposed KODE-CKX further extends the linear ODE to a general class of nonlinear and non-additive ODE. 

\begin{figure}[t!]
\centering
\includegraphics[width=\textwidth, height=2.25in]{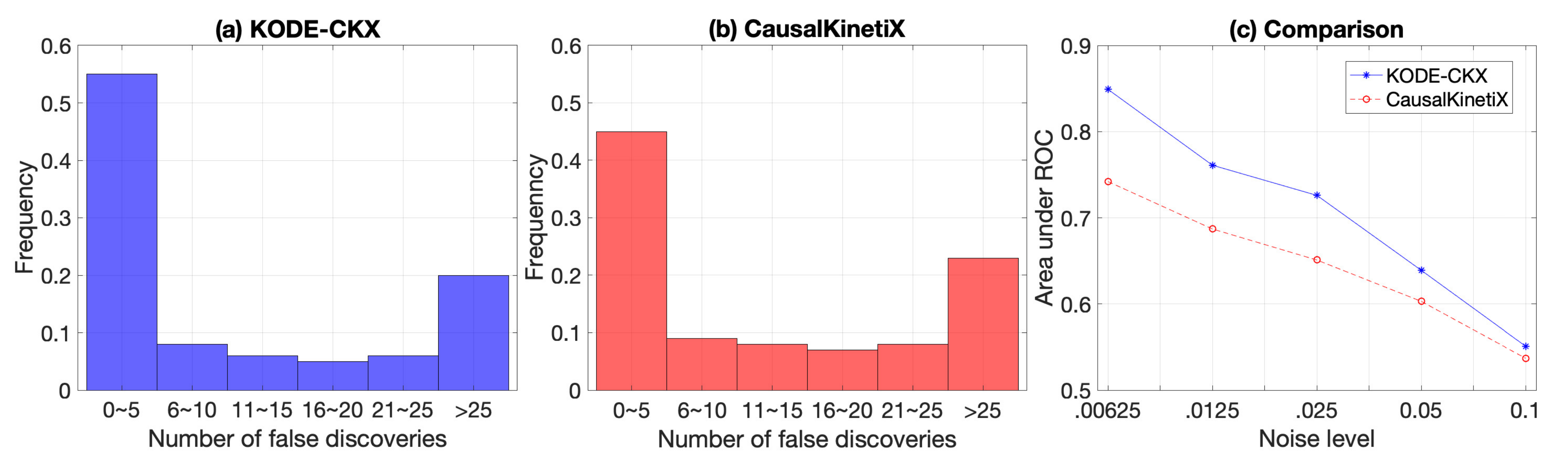}
\caption{The selection performance of  KODE-CKX and CausalKinetiX. The results are averaged over 100 data replications. (a) Number of false discoveries in the estimated model based on KODE-CKX; (b) Number of false discoveries in the estimated model based on CausalKinetiX; (c) Area under ROC under different noise levels.}
\label{fig:causalkinetix}
\end{figure}

To verify the empirical performance of KODE-CKX and to compare with CausalKinetiX, we consider the $100$-node \emph{E.coli1} gene regulatory network example in  Section \ref{sec:application}. Figure \ref{fig:causalkinetix} compares the models with the smallest noninvariance score from  KODE-CKX and CausalKinetiX, respectively, based on 100 data replications. Comparing Figure \ref{fig:causalkinetix}(a) and (b), it is seen that in the majority of cases, KODE-CKX is able to recover the causal parents, and it outperforms CausalKinetiX in terms of the number of false discoveries. Here the measurement errors were drawn from a normal distribution with mean zero and standard deviation $0.025$, the same setup as in Section \ref{sec:application}. We next further evaluate the performance of the two methods when we vary the standard deviation of the measurement errors. Figure \ref{fig:causalkinetix}(c) reports the AUC averaged over 100 data replications. It is seen again that, for all noise levels, KODE-CKX performs better than CausalKinetiX. 

In summary, our proposed KODE is readily applicable to numerous scenarios to facilitate the understanding of the regulatory causal mechanisms within a dynamic system from noisy data under interventions.

\baselineskip=17pt
\bibliographystyle{apa}
\bibliography{ref_kernel}

\begin{thebibliography}{}

\bibitem[\protect\astroncite{Aronszajn}{1950}]{Aronszajn1950}
Aronszajn, N. (1950).
\newblock Theory of reproducing kernels.
\newblock {\em Transactions of the American Mathematical Society}, 68:337--404.

\bibitem[\protect\astroncite{Bach}{2017}]{Bach2017}
Bach, F. (2017).
\newblock On the equivalence between kernel quadrature rules and random feature
  expansions.
\newblock {\em Journal of Machine Learning Research}, 18:714--751.

\bibitem[\protect\astroncite{Bachoc et~al.}{2019}]{Bachoc2019}
Bachoc, F., Leeb, H., and Pötscher, B.~M. (2019).
\newblock Valid confidence intervals for post-model-selection predictors.
\newblock {\em Annals of Statistics}, 47:1475--1504.

\bibitem[\protect\astroncite{Berk et~al.}{2013}]{Berk2013}
Berk, R., Brown, L., Buja, A., Zhang, K., and Zhao, L. (2013).
\newblock Valid post-selection inference.
\newblock {\em Annals of Statistics}, 41:802--837.

\bibitem[\protect\astroncite{Brunton et~al.}{2016}]{Brunton2016}
Brunton, S.~L., Proctor, J.~L., and Kutz, J.~N. (2016).
\newblock Discovering governing equations from data by sparse identification of
  nonlinear dynamical systems.
\newblock {\em Proceedings of the National Academy of Sciences},
  113:3932--3937.

\bibitem[\protect\astroncite{Buxton et~al.}{2004}]{buxton2004modeling}
Buxton, R.~B., Uluda{\u{g}}, K., Dubowitz, D.~J., and Liu, T.~T. (2004).
\newblock Modeling the hemodynamic response to brain activation.
\newblock {\em Neuroimage}, 23:S220--S233.

\bibitem[\protect\astroncite{Cao and Zhao}{2008}]{CaoZhao2008}
Cao, J. and Zhao, H. (2008).
\newblock {Estimating dynamic models for gene regulation networks}.
\newblock {\em Bioinformatics}, 24:1619--1624.

\bibitem[\protect\astroncite{Cao et~al.}{2019}]{CaoLuo2019}
Cao, X., Sandstede, B., and Luo, X. (2019).
\newblock A functional data method for causal dynamic network modeling of
  task-related fmri.
\newblock {\em Frontiers in Neuroscience}, 13:127.

\bibitem[\protect\astroncite{Chen et~al.}{2017}]{Chen2017}
Chen, S., Shojaie, A., and Witten, D.~M. (2017).
\newblock Network reconstruction from high-dimensional ordinary differential
  equations.
\newblock {\em Journal of the American Statistical Association},
  112:1697--1707.

\bibitem[\protect\astroncite{Chernozhukov et~al.}{2015}]{Chernozhukov2015}
Chernozhukov, V., Hansen, C., and Spindler, M. (2015).
\newblock Valid post-selection and post-regularization inference: An
  elementary, general approach.
\newblock {\em Annual Review of Economics}, 7:649--688.

\bibitem[\protect\astroncite{Chou and Voit}{2009}]{Chou2009}
Chou, I.-C. and Voit, E.~O. (2009).
\newblock Recent developments in parameter estimation and structure
  identification of biochemical and genomic systems.
\newblock {\em Mathematical Biosciences}, 219:57--83.

\bibitem[\protect\astroncite{Cox}{1983}]{Cox1983}
Cox, D.~D. (1983).
\newblock Asymptotics for m-type smoothing splines.
\newblock {\em Annals of Statistics}, 11:530--551.

\bibitem[\protect\astroncite{Cucker and Smale}{2002}]{Cucker2002}
Cucker, F. and Smale, S. (2002).
\newblock On the mathematical foundations of learning.
\newblock {\em American Mathematical Society Bulletin}, 39:1--49.

\bibitem[\protect\astroncite{Dattner and Klaassen}{2015}]{Dattner2015}
Dattner, I. and Klaassen, C. A.~J. (2015).
\newblock Optimal rate of direct estimators in systems of ordinary differential
  equations linear in functions of the parameters.
\newblock {\em Electronic Journal of Statistics}, 9:1939--1973.

\bibitem[\protect\astroncite{Duan et~al.}{2003}]{duan2003evaluation}
Duan, K., Keerthi, S.~S., and Poo, A.~N. (2003).
\newblock Evaluation of simple performance measures for tuning svm
  hyperparameters.
\newblock {\em Neurocomputing}, 51:41--59.

\bibitem[\protect\astroncite{Friston et~al.}{2003}]{friston2003dynamic}
Friston, K.~J., Harrison, L., and Penny, W. (2003).
\newblock Dynamic causal modelling.
\newblock {\em Neuroimage}, 19(4):1273--1302.

\bibitem[\protect\astroncite{Friston et~al.}{2019}]{friston2019dynamic}
Friston, K.~J., Preller, K.~H., Mathys, C., Cagnan, H., Heinzle, J., Razi, A.,
  and Zeidman, P. (2019).
\newblock Dynamic causal modelling revisited.
\newblock {\em Neuroimage}, 199:730--744.

\bibitem[\protect\astroncite{Gneiting et~al.}{2010}]{gneiting2010matern}
Gneiting, T., Kleiber, W., and Schlather, M. (2010).
\newblock Mat{\'e}rn cross-covariance functions for multivariate random fields.
\newblock {\em Journal of the American Statistical Association},
  105(491):1167--1177.

\bibitem[\protect\astroncite{Gonz{\'a}lez
  et~al.}{2014}]{gonzalez2014reproducing}
Gonz{\'a}lez, J., Vuja{\v{c}}i{\'c}, I., and Wit, E. (2014).
\newblock Reproducing kernel hilbert space based estimation of systems of
  ordinary differential equations.
\newblock {\em Pattern Recognition Letters}, 45:26--32.

\bibitem[\protect\astroncite{Gu}{2013}]{gu2013}
Gu, C. (2013).
\newblock {\em Smoothing Spline ANOVA Models}.
\newblock Springer Science \& Business Media.

\bibitem[\protect\astroncite{Hall and Marron}{1988}]{hall1988choice}
Hall, P. and Marron, J. (1988).
\newblock Choice of kernel order in density estimation.
\newblock {\em Annals of Statistics}, pages 161--173.

\bibitem[\protect\astroncite{Henderson and
  Michailidis}{2014}]{HendersonMichailidis2014}
Henderson, J. and Michailidis, G. (2014).
\newblock Network reconstruction using nonparametric additive ode models.
\newblock {\em PLOS ONE}, 9:1--15.

\bibitem[\protect\astroncite{Huang}{1998}]{huang1998projection}
Huang, J.~Z. (1998).
\newblock Projection estimation in multiple regression with application to
  functional anova models.
\newblock {\em Annals of Statistics}, 26(1):242--272.

\bibitem[\protect\astroncite{Izhikevich}{2007}]{Izhikevich2007}
Izhikevich, E. (2007).
\newblock {\em Dynamical Systems In Neuroscience}.
\newblock MIT Press.

\bibitem[\protect\astroncite{Javanmard and Montanari}{2014}]{Javanmard2014}
Javanmard, A. and Montanari, A. (2014).
\newblock Confidence intervals and hypothesis testing for high-dimensional
  regression.
\newblock {\em Journal of Machine Learning Research}, 15:2869--2909.

\bibitem[\protect\astroncite{Koltchinskii and Yuan}{2010}]{Koltchinskii2010}
Koltchinskii, V. and Yuan, M. (2010).
\newblock Sparsity in multiple kernel learning.
\newblock {\em Annals of Statistics}, 38:3660--3695.

\bibitem[\protect\astroncite{Liang and Wu}{2008}]{Liang2008}
Liang, H. and Wu, H. (2008).
\newblock Parameter estimation for differential equation models using a
  framework of measurement error in regression models.
\newblock {\em Journal of the American Statistical Association},
  103:1570--1583.

\bibitem[\protect\astroncite{Lin}{2000}]{Lin2000}
Lin, Y. (2000).
\newblock Tensor product space anova models.
\newblock {\em Annals of Statistics}, 28:734--755.

\bibitem[\protect\astroncite{Lin and Brown}{2004}]{lin2004statistical}
Lin, Y. and Brown, L.~D. (2004).
\newblock Statistical properties of the method of regularization with periodic
  gaussian reproducing kernel.
\newblock {\em Annals of Statistics}, 32(4):1723--1743.

\bibitem[\protect\astroncite{Lin and Zhang}{2006}]{LinZhang2006}
Lin, Y. and Zhang, H.~H. (2006).
\newblock Component selection and smoothing in multivariate nonparametric
  regression.
\newblock {\em Annals of Statistics}, 34:2272--2297.

\bibitem[\protect\astroncite{Loh and Wainwright}{2012}]{loh2012}
Loh, P.-L. and Wainwright, M.~J. (2012).
\newblock High-dimensional regression with noisy and missing data: Provable
  guarantees with nonconvexity.
\newblock {\em Annals of Statistics}, 40:1637--1664.

\bibitem[\protect\astroncite{Lu et~al.}{2020}]{Lu2020}
Lu, J., Kolar, M., and Liu, H. (2020).
\newblock Kernel meets sieve: Post-regularization confidence bands for sparse
  additive model.
\newblock {\em Journal of the American Statistical Association}, 0(0):1--16.

\bibitem[\protect\astroncite{Lu et~al.}{2011}]{LuLiang2011}
Lu, T., Liang, H., Li, H., and Wu, H. (2011).
\newblock High-dimensional {ODE}s coupled with mixed-effects modeling
  techniques for dynamic gene regulatory network identification.
\newblock {\em Journal of the American Statistical Association},
  106:1242--1258.

\bibitem[\protect\astroncite{Ma et~al.}{2009}]{Ma2009}
Ma, W., Trusina, A., El-Samad, H., Lim, W.~A., and Tang, C. (2009).
\newblock Defining network topologies that can achieve biochemical adaptation.
\newblock {\em Cell}, 138:760--773.

\bibitem[\protect\astroncite{Marbach et~al.}{2010}]{Marbach2010}
Marbach, D., Prill, R.~J., Schaffter, T., Mattiussi, C., Floreano, D., and
  Stolovitzky, G. (2010).
\newblock Revealing strengths and weaknesses of methods for gene network
  inference.
\newblock {\em Proceedings of the National Academy of Sciences,},
  107:6286--6291.

\bibitem[\protect\astroncite{Marbach et~al.}{2009}]{Marbach2009}
Marbach, D., Schaffter, T., Mattiussi, C., and Floreano, D. (2009).
\newblock Generating realistic in silico gene networks for performance
  assessment of reverse engineering methods.
\newblock {\em Journal of Computational Biology}, 16:229--239.

\bibitem[\protect\astroncite{Meinshausen et~al.}{2006}]{meinshausen2006high}
Meinshausen, N., B{\"u}hlmann, P., et~al. (2006).
\newblock High-dimensional graphs and variable selection with the lasso.
\newblock {\em Annals of Statistics}, 34(3):1436--1462.

\bibitem[\protect\astroncite{Meyer et~al.}{2003}]{meyer2003support}
Meyer, D., Leisch, F., and Hornik, K. (2003).
\newblock The support vector machine under test.
\newblock {\em Neurocomputing}, 55(1-2):169--186.

\bibitem[\protect\astroncite{Mikkelsen and
  Hansen}{2017}]{mikkelsen2017learning}
Mikkelsen, F.~V. and Hansen, N.~R. (2017).
\newblock Learning large scale ordinary differential equation systems.
\newblock {\em arXiv preprint arXiv:1710.09308}.

\bibitem[\protect\astroncite{Opsomer and Ruppert}{1997}]{Opsomer1997}
Opsomer, J.~D. and Ruppert, D. (1997).
\newblock Fitting a bivariate additive model by local polynomial regression.
\newblock {\em Annals of Statistics}, 25:186--211.

\bibitem[\protect\astroncite{Pfister et~al.}{2019}]{pfister2019learning}
Pfister, N., Bauer, S., and Peters, J. (2019).
\newblock Learning stable and predictive structures in kinetic systems.
\newblock {\em Proceedings of the National Academy of Sciences},
  116(51):25405--25411.

\bibitem[\protect\astroncite{Raskutti et~al.}{2011}]{Raskutti2011}
Raskutti, G., Wainwright, M.~J., and Yu, B. (2011).
\newblock Minimax rates of estimation for high-dimensional linear regression
  over $\ell_q $-balls.
\newblock {\em IEEE Transactions on Information Theory}, 57:6976--6994.

\bibitem[\protect\astroncite{Ravikumar et~al.}{2010}]{Ravikumar2010}
Ravikumar, P., Wainwright, M.~J., and Lafferty, J. (2010).
\newblock High-dimensional ising model selection using $l_1$-regularized
  logistic regression.
\newblock {\em Annals of Statistics}, 38:1287--1319.

\bibitem[\protect\astroncite{Rubenstein
  et~al.}{2018}]{rubenstein2018deterministic}
Rubenstein, P.~K., Bongers, S., Sch{\"o}lkopf, B., and Mooij, J.~M. (2018).
\newblock From deterministic odes to dynamic structural causal models.
\newblock {\em Proceedings of the 34th Conference Annual Conference on
  Uncertainty in Artificial Intelligence (UAI)}.

\bibitem[\protect\astroncite{Schaffter et~al.}{2011}]{Schaffter2011}
Schaffter, T., Marbach, D., and Floreano, D. (2011).
\newblock Genenetweaver: in silico benchmark generation and performance
  profiling of network inference methods.
\newblock {\em Bioinformatics}, 27:2263--2270.

\bibitem[\protect\astroncite{Silverman}{1985}]{Silverman1985}
Silverman, B.~W. (1985).
\newblock Some aspects of the spline smoothing approach to non‐parametric
  regression curve fitting.
\newblock {\em Journal of the Royal Statistical Society. Series B (Statistical
  Methodology)}, 47:1--21.

\bibitem[\protect\astroncite{Talagrand}{1996}]{Talagrand1996}
Talagrand, M. (1996).
\newblock New concentration inequalities in product spaces.
\newblock {\em Inventiones Mathematicae}, 126:505--563.

\bibitem[\protect\astroncite{Tsybakov}{2009}]{Tsybakov2009}
Tsybakov, A.~B. (2009).
\newblock {\em Introduction to Nonparametric Estimation}.
\newblock Springer Science \& Business Media.

\bibitem[\protect\astroncite{Tzafriri}{2003}]{Tzafriri2003}
Tzafriri, A.~R. (2003).
\newblock Michaelis-menten kinetics at high enzyme concentrations.
\newblock {\em Bulletin of Mathematical Biology}, 65:1111--1129.

\bibitem[\protect\astroncite{van~de Geer}{2000}]{Vandegeer2000}
van~de Geer, S. (2000).
\newblock {\em Empirical Processes in $M$-Estimation}.
\newblock Cambridge University Press.

\bibitem[\protect\astroncite{van~der Vaart and Wellner}{1996}]{Vandevaart1996}
van~der Vaart, A.~W. and Wellner, J.~A. (1996).
\newblock {\em Weak Convergence and Empirical Processes}.
\newblock Springer-Verlag, New York.

\bibitem[\protect\astroncite{Varah}{1982}]{Varah1982}
Varah, J.~M. (1982).
\newblock A spline least squares method for numerical parameter estimation in
  differential equations.
\newblock {\em SIAM Journal on Scientific and Statistical Computing}, 3:28--46.

\bibitem[\protect\astroncite{Volterra}{1928}]{Volterra1928}
Volterra, V. (1928).
\newblock {Variations and Fluctuations of the Number of Individuals in Animal
  Species living together}.
\newblock {\em ICES Journal of Marine Science}, 3:3--51.

\bibitem[\protect\astroncite{Wahba}{1983}]{wahba1983}
Wahba, G. (1983).
\newblock Bayesian ``confidence intervals'' for the cross-validated smoothing
  spline.
\newblock {\em Journal of the Royal Statistical Society. Series B (Statistical
  Methodology)}, 45:133--150.

\bibitem[\protect\astroncite{Wahba}{1990}]{wahba1990}
Wahba, G. (1990).
\newblock {\em Spline Models for Observational Data}.
\newblock SIAM, Philadelphia.

\bibitem[\protect\astroncite{Wahba et~al.}{1995}]{Wahba1995}
Wahba, G., Wang, Y., Gu, C., Klein, R., and Klein, B. (1995).
\newblock Smoothing spline {ANOVA} for exponential families, with application
  to the {W}isconsin {E}pidemiological {S}tudy of {D}iabetic {R}etinopathy.
\newblock {\em Annals of Statistics}, 23:1865--1895.

\bibitem[\protect\astroncite{Wang et~al.}{2009}]{Wang2009}
Wang, S., Nan, B., Zhu, N., and Zhu, J. (2009).
\newblock {Hierarchically penalized Cox regression with grouped variables}.
\newblock {\em Biometrika}, 96(2):307--322.

\bibitem[\protect\astroncite{Wu et~al.}{2014}]{Wu2014}
Wu, H., Lu, T., Xue, H., and Liang, H. (2014).
\newblock Sparse additive ordinary differential equations for dynamic gene
  regulatory network modeling.
\newblock {\em Journal of the American Statistical Association}, 109:700--716.

\bibitem[\protect\astroncite{Yuan and Zhou}{2016}]{Yuan2016}
Yuan, M. and Zhou, D.-X. (2016).
\newblock Minimax optimal rates of estimation in high dimensional additive
  models.
\newblock {\em Annals of Statistics}, 44(6):2564--2593.

\bibitem[\protect\astroncite{Zhang and Zhang}{2014}]{Zhang2014CI}
Zhang, C.-H. and Zhang, S.~S. (2014).
\newblock Confidence intervals for low dimensional parameters in high
  dimensional linear models.
\newblock {\em Journal of the Royal Statistical Society. Series B.},
  76(1):217--242.

\bibitem[\protect\astroncite{Zhang et~al.}{2015a}]{Zhang2015}
Zhang, T., Wu, J., Li, F., Caffo, B., and Boatman-Reich, D. (2015a).
\newblock A dynamic directional model for effective brain connectivity using
  electrocorticographic {(ECoG)} time series.
\newblock {\em Journal of the American Statistical Association}, 110:93--106.

\bibitem[\protect\astroncite{Zhang et~al.}{2017}]{Zhang2017}
Zhang, T., Yin, Q., Caffo, B., Sun, Y., and Boatman-Reich, D. (2017).
\newblock Bayesian inference of high-dimensional, cluster-structured ordinary
  differential equation models with applications to brain connectivity studies.
\newblock {\em Annals of Applied Statistics}, 11:868--897.

\bibitem[\protect\astroncite{Zhang et~al.}{2015b}]{zhang2015selection}
Zhang, X., Cao, J., and Carroll, R.~J. (2015b).
\newblock On the selection of ordinary differential equation models with
  application to predator-prey dynamical models.
\newblock {\em Biometrics}, 71(1):131--138.

\bibitem[\protect\astroncite{Zhao and Yu}{2006}]{Zhao2006}
Zhao, P. and Yu, B. (2006).
\newblock On model selection consistency of lasso.
\newblock {\em Journal of Machine Learning Research}, 7:2541--2563.

\bibitem[\protect\astroncite{Zhu et~al.}{2014}]{Zhu2014}
Zhu, H., Yao, F., and Zhang, H.~H. (2014).
\newblock Structured functional additive regression in reproducing kernel
  hilbert spaces.
\newblock {\em Journal of the Royal Statistical Society. Series B (Statistical
  Methodology)}, 76:581--603.

\end{thebibliography}

\appendix

\section{Proofs}

\subsection{Proof of Theorem \ref{thm:representer}}

\noindent
Denote the KODE objective in (\ref{eqn:kode}) by $\AA(F_j)$:
\begin{equation*}
\AA(F_j)\equiv \frac{1}{n}\sum_{i=1}^n\left\{y_{ij} - \theta_{j0} - \int_0^{t_i}F_j(\widehat{\xbf}(t))dt\right\}^2+ \tau_{nj} \left( \sum_{k=1}^p\|\PP^k F_j\|_{\HH} + \sum_{k\neq l, k=1}^p \sum_{l=1}^p \|\PP^{kl} F_j\|_\HH \right).
\end{equation*} 
Without loss of generality, let $\tau_{nj}=1$. Write $\mathcal H = \HH^{(0)} \oplus\HH^{(1)}$, where $\HH^{(0)} \equiv \{1\}$ and $\HH^{(1)} \equiv \sum_{k=1}^p\mathcal H_k \oplus\sum_{k\neq l,k=1}^p\sum_{l=1}^p\left[\mathcal H_{k}\otimes\mathcal H_{l}\right]$, where for any $F_j\in\HH$ \citep{Wahba1995}, 
\begin{equation*}
\|F_j\|_\HH^2 = \|F_j\|_{\HH^{(0)}}^2 + \|F_j\|_{\HH^{(1)}}^2, \quad \textrm{ and } \quad \|F_j\|_{\HH^{(1)}}^2 =  \sum_{k=1}^p\|\PP^kF_j\|_\HH^2+\sum_{k\neq l,k=1}^p\sum_{l=1}^p\|\PP^{kl}F_j\|_\HH^2.
\end{equation*}
Note that,
\begin{equation*}
\begin{aligned}
&\frac{p(p+1)}{2}\left( \sum_{k=1}^p\|\PP^kF_j\|_\HH^2+\sum_{k\neq l,k=1}^p\sum_{l=1}^p\|\PP^{kl}F_j\|_\HH^2 \right) \\
&\quad\quad\quad\geq J_2^2(F_j) \geq \sum_{k=1}^p\|\PP^kF_j\|_\HH^2+\sum_{k\neq l,k=1}^p\sum_{l=1}^p\|\PP^{kl}F_j\|_\HH^2.
\end{aligned}
\end{equation*}
Henceforth, for any $F_j\in\HH^{(1)}$, 
\begin{equation} \label{eqn:jfj}
J_2(F_j) \geq \|F_j\|_{\HH}.
\end{equation}
We next show the existence of the minimizer in three cases.
 
First, denote $\rho_j=\max_{i=1}^n(y_{ij}^2+|y_{ij}|+1)$. Let $K(\cdot,\cdot)$ be the reproducing kernel of $\mathcal H^{(1)}$, and let $\langle\cdot,\cdot\rangle_{\mathcal H^{(1)}}$ be the inner product in $\mathcal H^{(1)}$.  Write $a = \sup_{t\in\TT}K^{1/2}(\widehat{\xbf}(t),\widehat{\xbf}(t))$, where $\widehat{\xbf}$ is obtained from (\ref{eqn:sshatxj}). Consider the set
\begin{equation*}
\begin{aligned}
\DD_j  = \left\{ F_j\in\mathcal H: F_j =  b_j+F^{(1)}_{j}, b_j\in\{1\},F_j^{(1)}\in\mathcal H^{(1)}, J_2(F_j) \leq \rho_j,\ |b_j|\leq \rho^{1/2}+(a+1)\rho_j \right\}.
\end{aligned}
\end{equation*}
Then $\DD_j$ is a closed and convex compact set. Note that both $J_2(F_j)$ and the functional $n^{-1} \sum_{i=1}^n \left\{ y_{ij}-\theta_{j0}- \int_0^{t_i}F_j(\widehat{\xbf}(u))du \right\}^2$ are convex in $F_j$, and thus $\AA(F_j)$ is convex. Therefore, there exists a minimizer of the convex optimization problem \eqref{eqn:kode} in the convex set $\DD_j$. Denote the minimizer by $\widehat{F}_j\in \DD_j$. Then $\mathcal A(\widehat{F}_j) \leq \mathcal A(0) \leq n^{-1} \sum_{i=1}^ny_{ij}^2 < \rho_j$.

Second, for any $F_j\in\mathcal H$ with $J(F_j)>\rho_j$, then $F_j \not\in \mathcal D_j$. However, $\mathcal A(F_j)\geq J(F_j)>\rho_j$, which implies that $\mathcal A(F_j)>\mathcal A(\widehat{F}_j)$.

Third, for any $F_j\in\mathcal H$ with $J_2(F_j) \leq \rho_j$, $F_j=b_j+F_j^{(1)}$ with $b_j\in\{1\}$, $F_j^{(1)}\in \mathcal H^{(1)}$, and $|b_j|>\rho_j^{1/2}+(a+1)\rho_j$. By the reproducing property, for any $F^{(1)}_j\in\mathcal H^{(1)}$ and $t\in\TT$,
\begin{equation*}
\begin{aligned}
\left| F^{(1)}_j(\widehat{\xbf}(t)) \right| & = \left| \left\langle F_j^{(1)}(\cdot),K(\widehat{\xbf}(t),\cdot) \right\rangle_{\mathcal H^{(1)}} \right| \leq \left\| F_j^{(1)} \right\|_{\HH^{(1)}} \left\langle K(\widehat{\xbf}(t),\cdot),K(\widehat{\xbf}(t),\cdot) \right\rangle_{\mathcal H^{(1)}}^{1/2}\\
& = \left\| F_j^{(1)} \right\|_{\HH^{(1)}}K^{1/2}\left( \widehat{\xbf}(t),\widehat{\xbf}(t) \right) \leq  a J_2\left( F_j^{(1)} \right),
\end{aligned}
\end{equation*}
where the last step is by (\ref{eqn:jfj}) and the definition of $a$. Hence, for any $i=1,\ldots,n$, $t_i\in\TT$,
\begin{equation*}
\begin{aligned}
\min_{C_{j0}}\left|C_{j0} +b_jt_i+\int_0^{t_i}F_j^{(1)}(\widehat{\xbf}(u))du-y_{ij}\right| & \geq \left|b_jt_i+\int_0^{t_i}F_j^{(1)}(\widehat{\xbf}(u))du-y_{ij}\right|\\
& > [\rho_j^{1/2}+(a+1)\rho_j]-a\rho_j-\rho_j = \rho_j^{1/2}.
\end{aligned}
\end{equation*}
Therefore, $\mathcal A(F_j)>\rho$, and $\mathcal A(F_j)>\mathcal A(\widehat{F}_j)$. Consequently, for any $F_j\not\in \mathcal D_j$, $\mathcal A(F_j)>\mathcal A(\widehat{F}_j)$, and $\widehat{F}_j$ is a minimizer of  (\ref{eqn:kode})  in $\mathcal H$.

Next, we show that the minimizer $\widehat{F}_j$ is in a finite-dimensional space. Let $K_k(\cdot,\cdot)$ be the reproducing kernel of $\HH_k$. Then $K_{kl}\equiv K_kK_l$ is the reproducing kernel of $\HH_k\otimes \HH_l$ \citep{Aronszajn1950}. Write $\widehat{F}_j = \widehat{b}_j + \sum_{k=1}^p\widehat{F}_{jk}+ \sum_{k\neq l}\widehat{F}_{jkl}$,  where $\widehat{F}_{jk}\in\HH_k$, and $\widehat{F}_{jkl} \in \mathcal H_{k}\otimes\mathcal H_{l}$. Write $T_i(t) = \mathbf{1}\{0\leq t\leq t_i\}$, and $\bar{T}(t) = n^{-1} \sum_{i=1}^nT_i(t)$.  We have $ \int_\TT K(\widehat{\xbf}(s),\widehat{\xbf}(t))T_i(t)dt\in\HH$ \citep{Cucker2002}. Besides, 
\begin{equation*}
\begin{aligned}
\left\langle \int_\TT K(\widehat{\xbf}(s),\widehat{\xbf}(t))T_i(t)dt, \ F_j(\widehat{\xbf}(s))\right\rangle_\HH  
& =  \int_\TT \left\langle K(\widehat{\xbf}(s),\xbf(t)),  F_j(\widehat{\xbf}(s))\right\rangle_\HH T_i(t)dt \\
&= \int_\TT F_j(\widehat{\xbf}(t))T_i(t)dt.
\end{aligned}
\end{equation*}
Denote the projection of $\widehat{F}_{jk}$ onto the finitely spanned space 
\begin{equation*}
\left\{ \int_\TT K_k(\widehat{x}_k(t),\cdot)T_i(t)dt,i=1,\ldots,n \right\} \subset \HH_k
\end{equation*} as $\widehat{g}_{jk}$, and its orthogonal complement in $\HH_k$ as $\widehat{h}_{jk}$. Similarly, denote the projection of $\widehat{F}_{jkl}$ onto the finitely spanned space 
\begin{equation*}
\left\{ \int_\TT K_k(\widehat{x}_k(t),\cdot) K_l(\widehat{x}_l(t),\cdot) T_i(t)dt, i=1,\ldots,n \right\}\subset\HH_k\otimes \HH_l
\end{equation*} as $\widehat{g}_{kl}$, and its orthogonal complement in $\HH_k\otimes\HH_l$ as $\widehat{h}_{kl}$. Then $\widehat{F}_{jk} = \widehat{g}_{jk}+\widehat{h}_{jk}$, and $\widehat{F}_{jkl} = \widehat{g}_{jkl} +\widehat{h}_{jkl}$. Besides, $\|\widehat{F}_{jk}\|_\HH^2 = \|\widehat{g}_{jk}\|_\HH^2 + \|\widehat{h}_{jk}\|_\HH^2$, and $\|\widehat{F}_{jkl}\|_\HH^2 = \|\widehat{g}_{jkl}\|_\HH^2 + \|\widehat{h}_{jkl}\|_\HH^2$, for $k,l=1,\ldots,p, k\neq l$. Since $K = 1 + \sum_{k=1}^pK_k + \sum_{k\neq l}K_{kl}$ is the reproducing kernel of $\HH$, by the orthogonal structure,
\begin{equation*}
\begin{aligned}
& \int_\TT \widehat{F}_j(\widehat{\xbf}(t))T_i(t)dt = \left\langle \int_\TT \left\{1+ \sum_{k=1}^pK_k(\widehat{x}_k(t),\cdot) + \sum_{k\neq l}K_k(\widehat{x}_k(t),\cdot)K_l(\widehat{x}_l(t),\cdot) \right\} T_i(t)dt,\right.\\
& \quad\quad \left. b_j + \sum_{k=1}^p \left\{ \widehat{g}_{jk}(\widehat{x}_k(t))+\widehat{h}_{jk}(\widehat{x}_k(t)) \right\} +\sum_{k\neq l} \left\{ \widehat{g}_{jkl}(\widehat{x}_k(t),\widehat{x}_l(t))+\widehat{h}_{jkl}(\widehat{x}_k(t),\widehat{x}_l(t)) \right\}\right\rangle_\HH \\
= \; & b_j\int_\TT T_i(t)dt + \sum_{k=1}^p\left\langle \int_\TT K_k(\widehat{x}_k(t),\cdot)T_i(t)dt,\ \widehat{g}_{jk}(\widehat{x}_k(t))\right\rangle_\HH \\
& \quad\quad\quad\quad\quad\; + \sum_{k\neq l}\left\langle\int_\TT K_k(\widehat{x}_k(t),\cdot)K_l(\widehat{x}_l(t),\cdot)T_i(t)dt,\ \widehat{g}_{jkl}(\widehat{x}_k(t),\widehat{x}_l(t))\right\rangle_\HH.
\end{aligned}
\end{equation*}
Recall $\bar{y}_j = n^{-1} \sum_{i=1}^ny_{ij}$. Therefore, \eqref{eqn:kode} can be written as 
\begin{equation*}
\begin{aligned}
\frac{1}{n}\sum_{i=1}^n & \Bigg\{ (y_{ij}-\bar{y}_j)-b_j\int_\TT[T_i(t)-\bar{T}(t)]dt \\
& - \sum_{k=1}^p\left\langle \int_\TT K_k\left(\widehat{x}_k(s),\widehat{x}_k(t)\right)[T_i(t)-\bar{T}(t)]dt, \ \widehat{g}_{jk}(\widehat{x}_k(s))\right\rangle_\HH\\
& -\sum_{k\neq l}\left\langle \int_\TT K_k\left(\widehat{x}_k(s),\widehat{x}_k(t)\right)K_l\left(\widehat{x}_l(s),\widehat{x}_l(t)\right)[T_i(t)-\bar{T}(t)]dt, \ \widehat{g}_{jkl}\left(\widehat{x}_k(s),\widehat{x}_l(s)\right)\right\rangle_\HH \Bigg\}^2 \\
& + \tau_{nj}\left\{\sum_{k=1}^p(\|\widehat{g}_{jk}(\widehat{x}_k)\|_\HH^2+\|\widehat{h}_{jk}(\widehat{x}_k)\|_\HH^2)^{1/2}+\sum_{k\neq l}(\|\widehat{g}_{jkl}(\widehat{x}_k,\widehat{x}_l)\|_\HH^2+\|\widehat{h}_{jkl}(\widehat{x}_k,\widehat{x}_l)\|_\HH^2)^{1/2}\right\}.
\end{aligned}
\end{equation*}
Therefore, the minimizer $\widehat{F}_j$ of \eqref{eqn:kode} satisfies that $\widehat{h}_{jk}=\widehat{h}_{jkl}=0$, for any $k,l = 1, \ldots, p$ and $k \neq l$. This completes the proof of Theorem \ref{thm:representer}. 
\eop

\subsection{Proof of Theorem \ref{thm:postselection}}

\noindent
We first prove that $c_0(\xbf)$ does not depend on the true but unknown functional $F_j$. Consider
\begin{equation*}
\begin{aligned}
\widetilde{F}_j(\xbf) & =  \theta_{j0} + \rho_j^{1/2}\mathcal B(\xbf),\quad \xbf=\xbf(t),\\
Y_{ij} & = \mathcal L_{i}\widetilde{F}_j(\xbf)+\epsilon_{ij}, \quad  t\in\TT.
\end{aligned}
\end{equation*}
where $\theta_{j0}\sim\mathcal N(0,aI)$ and  
$\epsilon_{ij}\sim\mathcal N(0,\sigma_j^2)$. The parameter $\rho_j = \sigma_j^2/n\eta_{nj}$. The stochastic process $\mathcal B(\cdot)$ is a zero-mean Gaussian process with covariance $K_{\widehat{\thetabf}_{M_j}} = \sum_{k=1}^p\widehat{\theta}_{jk}K_k + \sum_{k\neq l}\widehat{\theta}_{jkl}K_{kl}$. The bounded  operator takes the form: $L_{i}\widetilde{F}_j(\xbf) \equiv \int_\TT \left\{ T_i(t) - \bar{T}(t) \right\} \widetilde{F}_j(\xbf(t))dt$, for any $\widetilde{F}_j\in\HH$. It is shown that \citep{wahba1990}, 
\begin{eqnarray*}
\lim_{a\to\infty}\E\left\{ \widetilde{F}_j(\xbf)|Y_{ij} = y_{ij} - \bar{y}_j, i = 1,\ldots,n \right\} = \widehat{F}_{j,\widehat{\thetabf}_{M_j}}(\xbf),
\end{eqnarray*} 
and the covariance matrix of $\left( \mathcal L_1\widehat{F}_j,\ldots,\mathcal L_n\widehat{F}_j \right)$ is $\text{Cov}(\mathcal L_1\widehat{F}_j,\ldots,\mathcal L_n\widehat{F}_j) = \sigma_j^2 A_{M_j}$, where $A_{M_j}$ is the smoothing matrix as defined in \eqref{eqn:smoothmatrix}  with the kernel corresponding to $\widehat{\thetabf}_{M_j}$ \citep{wahba1983, Silverman1985}. Consequently, the collection of all the quantities $A_{M_j}( \ybf_j -\bar{y}_j )$ are jointly distributed as $N(0,\sigma_j^2 A_{M_j})$, where $A_{M_j}$ is independent of $F_j$. Henceforth, the joint distribution of the collection of ratios $|A_{M_j}( \ybf_j -\bar{y}_j ) | / \sigma_j$ is independent of $F_j$.

Next, we prove the coverage property. Observe that, for any $i=1,\ldots,n$,
\begin{equation*}
\widehat{F}_{j,\widehat{\thetabf}_{M_j}}(\widehat{\xbf}(t_i)) -\E\left\{ \widehat{F}_{j,\widehat{\thetabf}_{M_j}}(\widehat{\xbf}(t_i)) \right\} = \{A_{M_j}\}_{i\cdot}( \ybf_j -\bar{y}_j ).
\end{equation*}
We then have the following upper bound, 
\begin{equation*}
\left|\{\widetilde{A}_{M_j}\}_{i\cdot}( \ybf_j -\bar{y}_j )\right| / \sigma_j \leq \max_{M_j\subseteq\mathcal M}\left|\{\widetilde{A}_{M_j}\}_{i\cdot}( \ybf_j -\bar{y}_j ) \right| / \sigma_j.
\end{equation*}
By the choice of $c_0(\widehat{\xbf})$ in (\ref{eqn:defofc0}), the coverage property holds.

Lastly, we show that there exists a unique $c_0(\widehat{\xbf}_j(t_i))$ satisfying \eqref{eqn:defofc0}. Consider the maximum statistic, $\max_{M_j\subseteq\mathcal M} \left| \{\widetilde{A}_{M_j}\}_{i\cdot}( \ybf_j -\bar{y}_j )\right|/\sigma_j$, with the corresponding distribution $H(t) = \mathbb P\left[ \max_{M_j\subseteq\mathcal M}\left| \{\widetilde{A}_{M_j}\}_{i\cdot}( \ybf_j -\bar{y}_j ) \right| / \sigma_j \leq t \right]$. We show that $H(t)=0$ for $t\leq 0$,  is continuous on $\R$, and is strictly increasing in $t\geq 0$. 

Note that, for $t<0$, the event $\left\{ \max_{M_j\subseteq\mathcal M}\left| \{\widetilde{A}_{M_j}\}_{i\cdot}( \ybf_j -\bar{y}_j ) \right| / \sigma_j \leq t \right\}$ is empty. For $t=0$, this event is an intersection of the sets $\left\{ \{\widetilde{A}_{M_j}\}_{i\cdot}( \ybf_j -\bar{y}_j ) = 0 \right\}$ for any $M_j \subseteq \mathcal M$, where at least one of these sets has a probability zero, given $\ybf_j \neq \bar{y}_j$. Henceforth,  $H(t)=0$ for $t\leq 0$. To prove the continuity of $H$ on $(0,\infty)$, we note that, for any $M_j\subseteq\mathcal M$ and $t\geq 0$, $\mathbb P\left[ \{\widetilde{A}_{M_j}\}_{i\cdot}( \ybf_j -\bar{y}_j ) / \sigma_j = t \right] = 0$, since $\ybf_j$ is a continuous variable. Finally, we show the strict monotonicity that $H(t_1)<H(t_2)$ for any $0<t_1<t_2$. Toward that goal, suppose there exists $\widetilde{\ybf}_j \in \R^n$ such that $\max_{M_j\subseteq\mathcal M}\left| \{\widetilde{A}_{M_j}\}_{i\cdot} \widetilde{\ybf}_j  \right| / \sigma_j = t_1$. There exists $M_j^\ddagger\subseteq\mathcal M$, such that $\{\widetilde{A}_{M_j^\ddagger}\}_{i\cdot} \widetilde{\ybf}_j  / \sigma_j = t_1$, which is obtained without loss of generality by changing the sign of $t_1$.  Let $\mathcal Y$ be the set of all $\ybf_j \in \R^n$ such that $\{\widetilde{A}_{M_j^\ddagger}\}_{i\cdot} \left[ ( \ybf_j -\bar{y}_j ) - \widetilde{\ybf}_j  \right] / \sigma_j > 0$, and $\left| \{\widetilde{A}_{M_j}\}_{i\cdot} \left[( \ybf_j -\bar{y}_j )-  \widetilde{\ybf}_j \right] \right| / \sigma_j < (t_2-t_1)/2$ for any $M_j \subseteq \mathcal M$. Then for any $\ybf_j \in \mathcal Y$, 
\vspace{-0.1in}
\begin{equation*}
\begin{aligned}
& \max_{M_j\subseteq\mathcal M} \left| \{\widetilde{A}_{M_j}\}_{i\cdot}( \ybf_j -\bar{y}_j ) \right| / \sigma_j \\
\leq \;\; & \max_{M_j\subseteq\mathcal M} \left| \{\widetilde{A}_{M_j}\}_{i\cdot}\left[( \ybf_j -\bar{y}_j ) -  \widetilde{\ybf}_j  \right] \right| / \sigma_j  + \max_{M_j\subseteq\mathcal M} \left| \{\widetilde{A}_{M_j}\}_{i\cdot}  \widetilde{\ybf}_j  \right| / \sigma_j \\
< \;\; & (t_2-t_1)/2+t_1 < t_2.
\end{aligned}
\end{equation*}
Moreover, $\{\widetilde{A}_{M_j^\ddagger}\}_{i\cdot}( \ybf_j -\bar{y}_j )/ \sigma_j > \{\widetilde{A}_{M_j^\ddagger}\}_{i\cdot}  \widetilde{\ybf}_j  / \sigma_j =t_1>0$. Therefore, 
\begin{equation*}
\max_{M_j\subseteq\mathcal M} \left| \{\widetilde{A}_{M_j}\}_{i\cdot}( \ybf_j -\bar{y}_j )\right| / \sigma_j \geq \left| \{\widetilde{A}_{M_j^\ddagger}\}_{i\cdot}( \ybf_j -\bar{y}_j ) \right| / \sigma_j > t_1,
\end{equation*}
which implies that
$H(t_2)-H(t_1)\geq \mathbb P(\mathcal Y)>0$. Consequently, there exists a unique $c_0(\widehat{\xbf}_j(t_i))$ satisfying \eqref{eqn:defofc0}, which is the $(1-\alpha)$th quantile of the distribution of $\max_{M_j\subseteq\mathcal M} \big| \{\widetilde{A}_{M_j}\}_{i\cdot}( \ybf_j -\bar{y}_j )\big| / \sigma_j$. This completes the proof of Theorem \ref{thm:postselection}.
\eop

\subsection{Proof of Proposition \ref{lem:findingc0}}

\noindent 
Note that, for any $t\geq 0$,
\begin{equation*}
\begin{aligned}
&\mathbb P_{n,F_j,\sigma_j}\left[ \max_{M_j\subseteq\mathcal M} \left| \{\widetilde{A}_{M_j}\}_{i\cdot}( \ybf_j -\bar{y}_j ) \right| / \sigma_j > t \right] \\
= \;\; & \mathbb P_{n,F_j,\sigma_j}\left[ \max_{M_j\subseteq\mathcal M} \left| \{\widetilde{A}_{M_j}\}_{i\cdot}( \ybf_j -\bar{y}_j ) \right| / \| \ybf_j -\bar{y}_j  \|_{l_2} > \left( \sigma_j / \| \ybf_j -\bar{y}_j \|_{l_2} \right) t \right] \\
= \;\; & \mathbb P_{n,F_j,\sigma_j}\left(\max_{M_j\subseteq\mathcal M}|\{\widetilde{A}_{M_j}\}_{i\cdot}V|>t/U\right),
\end{aligned}
\end{equation*}
where $V$ is uniformly distributed on the unit sphere in $\mathbb R^n$, and $U$ is a nonnegative random variable such that $U^2$ follows an $\chi^2(n)$-distribution. Combining this result with the definition in \eqref{eqn:defofc0} completes the proof.
\eop

\subsection{Proof of Theorem \ref{thm:optimalestofpredictor}}

\noindent 
The upper bound of the convergence rate can be established following \citet{Cox1983}. The minimax lower bound of the convergence rate can be established following \citet{Tsybakov2009}. Moreover, the results hold for both fixed and random designs of $t\in\TT$. 
\eop

\subsection{Proof of Theorem \ref{thm:optimalestoffunctional}}

\noindent 
We divide the proof of this theorem to three parts. To establish the minimax rate, we first prove the upper bound in Section \ref{sec:uppbdfunctional}, then prove the lower bound in Section \ref{sec:lowbdfunctional}. We give two auxiliary lemmas that are useful for the proof in Section \ref{sec:auxlem}.

\subsubsection{Upper bound}
\label{sec:uppbdfunctional}

\noindent
For $j=1,\ldots,p$, write $\widehat{F}_j(\widehat{x}) = \widehat{b}_j + \sum_{k=1}^p\widehat{F}_{jk}(\widehat{x}) + \sum_{k\neq l}\widehat{F}_{jkl}(\widehat{x})$, where $\sum_{i=1}^n\widehat{F}_{jk}(\widehat{x}_k(t_{i}))=0$, and $\sum_{i=1}^n\widehat{F}_{jkl}(\widehat{x}_k(t_{i}),\widehat{x}_l(t_{i}))=0$. Write $F_j(x) = b_j + \sum_{k=1}^pF_{jk}(x) + \sum_{k\neq l}F_{jkl}(x)$, where $\sum_{i=1}^nF_{jk}(x_k(t_{i}))=0$, and $\sum_{i=1}^nF_{jkl}(x_k(t_{i}),x_l(t_{i}))=0$. In light of the fact that $\widehat{\theta}_{j0}$ that minimizes \eqref{eqn:kode} is given by $\widehat{\theta}_{j0} = \bar{y}_j - \int_\TT\bar{T}(t)\widehat{F}_j(\widehat{\xbf}(t))dt$, we focus our attention on $\widehat{F}_j(\widehat{\xbf}(t))$ in the following proof, while the convergence rate of $\widehat{\theta}_{j0}$ is the same as that of $\widehat{F}_j(\widehat{\xbf}(t))$. 

Consider that $\widehat{F}_j$ is obtained from 
\begin{equation*}
\underset{F_j\in\HH}{\min} 
\left[\frac{1}{n}\sum_{i=1}^n\left\{ y_{ij} - \int_0^{t_i}F_j(\widehat{\xbf}(t))dt \right\}^2 + \tau_{nj} J_2(F_j)\right],
\end{equation*}
which implies that
\begin{equation*}
\begin{aligned}
& \frac{1}{n}\sum_{i=1}^n\left\{\int_0^{t_i}F_j(\xbf(t))dt+\epsilonbf_{ij} - \int_0^{t_i}\widehat{F}_j(\widehat{\xbf}(t))dt\right\}^2+ \tau_{nj} J_2(\widehat{F}_j)\\
\leq \;\; &   
\frac{1}{n}\sum_{i=1}^n\left\{\int_0^{t_i}F_j(\xbf(t))dt+\epsilonbf_{ij} - \int_0^{t_i}F_j(\widehat{\xbf}(t))dt\right\}^2+ \tau_{nj} J_2(F_j).
\end{aligned}
\end{equation*}
With rearrangement of the terms, we have, 
\begin{equation}
\label{eqn:totaldiff}
\begin{aligned}
& \frac{1}{n}\sum_{i=1}^n\left[ \int_0^{t_i}\left\{ F_j(\xbf(t))-\widehat{F}_j(\widehat{\xbf}(t)) \right\} dt \right]^2+ \tau_{nj} J_2(\widehat{F}_j)\\
\leq \;\; & \frac{2}{n}\sum_{i=1}^n\epsilonbf_{ij}\left[ \int_0^{t_i} \left\{ \widehat{F}_j(\widehat{\xbf}(t))-F_j(\widehat{\xbf}(t)) \right\}dt \right] \\
&\quad\quad\quad +
\frac{1}{n}\sum_{i=1}^n \left[\int_0^{t_i}\left\{ F_j(\xbf(t))-F_j(\widehat{\xbf}(t)) \right\} dt \right]^2+ \tau_{nj} J_2(F_j).
\end{aligned}
\end{equation}
By Assumption \ref{assump:complexity} and the Taylor expansion, 
\begin{equation*}
(\widehat{F}_j-F_j)(\widehat{x}) =(\widehat{F}_j-F_j)(x) + \frac{\partial}{\partial t}(\widehat{F}_j-F_j)(x)(\widehat{x}-x) + o_p\left(\max_{k=1,\ldots,p}\|\widehat{x}_k-x_k\|_{L_2}\right), 
\end{equation*}
where the Fr\'echet derivative of any  $g(\cdot)\in\HH$ is defined as,
\begin{equation*}
\frac{\partial}{\partial t} g(x)(\widehat{x}-x) = \sum_{k=1}^p\frac{\partial g(x)}{\partial x_k}(\widehat{x}_k-x_k).
\end{equation*}
Then the first term on the right-hand-side of \eqref{eqn:totaldiff} can be written as,
\begin{equation*}
\begin{aligned}
\frac{2}{n}\sum_{i=1}^n\epsilonbf_{ij} & \left[\int_0^{t_i}\left\{ \widehat{F}_j(\widehat{\xbf}(t))-F_j(\widehat{\xbf}(t)) \right\}dt \right] 
= \frac{2}{n}\sum_{i=1}^n\epsilonbf_{ij}\left\{\int_0^{t_i}(\widehat{F}_j-F_j)(x(t))dt\right\} \\
& + \frac{2}{n}\sum_{i=1}^n\epsilon_{ij}\left[ \int_0^{t_i} \frac{\partial}{\partial t}(\widehat{F}_j-F_j)(x(t))\{\widehat{x}(t)-x(t)\}dt + o_p\left(\max_{k=1,\ldots,p}\|\widehat{x}_k-x_k\|_{L_2}\right) \right]. 
\end{aligned}
\end{equation*}
Meanwhile, by the Taylor expansion, the first term on the left-hand-side of \eqref{eqn:totaldiff} can be written as,
\begin{equation*}
\begin{aligned}
& \frac{1}{n}\sum_{i=1}^n\left[ \int_0^{t_i}\left\{ F_j(\xbf(t))-\widehat{F}_j(\widehat{\xbf}(t)) \right\}dt \right]^2 \\
= \;\; & \frac{1}{n}\sum_{i=1}^n\left[ \int_0^{t_i}\left\{ \widehat{F}_j(x(t))-F_j(x(t)) \right\}dt + \int_0^{t_1} \frac{\partial}{\partial t}\widehat{F}_j(x(t))\{\widehat{x}(t)-x(t)\}dt \right.\\
&\quad\quad\quad\quad\quad\quad\quad\quad\quad\quad\quad\quad\quad\quad\quad\quad\quad\quad\quad\quad \left.- o_p\left( \max_{k=1,\ldots,p}\|\widehat{x}_k-x_k\|_{L_2} \right)\right]^2\\
= \;\; & \frac{1}{n}\sum_{i=1}^n\left[ \int_0^{t_i}\left\{ F_j(\xbf(t))-\widehat{F}_j(\xbf(t)) \right\}dt\right]^2 + \frac{1}{n}\sum_{i=1}^n\left[ \int_0^{t_i} \frac{\partial}{\partial t} \widehat{F}_j(x(t))\{\widehat{x}(t)-x(t)\}dt \right]^2\\
&\quad -\frac{2}{n}\sum_{i=1}^n\int_0^{t_i}\left\{ F_j(\xbf(t))-\widehat{F}_j(\xbf(t)) \right\}dt \int_0^{t_i} \frac{\partial}{\partial t}\widehat{F}_j(x(t))\{ \widehat{x}+(t)-x(t) \}dt + \Rbf_1,
\end{aligned}
\end{equation*}
where the remainder term $\Rbf_1$ is of the form, 
\begin{equation*}
\begin{aligned}
\Rbf_1 & = \frac{1}{n}\sum_{i=1}^n\left( o_p\left( \max_{k=1,\ldots,p}\|\widehat{x}_k-x_k\|^2_{L_2} \right) \right. \\
&\quad \left. - o_p\left( \max_{k=1,\ldots,p}\|\widehat{x}_k-x_k\|_{L_2} \right) \int_0^{t_i}\left[ F_j(x(t))-\widehat{F}_j(x(t)) - \frac{\partial}{\partial t} \widehat{F}_j(x(t))\{ \widehat{x}(t)-x(t) \} \right] dt \right).
\end{aligned}
\end{equation*}
Therefore, the inequality \eqref{eqn:totaldiff} is equivalent to
\begin{equation} \label{eqn:newtotaldiff}
\begin{aligned}
& \frac{1}{n}\sum_{i=1}^n \left[ \int_0^{t_i}\left\{ F_j(\xbf(t))-\widehat{F}_j(\xbf(t)) \right\}dt \right]^2 + \frac{1}{n}\sum_{i=1}^n\left[\int_0^{t_i}\frac{\partial}{\partial t} \widehat{F}_j(x(t))\{ \widehat{x}(t)-x(t) \}dt \right]^2  \\
& + \frac{2}{n}\sum_{i=1}^n\int_0^{t_i}\left\{ \widehat{F}_j(\xbf(t))-F_j(\xbf(t)) \right\}dt \int_0^{t_i} \frac{\partial}{\partial t}\widehat{F}_j(x(t))\{\widehat{x}(t)-x(t)\}dt + \Rbf_1 + \tau_{nj} J_2(\widehat{F}_j)  \\
\leq \;\; & \frac{2}{n}\sum_{i=1}^n\epsilonbf_{ij}\left\{\int_0^{t_i}(\widehat{F}_j-F_j)(x(t))dt\right\} +
\frac{1}{n}\sum_{i=1}^n\left[ \int_0^{t_i}\left\{ F_j(\xbf(t))-F_j(\widehat{\xbf}(t)) \right\}dt \right]^2  \\
 & + \frac{2}{n}\sum_{i=1}^n\epsilon_{ij}\left[ \int_0^{t_i} \frac{\partial}{\partial t}(\widehat{F}_j-F_j)(x)\{ \widehat{x}(t)-x(t) \}dt +  o_p\left( \max_{k=1,\ldots,p}\|\widehat{x}_k-x_k\|_{L_2} \right) \right] + \tau_{nj} J_2(F_j)  \\
\end{aligned}
\end{equation}
Write the left-hand side of \eqref{eqn:newtotaldiff} as $\widetilde\Delta_1 + \widetilde\Delta_2 + \widetilde\Delta_3 + R_1 + \tau_{nj} J_2(\widehat{F}_j)$, and the right-hand side of \eqref{eqn:newtotaldiff} as $\Delta_1 + \Delta_2 + \Delta_3 + \tau_{nj} J_2(F_j)$. Our proof strategy is to derive the upper and lower bounds for the left and right-hand sides of \eqref{eqn:newtotaldiff}, respectively, then put them together.

\bigskip
\noindent
\textbf{Step 1: Bounding the right-hand-side of (\ref{eqn:newtotaldiff})}. 
We first bound the three terms $\Delta_1, \Delta_2, \Delta_3$ on the right-hand-side of \eqref{eqn:newtotaldiff}. 

For $\Delta_1$, by Lemma \ref{lem:bdonradamacher1} and the Minkowski inequality, we have, 
\begin{equation*}
\begin{aligned}
\Delta_1 \leq \; & O_p \bigg\{ \left\| \widehat{F}_j(x(t))-F_j(x(t)) \right\|_{L_2}^2 \log^{-2}\left\| \widehat{F}_j(x(t))-F_j(x(t)) \right\|_{L_2} \\
& + \left(\frac{n}{\log n}\right)^{-\frac{2\beta_2}{2\beta_2+1}} + \frac{\log p}{n} + \sqrt{\frac{\log p}{n}} \left\| \widehat{F}_j(x(t))-F_j(x(t)) \right\|_{L_2} \bigg\}.
\end{aligned}
\end{equation*}

For $\Delta_2$, by the Taylor expansion and Assumption \ref{assump:complexity}, we have, 
\begin{equation}
\label{eqn:bdonerrorinx}
\begin{aligned}
\Delta_2 \leq \; & \frac{c}{n}\sum_{i=1}^n \left[ \int_0^{t_i} \frac{\partial}{\partial t} F_j(\xbf(t))\{ \xbf(t)-\widehat{\xbf}(t) \} + o_p\left( \max_{k=1,\ldots,p}\|x_k-\widehat{x}_k\|^2_{L_2} \right)dt \right]^2\\
\leq \; & \|F_j\|_\HH^2\max_{k=1,\ldots,p}\|x_k-\widehat{x}_k\|_{L_2}^2 + o_p\left(\max_{k=1,\ldots,p}\|x_k-\widehat{x}_k\|^2_{L_2}\right) 
= O_p\left( n^{\frac{-2\beta_1}{2\beta_1+1}} \right).
\end{aligned}
\end{equation}
for some constant $c$, where the second step is by the Jensen's inequality, and the last step is due to Theorem \ref{thm:optimalestofpredictor}.

For $\Delta_3$, since $\beta_2>1$, $\partial K(x,\cdot)/\partial x_k\in\HH$, and by the reproducing property, we have, 
\begin{equation*}
\frac{\partial (\widehat{F}_j-F_j)(x)}{\partial x_k} = \left\langle\widehat{F}_j-F_j,\frac{\partial K(x,\cdot)}{\partial x_k}\right\rangle_\HH\leq \|\widehat{F}_j-F_j\|_\HH^{1/2} \left\|\frac{\partial K(x,\cdot)}{\partial x_k}\right\|^{1/2}_\HH<\infty.
\end{equation*}
Hence, $\partial (\widehat{F}_j-F_j)(x)/\partial x_k\in\HH$, and for any $x$, $|\partial (\widehat{F}_j-F_j)(x)/\partial x_k|\leq\|\partial (\widehat{F}_j-F_j)(x)/\partial x_k\|_\HH<\infty$, which together with Assumption \ref{assump:fluctuation}, implies that $\max_{k}\left\{|\partial (\widehat{F}_j-F_j)(x)/\partial x_k|\right\} \leq C\|\widehat{F}_j-F_j\|_{L_2}$ almost surely. By Assumption \ref{assump:complexity} and the Cauchy-Schwarz inequality, we have, 
\begin{equation*}
\begin{aligned}
\Delta_3 \leq \;\; & \frac{2c}{n}\sum_{i=1}^n|\epsilon_{ij}|\int_0^{t_i} C\|\widehat{F}_j(x(t))-F_j(x(t))\|_{L_2} \max_{k=1,\ldots,p}|\widehat{x}_k(t)-x_k(t)|dt \\
&\quad\quad\quad\quad\quad\quad\quad\quad\quad\quad\quad\quad\quad\quad\quad\quad + o_p\left( n^{-1/2}\max_{k=1,\ldots,p}\|x_k-\widehat{x}_k\|^2_{L_2} \right) \\
\leq \;\; & 2c\max_{k=1,\ldots,p}\|\widehat{x}_k-x_k\|_{L_2} \left\| \widehat{F}_j(x(t))-F_j(x(t)) \right\|_{L_2} \frac{1}{n}\sum_{i=1}^n|\epsilon_{ij} C| \\
&\quad\quad\quad\quad\quad\quad\quad\quad\quad\quad\quad\quad\quad\quad\quad\quad+ o_p\left(n^{-1/2}\max_{k=1,\ldots,p}\|x_k-\widehat{x}_k\|^2_{L_2}\right)\\
= \;\; & O_p\left( n^{\frac{-\beta_1}{2\beta_1+1}}\|\widehat{F}_j(x(t))-F_j(x(t))\|_{L_2} \right),
\end{aligned}
\end{equation*}
for some constant $c$, where the last step is due to the strong law of large numbers.

\bigskip
\noindent
\textbf{Step 2: Bounding the left-hand-side of \eqref{eqn:newtotaldiff}}. 
We next bound the terms $\widetilde\Delta_1, \widetilde\Delta_2, \widetilde\Delta_3$ and $R_1$ on the left-hand-side of \eqref{eqn:newtotaldiff}. 

For $\widetilde\Delta_1$, by Lemma \ref{lem:bdonradamacher}, with probability at least $1-2p^{-c_1}$, for some constant $C>0$, 
\begin{equation}
\label{eqn:step2term1}
\begin{aligned}
\widetilde\Delta_1 \geq & \left\| F_j(x(t))-\widehat{F}_j(x(t)) \right\|_{L_2}^2 - C \bigg\{ \left\| F_j(x(t))-\widehat{F}_j(x(t)) \right\|^2_{L_2}\log^{-2} \left\|F_j(x(t))-\widehat{F}_j(x(t)) \right\|_{L_2} \\
+ & \left(\frac{n}{\log n}\right)^{-\frac{2\beta_2}{2\beta_2+1}} + (c_1+1)\frac{\log p}{n} + \sqrt{(c_1+1)\frac{\log p}{n}} \left\|F_j(x(t))-\widehat{F}_j(x(t))\right\|_{L_2} + n^{-1/2}e^{-p} \bigg\}.
\end{aligned}
\end{equation}

For $\widetilde\Delta_2$, we can drop this term, because $\widetilde\Delta_2 \geq 0$. 

For $\widetilde\Delta_3$, by the Cauchy-Schwarz inequality,
\begin{equation*}
\begin{aligned}
\widetilde\Delta_3 \geq \; & -2\left( \frac{1}{n}\sum_{i=1}^n\left[ \int_0^{t_i}\left\{ \widehat{F}_j(\xbf(t))-F_j(\xbf(t)) \right\} dt \right]^2 \right)^{1/2} \\
& \times \left( \frac{1}{n}\sum_{i=1}^n\left[ \int_0^{t_i} \frac{\partial}{\partial t} \widehat{F}_j(x(t))\{ \widehat{x}(t)-x(t) \}dt \right]^2 \right)^{1/2}\\
\geq \; & -2 \left\| \widehat{F}_j(x(t))-F_j(x(t)) \right\|_{L_2} \|F_j\|_\HH\max_{k=1,\ldots,p}\|x_k-\widehat{x}_k\|_{L_2}\\
= \; & O_p\left( n^{\frac{-\beta_1}{2\beta_1+1}} \left\| \widehat{F}_j(x(t))-F_j(x(t)) \right\|_{L_2}  \right),
\end{aligned}
\end{equation*}
where the second step is due to the Minkowski inequality.

For the remainder term $R_1$ on the left-hand-side of \eqref{eqn:newtotaldiff}, by Assumption \ref{assump:complexity} and the Cauchy-Schwarz inequality, we have, 
\begin{equation*}
\begin{aligned}
\Rbf_1 
& = o_p\left(\max_{k=1,\ldots,p}\|x_k-\widehat{x}_k\|_{L_2} \, \left\| \widehat{F}_j(x(t))-F_j(x(t)) \right\|_{L_2} + \max_{k=1,\ldots,p}\|x_k-\widehat{x}_k\|^2_{L_2} \, \|F_j\|_\HH\right)\\
& = o_p\left( n^{\frac{-\beta_1}{2\beta_1+1}} \left\| \widehat{F}_j(x(t))-F_j(x(t)) \right\|_{L_2} \right) + o_p\left( n^{\frac{-2\beta_1}{2\beta_1+1}} \right),
\end{aligned}
\end{equation*}
where the second step is again due to the Minkowski inequality.

\bigskip
\noindent
\textbf{Step 3: Putting the two bounds together}. 
Combining the bounds for each term in \eqref{eqn:newtotaldiff}, we obtain that, for any $c_1>0$ and $c_2>1$, with probability at least $1-4p^{-c_1}$, there exists a constant $C>0$, such that 
\begin{equation*}
\begin{aligned}
 & \|F_j(x(t))-\widehat{F}_j(x(t))\|_{L_2}^2\\
\leq \; & C \left[ c_2^{-\frac{4\beta_2}{2\beta_2-1}} \left\| F_j(x(t))-\widehat{F}_j(x(t)) \right\|^2_{L_2} \log^{-2} \left\| F_j(x(t))-\widehat{F}_j(x(t)) \right\|_{L_2} + c_2^{\frac{4\beta_2}{4\beta_2+1}} \left(\frac{n}{\log n}\right)^{-\frac{2\beta_2}{2\beta2+1}} \right. \\
& + (c_1+1) \frac{\log p}{n} + \sqrt{(c_1+1) \frac{\log p}{n}} \left\| F_j(x(t))-\widehat{F}_j(x(t)) \right\|_{L_2} + n^{-1/2}e^{-p}\\
& \left. + n^{\frac{-\beta_1}{2\beta_1+1}} \left\| \widehat{F}_j(x(t))-F_j(x(t)) \right\|_{L_2} + n^{\frac{-2\beta_1}{2\beta_1+1}}+ \tau_{nj} \left\{ J_2(F_j)-J_2(\widehat{F}_j) \right\} \right].
\end{aligned}
\end{equation*}
Taking $c_2$ large enough such that $C c_2^{-4\beta_2/(2\beta_2-1)} \leq 1/2$, then 
\begin{equation*}
\begin{aligned}
& \left\|F_j(x(t)) - \widehat{F}_j(x(t)) \right\|_{L_2}^2 \log^{-2} \left\| F_j(x(t))-\widehat{F}_j(x(t)) \right\|_{L_2} \leq 2C \left[ c_2^{\frac{4\beta_2}{4\beta_2+1}}\left(\frac{n}{\log n}\right)^{-\frac{2\beta_2}{2\beta2+1}} \right. \\
& \quad\quad\quad\quad\; +(c_1+1)\frac{\log p}{n} + \sqrt{(c_1+1)\frac{\log p}{n}} \left\| F_j(x(t))-\widehat{F}_j(x(t)) \right\|_{L_2} \\
& \left. \quad\quad\quad\quad\; + n^{-1/2}e^{-p}+  n^{\frac{-\beta_1}{2\beta_1+1}} \left\| \widehat{F}_j(x(t))-F_j(x(t)) \right\|_{L_2} + n^{\frac{-2\beta_1}{2\beta_1+1}}+\tau_{nj} \left\{ J_2(F_j)-J_2(\widehat{F}_j) \right\} \right].
\end{aligned}
\end{equation*}
Therefore, 
\begin{equation*} \label{eqn:bdonl2}
\begin{aligned}
\left\| F_j(x(t))-\widehat{F}_j(x(t)) \right\|_{L_2}^2 = O_p\left\{ \left( \frac{n}{\log n} \right)^{-\frac{2\beta_2}{2\beta2+1}} + \frac{\log p}{n}  + n^{-\frac{2\beta_1}{2\beta_1+1}} \right\}.
\end{aligned}
\end{equation*}
This leads to the desired upper bound.

\subsubsection{Lower bound}
\label{sec:lowbdfunctional}

\noindent
We first construct a matrix $A_j \in \R^{p^2 \times \widetilde{N}}$ for each $j=1,\ldots,p$, whose entry is chosen from $\{\pm 1,0\}$, and is used to index a set of functions for establishing the lower bound. Here, the value of $\widetilde{N}$ is to be specified later. We choose $1\leq s_j<\infty$ rows of $A_j$ to be nonzero. By the Vershamov-Gilbert Lemma \citep{Tsybakov2009}, there exist a set $\{\zeta_1,\ldots,\zeta_{m_1}\}\subset\{0,1\}^{p^2}$ such that, (a) $\|\zeta_k\|_{l_1}=s_j$, for $k=1, \ldots, m_1$; (b) $\|\zeta_{k_1}-\zeta_{k_2}\|_{l_1}\geq s_j/2$, for $k_1\neq k_2$; and (c) $4 \log m_1\geq s_j\log(p^2/s_j)$. By the same lemma, there exist a set $\{\zeta_1^\dagger,\ldots,\zeta_{m_2}^\dagger\}\in\{-1,1\}^{s_j \times \widetilde{N}}$ such that, (a$'$) $\|\zeta^\dagger_{k_1} - \zeta^\dagger_{k_2}\|_{F} \geq \widetilde{N}s_j/2$, for $k_1\neq k_2$; and (b$'$) $8 \log m_2 \geq \widetilde{N} s_j$. We set the zero rows of $A_j$ according to $\zeta_k$, and set the nonzero rows of $A_j$ according to $\zeta_k^\dagger$. As such, the matrix $A_j$ is chosen from the set, 
\begin{equation*}
\A = \left\{ A_j(\zeta_{k_1},\zeta^\dagger_{k_2})\in\R^{p^2\times \widetilde{N}}:k_1=1,\ldots,m_1, k_2=1,\ldots,m_2 \right\},
\end{equation*}
where $\text{card}(\A) = m_1m_2$. By the above constructions (c) and (b$'$), we have that, 
\begin{equation*}
\log \text{card}(\A) \geq \frac{1}{4}s_j\log(p^2/s_j) + \frac{1}{8} \widetilde{N} s_j.
\end{equation*}

Next, we define functions of the form $g_{A_j}$ with $A_j\in\A $. Note that, by the spectral theorem, the reproducing kernel $K_j$ of the RKHS $\HH_j$ admits the eigenvalue decomposition
\begin{equation*}
K_j(\widehat{x}_j,\widehat{x}_j')=\sum_{\nu\ge 1}\gamma_{j\nu}\phi_{j\nu}(\widehat{x}_j)\phi_{j\nu}(\widehat{x}_j')
\end{equation*}
where $\gamma_{1}\geq \gamma_2\geq \cdots\geq 0$ are its eigenvalues, and $\{\phi_\nu:\nu\geq 1\}$ are the corresponding eigenfunctions that are orthonormal in $L_2$. Since $\HH_j$ is embedded to a $\beta_2$th-order Sobolev space, the eigenvalues decays as $\gamma_{j\nu}\asymp\nu^{-2\beta_2}$ \citep{wahba1990}. We define the function, 
\begin{equation*}
\begin{aligned}
g_{A_j}(\widehat{x}_1,\ldots,\widehat{x}_p) = \; & \widetilde{N}^{-1/2}\sum_{j=1}^p\sum_{\nu=1}^{\widetilde{N}} a_{j^2,\nu}\gamma^{1/2}_{j,\widetilde{N}+\nu}\phi_{j,\widetilde{N}+\nu}(\widehat{x}_j)\\
& + \widetilde{N}^{-1/2}\sum_{j, k=1,\ldots,p;j\neq k}\sum_{\nu=1}^{\widetilde{N}} a_{j\cdot k,\nu}\gamma^{1/2}_{j,\widetilde{N}+\nu}\gamma_{k,\widetilde{N}+\nu}^{1/2}\phi_{j,\widetilde{N}+\nu}(\widehat{x}_j)\phi_{k,\widetilde{N}+\nu}(\widehat{x}_k),\quad A_j\in\A,
\end{aligned}
\end{equation*}
Let $\| \cdot \|^0_\HH$ denote the $\ell_0$-norm. Then, we have, 
\begin{equation*}
\begin{aligned}
\|g_{A_j}\|^0_\HH &\leq \sum_{j=1}^p\left\|\sum_{\nu=1}^{\tilde{N}} a_{j^2,\nu}\gamma_{j,\widetilde{N}+\nu}^{1/2}\phi_{j,\widetilde{N}+\nu}(\widehat{x}_j)\right\|_{\HH}^0\\
&\quad+\sum_{j,k=1,\ldots,p;j\neq k}\left\|\sum_{\nu=1}^Na_{j\cdot k,\nu}\gamma_{j,\widetilde{N}+\nu}^{1/2}\gamma_{k,\widetilde{N}+\nu}^{1/2}\phi_{j,\widetilde{N}+\nu}(\widehat{x}_j)\phi_{k,\widetilde{N}+\nu}(\widehat{x}_k)\right\|_{\HH}^0\leq s_j.
\end{aligned}
\end{equation*}
For any two matrices $A_j,B_j\in\A$, we have,
\begin{equation*}
\begin{aligned}
\|g_{A_j}-g_{B_j}\|_{L_2}^2 \geq \; & C_1\widetilde{N}^{-1}\sum_{j=1}^p\sum_{\nu=1}^{\widetilde{N}}\gamma_{j,\widetilde{N}+\nu}\left(a_{j^2,\nu}-b_{j^2,\nu}\right)^2 \\
& + \; C_1\widetilde{N}^{-1}\sum_{j,k=1,\ldots,p;j\neq k}\sum_{\nu=1}^{\widetilde{N}}\gamma_{j,\widetilde{N}+\nu}\gamma_{k,\widetilde{N}+\nu}\left(a_{j\cdot k,\nu}-b_{j\cdot k,\nu}\right)^2\\
\geq \; & C_2\widetilde{N}^{-1}(2\widetilde{N})^{-4\beta_2}\sum_{j,k=1,\ldots,p;j\neq k}\sum_{\nu=1}^{\widetilde{N}}\left(a_{j\cdot k,\nu}-b_{j\cdot k,\nu}\right)^2 \geq C_3s\widetilde{N}^{-4\beta_2},
\end{aligned}
\end{equation*}
for some constants $C_1,C_2,C_3> 0$, where the second and third steps are by the construction (a$'$). On the other hand, for any $A_j\in\A$, and by the Minkowski inequality,  
\begin{equation*}
\begin{aligned}
\|g_{A_j}\|_{L_2}^2 \leq \; & C_4\widetilde{N}^{-1}\sum_{j=1}^{p}\left\|\sum_{\nu=1}^Na_{j^2,\nu}\gamma_{j,\widetilde{N}+\nu}^{1/2}\phi_{j,\widetilde{N}+\nu}\right\|_{L_2}^2\\
& + \; C_4\widetilde{N}^{-1}\sum_{j,k=1,\ldots,p;j\neq k}\left\|\sum_{\nu=1}^Na_{j\cdot k,\nu}\gamma_{j,\widetilde{N}+\nu}^{1/2}\gamma_{k,\widetilde{N}+\nu}^{1/2}\phi_{j,\widetilde{N}+\nu}\phi_{k,\widetilde{N}+\nu}\right\|_{L_2}^2\\
\leq \; & C_4\widetilde{N}^{-1}\sum_{j=1}^p\sum_{\nu=1}^{\widetilde{N}} \gamma_{j,\widetilde{N}+\nu}a_{j^2,\nu}^2 
\leq C_5\widetilde{N}^{-1}\widetilde{N}^{-2\beta_2}\widetilde{N}s_j = C_5s_j\widetilde{N}^{-2\beta_2},
\end{aligned}
\end{equation*}
for some constants $C_4,C_5>0$, where the second and third steps are by (a).

We are now ready to derive the lower bound. Let $\mathcal Z$ denote a random variable uniformly distributed on $\{1,2,\ldots,\text{card}(\A)\}$. Then for any $j=1,\ldots,p$,
\begin{equation*}
\inf_{\widetilde{F}_j}\sup_{F_j\in\HH}\P\left\{\|\widetilde{F}_j(\widehat{x})-F_j(\widehat{x})\|_{L_2}^2\geq \frac{1}{4}\min_{A_j\neq B_j\in\A}\|g_{A_j}-g_{B_j}\|_{L_2}^2\right\}\geq \inf_{\widehat{\mathcal Z}}\P\left\{ \widehat{\mathcal Z} \neq \mathcal Z \right\},
\end{equation*}
where the infimum on the right-hand-side is taken over all decision rules that are measurable functions of the data \citep{Tsybakov2009}. By the Fano's Lemma, we have, 
\begin{equation*}
\P\left\{\widehat{\mathcal Z}\neq \mathcal Z|t_1,\ldots,t_n\right\}\geq 1-\frac{1}{\log\{\text{card}(\A)\}} \bigg[ \mathcal I_{t_1,\ldots,t_n}(y_{1j},\ldots,y_{nj};\mathcal Z)+\log 2 \bigg],
\end{equation*}
where $\mathcal I_{t_1,\ldots,t_n}(y_{1j},\ldots,y_{nj};\mathcal Z)$ is the mutual information between $\mathcal Z$ and $(y_{1j},\ldots,y_{nj})$ conditioning on $(t_1,\ldots,t_n)$. Note that
\begin{equation*}
\begin{aligned}
\E_{t_1,\ldots,t_n}\left[\mathcal I_{t_1,\ldots,t_n}(y_{1j},\ldots,y_{nj};\mathcal Z)\right] 
& \leq \frac{n}{\text{card}(\A)\{ \text{card}(\A)-1 \}}\sum_{A_j\neq B_j\in\A}\E_{t_1,\ldots,t_n}\|g_{A_j}-g_{B_j}\|^2_{L_2}\\
& \leq \frac{n}{2}\max_{A_j\neq B_j\in\A}\|g_{A_j}-g_{B_j}\|_{L_2}^2 
\leq 2n\max_{A_j\in\A}\|g_{A_j}\|_{L_2}^2\leq 2C_5ns_j \widetilde{N}^{-2\beta_2}.
\end{aligned}
\end{equation*}
Henceforth,
\begin{equation*}
\begin{aligned}
\inf_{\widetilde{F}_j}\sup_{F_j\in\HH}\P\left\{\|\widetilde{F}_j(\widehat{x})-F_j(\widehat{x})\|_{L_2}^2\geq C_3s_j \widetilde{N}^{-4\beta_2}\right\}\geq \inf_{\widehat{\mathcal Z}}\P\{\widehat{\mathcal Z}\neq \mathcal Z\} 
\geq 1- \frac{2C_5ns_j \widetilde{N}^{-2\beta_2}+\log 2}{\frac{1}{4}s_j\log(p^2/s_j)+\frac{1}{8} \widetilde{N} s_j}.
\end{aligned}
\end{equation*}
Taking $\widetilde{N}=1$ and $s_j =C_6n^{-1}\log p$ for a sufficiently small constant $C_6$ yields that
\begin{equation*}
\inf_{\widetilde{F}_j}\sup_{F_j\in\HH}\P\left\{\|\widetilde{F}_j(\widehat{x})-F_j(\widehat{x})|_{L_2}^2\geq C_7\frac{\log p}{n}\right\}\geq \frac{1}{2}, 
\end{equation*}
for some constant $C_7>0$. Meanwhile, taking $s_j=1$ and $\widetilde{N}=C_8(n\log n^{-1})^{\frac{1}{4\beta_2+2}}$ for a sufficiently small $C_8>0$ yields that
\begin{equation*}
\inf_{\widetilde{F}_j}\sup_{F_j\in\HH}\P\left\{\|\widetilde{F}_j(\widehat{x})-F_j(\widehat{x})\|_{L_2}^2\geq C_9(n\log^{-1} n)^{-\frac{2\beta_2}{2\beta_2+1}}\right\} \geq \frac{1}{2},
\end{equation*}
for some $C_9>0$. Therefore, we have
\begin{equation*}
\inf_{\widetilde{F}_j}\sup_{F_j\in\HH}\P\left[\|\widetilde{F}_j(\widehat{x})-F_j(\widehat{x})\|_{L_2}^2\geq C_{10}\left\{\frac{\log p}{n}+(n\log^{-1}n)^{-\frac{2\beta_2}{2\beta_2+1}}\right\}\right] \geq \frac{1}{2},
\end{equation*}
for some $C_{10}>0$.
Finally, note that, $\widehat{x}_j$ is an estimator of $x_j$ satisfying that $\|\widehat{x}_j-x_j\|_{L_2}^2 = O_p\left( n^{-\frac{2\beta_1}{2\beta_1+1}} \right)$. Then for any $F_j, \widetilde{F}_j \in \HH$,
\begin{equation*}
\P\left[ \min\left\{\|F_j(x)-F_j(\widehat{x})\|_{L_2}^2,\|\widetilde{F}_j(x)-\widetilde{F}_j(\widehat{x})\|_{L_2}^2\right\}\geq C_{11}n^{-\frac{2\beta_1}{2\beta_1+1}} \right] \geq \frac{1}{2},
\end{equation*}
Therefore,
\begin{equation*}
\inf_{\widetilde{F}_j}\sup_{F_j\in\HH}\P\left[ \|\widetilde{F}_j(x)-F_j(x)\|_{L_2}^2\geq C_{12}\left\{ \frac{\log p}{n}+(n\log^{-1}n)^{-\frac{2\beta_2}{2\beta_2+1}}+n^{-\frac{2\beta_1}{2\beta_1+1}} \right\} \right] \geq \frac{1}{2},
\end{equation*}
for some constant $C_{12}>0$, which completes the proof of Theorem \ref{thm:optimalestoffunctional}. 
\eop

\subsubsection{Auxiliary lemmas for Theorem \ref{thm:optimalestoffunctional}}
\label{sec:auxlem}

\noindent 
For any $g\in\HH$, define the norm, $\|g(x(t))\|_n =\sqrt{(1/n) \sum_{i=1}^ng^2(x(t_i))}$.

\begin{lemma}
\label{lem:bdonradamacher1}
Suppose that $F_j\in\HH$, and the errors $\{\epsilonbf_{ij}\}_{i=1}^n$ are i.i.d.\ Gaussian. Then there exists some constant $C>0$ such that, for any $c_1>0$ and $c_2>1$, with probability at least $1-2p^{-c_1}$,
\begin{equation*}
\begin{aligned}
& \frac{1}{n}\sum_{i=1}^n\epsilon_{ij}F_j(x(t_{i}))\\
\leq \; & C \left\{ c_2^{-\frac{4\beta_2}{2\beta_2-1}}\|F_j(x(t))\|_{L_2}^2 \log^{-2}\|F_j(x(t))\|_{L_2} + \left( c_2^{-\frac{4\beta_2}{2\beta_2-1}}+c_2^{\frac{4\beta_2}{4\beta_2+1}} \right) \left( \frac{n}{\log n} \right)^{-\frac{2\beta_2}{2\beta_2+1}} \right. \\
& \left. + \left( c_2^{-\frac{4\beta_2}{2\beta_2-1}} + c_1 + 1 \right) \frac{\log p}{n} + \sqrt{(c_1+1) \frac{\log p}{n} } \|F_j(x(t))\|_{L_2} + n^{-1/2}e^{-p} \right\}.
\end{aligned}
\end{equation*}
\end{lemma}

\noindent
\textbf{Proof of Lemma \ref{lem:bdonradamacher1}}:
Recall the RKHS $\HH$ defined in \eqref{eqn:spaceH}. For notational simplicity, we denote $F_{jk} \equiv F_{jkk}$ for $k=1,\ldots,p$.
It has been shown that the $\nu$th eigenvalue of the reproducing kernel of RKHS $\HH$ is of order $(\nu\log^{-1}\nu)^{-2\beta_2}$, for $\nu\geq 1$; see, e.g., \citet{Bach2017}. Since $\{\epsilon_{ij}\}_{i=1}^{n}$ are i.i.d.\ Gaussian, by Lemma 2.2 of \citet{Yuan2016} and
Corollary 8.3 of \citet{Vandegeer2000}, we have that, for any $c_1>0$, with probability at least $1-p^{-c_1}$,
\begin{equation}
\label{eqn:empmean}
\begin{aligned}
& \frac{1}{n}\sum_{i=1}^n\epsilon_{ij}F_j(x(t_{i})) \\
\leq \; & 2C_1n^{-1/2}\sum_{k,l=1}^p \left(\|F_{jkl}(x(t))\|_{n}\log^{-1}\|F_{jkl}(x(t))\|_{n}\right)^{1-\frac{1}{2\beta_2}}  \left( \|F_{jkl}\|_{\HH}\log^{-1}\|F_{jkl}\|_{\HH} \right)^{\frac{1}{2\beta_2}} \\
& + 2C_1n^{-1/2}\sqrt{(c_1+1)\log p}\sum_{k,l=1}^p  \|F_{jkl}(x(t))\|_{n}  + 2C_1n^{-1/2}e^{-p}\sum_{k,l=1}^p\|F_{jkl}\|_{\HH} \\ 
\equiv \; & 2C_1 (\Delta_4 + \Delta_5 + \Delta_6), 
\end{aligned}
\end{equation}
for some constant $C_1$. Next, we bound the three terms $\Delta_4, \Delta_5, \Delta_6$ on the right-hand-side of \eqref{eqn:empmean}, respectively. 

For $\Delta_4$, by the Young's inequality, for any $c_2>1$, we have, 
\begin{equation*}
\begin{aligned}
\Delta_4 \leq \; & c_2^{-\frac{4\beta_2}{2\beta_2-1}}\sum_{k,l=1}^p \left(\|F_{jkl}(x(t))\|_{n}\log^{-1}\|F_{jkl}(x(t))\|_{n}\right)^{2}\\
& + c_2^{\frac{4\beta_2}{4\beta_2+1}}n^{-\frac{2\beta_2}{2\beta2+1}}\sum_{k,l=1}^p \left( \|F_{jkl}\|_{\HH}\log^{-1}\|F_{jkl}\|_{\HH} \right)^{\frac{2}{2\beta_2+1}}.
\end{aligned}
\end{equation*}
Note that 
\begin{equation*}
\sum_{k,l=1}^d \left( \|F_{jk}\|_{\HH}\log^{-1}\|F_{jk}\|_{\HH} \right)^{\frac{2}{2\beta_2+1}} \leq C'_2 \sum_{k,l=1}^d \left( \|F_{jk}\|_{\HH}\log^{-1}\|F_{jk}\|_{\HH} \right)^{0}  \leq C_2, 
\end{equation*}
for some constants $C_2',C_2$, where the last step is due to Assumption \ref{assump:complexity} that the number of nonzero functional components of $F_j$ is bounded. Henceforth,
\begin{equation}
\label{eqn:firststepbd}
\begin{aligned}
& n^{-1/2}\sum_{k,l=1}^d\left(\|F_{jkl}(x(t))\|_{n}\log^{-1}\|F_{jkl}(x(t))\|_{n}\right)^{1-\frac{1}{2\beta_2}} \left(\|F_{jkl}\|_{\HH}\log^{-1}\|F_{jkl}\|_{\HH}\right)^{\frac{1}{2\beta_2}}\\
\leq \; & c_2^{-\frac{4\beta_2}{2\beta_2-1}}\sum_{k,l=1}^p\left(\|F_{jkl}(x(t))\|_{n}\log^{-1}\|F_{jkl}(x(t))\|_{n}\right)^{2} + c_2^{\frac{4\beta_2}{4\beta_2+1}}n^{-\frac{2\beta_2}{2\beta2+1}}C_2.
\end{aligned}
\end{equation}
By Theorem 4 of \citet{Koltchinskii2010}, there exists some constant $C_3>0$ such that, with probability at least $1-p^{-c_1}$,
\begin{equation*}
\begin{aligned}
\sum_{k,l=1}^p\left( \|F_{jkl}(x(t))\|_n\log^{-1}\|F_{jkl}(x(t))\|_{n} \right)^2
\leq 2C_3^2\sum_{k,l=1}^p \left( \|F_{jkl}(x(t))\|_{L_2}\log^{-1}\|F_{jkl}(x(t))\|_{L_2} \right)^2 + \\
+ \; 2C_3^2\left\{ \left( \frac{n}{\log n} \right)^{-\frac{2\beta_2}{2\beta_2+1}}+\frac{(c_1+1)\log d}{n} \right\} \sum_{k,l=1}^p \left( \|F_{jkl}\|_{\HH}\log^{-1} \|F_{jkl}\|_{\HH} \right)^2.
\end{aligned}
\end{equation*}
Note that there exists some constant $c_3>1$, such that 
\begin{equation*}
\sum_{k,l=1}^p \left( \|F_{jkl}\|_{L_2}\log^{-1}\|F_{jkl}\|_{L_2} \right)^2 \leq c_3 \left( \|F_j\|_{L_2}\log^{-1}\|F_j\|_{L_2} \right)^2,
\end{equation*} and 
\begin{equation*}\sum_{k,l=1}^p \left( \|F_{jkl}\|_{\HH}\log^{-1} \|F_{jkl}\|_{\HH} \right)^2 \leq \sum_{k,l=1}^p ( \|F_{jkl}\|_{\HH} \log^{-1} \|F_{jkl}\|_{\HH} )^0\leq C_2.
\end{equation*} Then, we have
\begin{equation*}
\begin{aligned}
\sum_{k,l=1}^p \left( \|F_{jkl}(x(t))\|_n\log^{-1}\|F_{jkl}(x(t))\|_{n} \right)^2 \leq 2C_3^2c_3 \left( \|F_j(x(t))\|_{L_2}\log^{-1}\|F_j(x(t))\|_{L_2} \right)^2 \\
+ \; 2C_2C_3^2\left\{ \left(\frac{n}{\log n} \right)^{-\frac{2\beta_2}{2\beta_2+1}} + \frac{(c_1+1)\log p}{n} \right\}.
\end{aligned}
\end{equation*}
Inserting into (\ref{eqn:firststepbd}) yields that
\begin{equation}
\label{eqn:firststepbdfinal}
\begin{aligned}
\Delta_4 \leq \; & 2C_3^2c_3c_2^{-\frac{4\beta_2}{2\beta_2-1}}(\|F_j(x(t))\|_{L_2}\log^{-1}\|F_j(x(t))\|_{L_2})^2 \\
& + 2C_2C_3^2c_2^{-\frac{4\beta_2}{2\beta_2-1}} \left\{ \left( \frac{n}{\log n} \right)^{-\frac{2\beta_2}{2\beta_2+1}}+\frac{(c_1+1)\log p}{n} \right\} +  C_2c_2^{\frac{4\beta_2}{4\beta_2+1}} \left( \frac{n}{\log n} \right)^{-\frac{2\beta_2}{2\beta2+1}}.
\end{aligned}
\end{equation}

For $\Delta_5$, by Theorem 4 of \citet{Koltchinskii2010} again, there exists a constant $C_4>0$, such that
\begin{equation*}
\begin{aligned}
& \sum_{k,l=1}^p\|F_{jkl}(x(t))\|_{n} \\
\leq \; & C_4\sum_{k,l=1}^p\|F_{jkl}(x(t))\|_{L_2}+C_4\left\{ \left(\frac{n}{\log n}\right)^{-\frac{\beta_2}{2\beta_2+1}}+\sqrt{\frac{(c_1+1)\log p}{n}} \right\}\sum_{k,l=1}^p\|F_{jkl}\|_{\HH}\\
\leq \; & C_4\sum_{k,l=1}^p\|F_{jkl}(x(t))\|_{L_2}+C_2C_4\left\{ \left(\frac{n}{\log n}\right)^{-\frac{\beta_2}{2\beta_2+1}}+\sqrt{\frac{(c_1+1)\log p}{n}}\right\}.
\end{aligned}
\end{equation*}
Define the set $\mathcal Q_1 \equiv \left\{ k,l=1,\ldots,p: \|F_{jkl}(x(t))\|_{L_2}>\sqrt{n^{-1}\log p} \right\}$. By the Cauchy-Schwartz inequality, we have, 
\begin{equation*}
\begin{aligned}
& \sum_{k,l\in\mathcal Q_1}\|F_{jkl}(x(t))\|_{L_2} \leq \text{card}^{1/2}(\mathcal Q_1)\cdot\left(\sum_{k,l\in\mathcal Q_1}\|F_{jkl}(x(t))\|^2_{L_2}\right)^{1/2}\\
\leq \; & \sum_{k,l=1}^p\|F_{jkl}\|_\HH^0\cdot\left(\sum_{k,l=1}^p\|F_{jkl}(x(t))\|^2_{L_2}\right)^{1/2}\leq C_2c_4\|F_j(x(t))\|_{L_2}, 
\end{aligned}
\end{equation*}
where $c_4>1$ satisfies that $\sum_{k,l=1}^p\|F_{jkl}(x(t))\|^2_{L_2}\leq c^2_4\|F_j(x(t))\|^2_{L_2}$. Next, define the set $\mathcal Q_2 \equiv \{k,l=1,\ldots,p: \|F_{jkl}(x(t))\|_{L_2}\leq\sqrt{n^{-1}\log p}\}$. By definition,
\begin{equation*}
\begin{aligned}
\sum_{k,l\in\mathcal Q_2}\|F_{jkl}(x(t))\|_{L_2}   \leq \sum_{k,l\in\mathcal Q_2}\|F_{jkl}(x(t))\|^0_{L_2}\sqrt{\frac{\log p}{n}} \sqrt{\frac{\log p}{n}}\sum_{k,l=1}^p\|F_{jkl}(x(t))\|^0_{L_2}\leq C_2 \sqrt{\frac{\log p}{n}}.
\end{aligned}
\end{equation*}
Combining $\mathcal Q_1$ and $\mathcal Q_2$ gives,
\begin{equation*}
\begin{aligned}
\sum_{k,l=1}^p\|F_{jkl}(x(t))\|_{L_2} &\leq \sum_{k,l\in\mathcal Q_1}\|F_{jkl}(x(t))\|_{L_2} + \sum_{k,l\in\mathcal Q_2}\|F_{jkl}(x(t))\|_{L_2}\\
&\leq C_2c_4\|F_j(x(t))\|_{L_2}+ C_2  \sqrt{\frac{\log p}{n}}.
\end{aligned}
\end{equation*}
Henceforth, we can bound $\Delta_5$ as,
\begin{equation*}
\begin{aligned}
\Delta_5 \leq C_2C_4c_4\sqrt{\frac{\log p}{n}}\|F_j(x(t))\|_{L_2} + C_2C_4\left(\frac{n}{\log n}\right)^{-\frac{\beta_2}{2\beta_2+1}}\sqrt{\frac{\log p}{n}} + 2C_2C_4\sqrt{(c_1+1)}\frac{\log p}{n}.
\end{aligned}
\end{equation*}

For $\Delta_6$, it can be bounded as,
\begin{equation*}
\Delta_6 \leq n^{-1/2}e^{-p}\sum_{k,l=1}^p\|F_{jk}\|^0_{\HH} \leq C_2n^{-1/2}e^{-p}.
\end{equation*}
Combining the bounds for $\Delta_4, \Delta_5, \Delta_6$, and applying the Cauchy-Schwarz inequality completes the proof of Lemma \ref{lem:bdonradamacher1}. 
\eop

\bigskip
\begin{lemma}
\label{lem:bdonradamacher}
Suppose that $F_j\in\HH$. Then there exists some constant $C>0$ such that, for any $c_1>0$ and $c_2>1$, with probability at least $1-2p^{-c_1}$,
\begin{equation*}
\begin{aligned}
\|F_j(x(t))\|_{L_2}^2 & \leq \|F_j(x(t))\|_{n}^2 + C\left\{ c_2^{-\frac{4\beta_2}{2\beta_2-1}}\|F_{j}(x(t))\|^2_{L_2}\log^{-2}\|F_{j}\|_{L_2} + c_2^{\frac{4\beta_2}{4\beta_2+1}}\left( \frac{n}{\log n} \right)^{-\frac{2\beta_2}{2\beta2+1}} \right. \\
& \left. + (c_1+1) \frac{\log p}{n} + \sqrt{(c_1+1) \frac{\log p}{n} } \|F_j(x(t))\|_{L_2}+n^{-1/2}e^{-p} \right\}.
\end{aligned}
\end{equation*}
\end{lemma}

\noindent
\textbf{Proof of Lemma \ref{lem:bdonradamacher}}:
Note that
\begin{equation*}
\|F_j(x(t))\|_{L_2}^2 - \|F_j(x(t))\|_n^2\leq \underset{\begin{subarray}{c} 
g\in\HH,\|g\|_\HH^0\leq \|F_j\|_\HH^0 \\ 
\|g\|_{L_2}\leq \|F_j\|_{L_2}
\end{subarray}}{\sup}\left(\|g\|_{L_2}^2-\|g\|_n^2\right).
\end{equation*}
By the Talagrand's concentration inequality \citep{Talagrand1996}, with probability at least $1-e^{-c_1}$,
\begin{equation}
\label{eqn:decomp1}
\begin{aligned}
\underset{\begin{subarray}{c}
g\in\HH,\|g\|_\HH^0\leq \|F_j\|_\HH^0 \\
\|g\|_{L_2}\leq \|F_j\|_{L_2}
\end{subarray}}{\sup}\left(\|g\|_{L_2}^2-\|g\|_n^2\right) 
\leq 2 \left\{\E \underset{\begin{subarray}{c}
g\in\HH,\|g\|_\HH^0\leq \|F_j\|_\HH^0 \\
\|g\|_{L_2}\leq \|F_j\|_{L_2}
\end{subarray}}{\sup}\left(\|g\|_{L_2}^2-\|g\|_n^2\right)+4\|F_j(x(t))\|_{L_2}\sqrt{\frac{c_1}{n}}+\frac{16c_1}{n}\right\}.
\end{aligned}
\end{equation}
By the symmetrization inequality for the Rademacher process \citep{Vandevaart1996}, there exists a constant $C_1>0$, such that
\begin{equation}
\label{eqn:decomp2}
\begin{aligned}
\E \underset{\begin{subarray}{c}
  g\in\HH,\|g\|_\HH^0\leq \|F_j\|_\HH^0 \\
  \|g\|_{L_2}\leq \|F_j\|_{L_2}
  \end{subarray}}{\sup}\left(\|g\|_{L_2}^2-\|g\|_n^2\right) &\leq \E \underset{\begin{subarray}{c}
  g\in\HH,\|g\|_\HH^0\leq \|F_j\|_\HH^0 \\
  \|g\|_{L_2}\leq \|F_j\|_{L_2}
  \end{subarray}}{\sup}\left\{ \frac{1}{n}\sum_{i=1}^n\omega_ig^2(x(t_i)) \right\} \\
  & \leq C_1\E \underset{\begin{subarray}{c}
  g\in\HH,\|g\|_\HH^0\leq \|F_j\|_\HH^0 \\
  \|g\|_{L_2}\leq \|F_j\|_{L_2}
  \end{subarray}}{\sup}\left\{ \frac{1}{n}\sum_{i=1}^n\omega_ig(x(t_i)) \right\},
\end{aligned}
\end{equation}
where $\omega,\ldots,\omega_n$ are independent random variables drawn from the Rademacher distribution; i.e., $\P(\omega_i = 1)=\P(\omega_i=-1)=1/2$, for $i=1,\ldots,n$. The last inequality in \eqref{eqn:decomp2} is due to the contraction inequality, and the fact that $g^2$ is a Lipschitz function. Henceforth, with the Talagrand's concentration inequality, there exists a constant $C_2>0$, such that,  with probability at least $1-e^{-c_1}$,
\begin{equation}
\label{eqn:decomp3}
\begin{aligned}
&\E \underset{\begin{subarray}{c}
  g\in\HH,\|g\|_\HH^0\leq \|F_j\|_\HH^0 \\
  \|g\|_{L_2}\leq \|F_j\|_{L_2}
  \end{subarray}}{\sup}\left\{ \frac{1}{n}\sum_{i=1}^n\omega_ig(x(t_i))\right\} \\
\leq \; & C_2\left[ \underset{\begin{subarray}{c}
  g=\sum_{k,l=1}^pg_{kl}\in\HH,\|g\|_\HH^0\leq \|F_j\|_\HH^0 \\
  \|g\|_{L_2}\leq \|F_j\|_{L_2}
  \end{subarray}}{\sup}\sum_{k,l=1}^p\left\{ \frac{1}{n}\sum_{i=1}^n\omega_ig_{kl}(x(t_i)) \right\} + \|F_j\|_{L_2}\sqrt{\frac{c_1}{n}}+\frac{c_1}{n}\right].
\end{aligned}
\end{equation}
By Lemma 2.2 of \citet{Yuan2016}, and the result that the $\nu$th eigenvalue of RKHS $\HH$ is of order $(\nu\log^{-1}\nu)^{-2\beta_2}$, for $\nu\geq 1$ \citep{Bach2017}, there exists a constant $C_3>0$, such that, with probability at least $1-d^{-c_1}$,
\begin{equation*}
\begin{aligned}
&\sum_{k,l=1}^p\left\{ \frac{1}{n}\sum_{i=1}^n\omega_ig_{kl}(x(t_i)) \right\} \\
\leq \; & C_3 n^{-1/2}\sum_{k,l=1}^p\left\{ \left( \|g_{kl}\|_{\HH}\log^{-1}\|g_{kl}\|_{\HH} \right)^{\frac{1}{2\beta_2}} \left(\|g_{kl}\|_{L_2}\log^{-1}\|g_{kl}\|_{L_2} \right)^{1-\frac{1}{2\beta_2}} \right. \\
&\left. + \|g_{kl}\|_{L_2}\sqrt{(c_1+1)\log p}+e^{-p}\|g_{kl}\|_\HH \right\},
\end{aligned}
\end{equation*}
Following the arguments for bounding $\Delta_4$ in \eqref{eqn:empmean}, there exists a constant $C_4>0$ and for any $c_2>1$, such that
\begin{equation*}
\begin{aligned}
& n^{-1/2}\underset{\begin{subarray}{c}
  g=\sum_{k,l=1}^pg_{kl}\in\HH,\|g\|_\HH^0\leq \|F_j\|_\HH^0 \\
  \|g\|_{L_2}\leq \|F_j\|_{L_2}
  \end{subarray}}{\sup} \sum_{k,l=1}^p\left( \|g_{jkl}\|_{\HH}\log^{-1}\|g_{jkl}\|_{\HH} \right)^{\frac{1}{2\beta_2}} \left( \|g_{kl}\|_{L_2}\log^{-1}\|g_{kl}\|_{L_2} \right)^{1-\frac{1}{2\beta_2}}\\
\leq \; & C_4c_2^{-\frac{4\beta_2}{2\beta_2-1}}\underset{\begin{subarray}{c}
  g=\sum_{k,l=1}^pg_{kl}\in\HH,\|g\|_\HH^0\leq \|F_j\|_\HH^0 \\
  \|g\|_{L_2}\leq \|F_j\|_{L_2}
  \end{subarray}}{\sup}\sum_{k,l=1}^p\left(\|g_{kl}\|_{L_2}\log^{-1}\|g_{kl}\|_{L_2}\right)^{2} + C_4c_2^{\frac{4\beta_2}{4\beta_2+1}}\left(\frac{n}{\log n}\right)^{-\frac{2\beta_2}{2\beta2+1}}\\
\leq \; & C_4c_2^{-\frac{4\beta_2}{2\beta_2-1}}C_5\|F_{j}(x(t))\|^2_{L_2}\log^{-2}\|F_{j}(x(t))\|_{L_2} +C_4c_2^{\frac{4\beta_2}{4\beta_2+1}}\left(\frac{n}{\log n}\right)^{-\frac{2\beta_2}{2\beta2+1}}, 
\end{aligned}
\end{equation*}
where the last step is due to  $\sum_{k,l=1}^p \left( \|F_{jkl}\|_{L_2}\log^{-1}\|F_{jkl}\|_{L_2} \right)^2 \leq C_5\left( \|F_j\|_{L_2}\log^{-1}\|F_j\|_{L_2} \right)^2$ for some constant $C_5>1$. Following the arguments for bounding $\Delta_5$ in  \eqref{eqn:empmean}, there exists a constant $C_6>0$, such that
\begin{equation*}
\begin{aligned}
\sum_{k,l=1}^p\|g_{kl}\|_{L_2} &\leq C_6\left\{ \sqrt{\frac{\log p}{n}}+\|F_{j}(x(t))\|_{L_2} \right\}.
\end{aligned}
\end{equation*}
Henceforth, for some constant $C_7>0$,
\begin{equation*}
\begin{aligned}
 &\underset{\begin{subarray}{c}
  g=\sum_{k,l=1}^pg_{kl}\in\HH,\|g\|_\HH^0\leq \|F_j\|_\HH^0 \\
  \|g\|_{L_2}\leq \|F_j\|_{L_2}
  \end{subarray}}{\sup}\sum_{k,l=1}^p\left\{\frac{1}{n}\sum_{i=1}^n\omega_ig_{kl}(x(t_i))\right\} \\
\leq \; & C_7\left\{ c_2^{-\frac{4\beta_2}{2\beta_2-1}}\|F_{j}(x(t))\|^2_{L_2}\log^{-2}\|F_{j}(x(t))\|_{L_2} +c_2^{\frac{4\beta_2}{4\beta_2+1}}\left(\frac{n}{\log n}\right)^{-\frac{2\beta_2}{2\beta2+1}} \right\}\\
& + C_7\sqrt{\frac{(c_1+1)\log p}{n}}\left\{ \sqrt{\frac{\log p}{n}}+\|F_{j}(x(t))\|_{L_2} \right\} + C_7n^{-1/2}e^{-p}.
\end{aligned}
\end{equation*}
Together with (\ref{eqn:decomp1}), (\ref{eqn:decomp2}), and (\ref{eqn:decomp3}), we have, with probability at least $1-2e^{-c_1}$,
\begin{equation*}
\begin{aligned}
&\underset{\begin{subarray}{c}
  g\in\HH,\|g\|_\HH^0\leq \|F_j\|_\HH^0 \\
  \|g\|_{L_2}\leq \|F_j\|_{L_2}
  \end{subarray}}{\sup}\left(\|g\|_{L_2}^2-\|g\|_n^2\right)\\
\leq \; & C_8\left\{ c_2^{-\frac{4\beta_2}{2\beta_2-1}}\|F_{j}(x(t))\|^2_{L_2}\log^{-2}\|F_{j}(x(t))\|_{L_2} +c_2^{\frac{4\beta_2}{4\beta_2+1}}\left(\frac{n}{\log n}\right)^{-\frac{2\beta_2}{2\beta2+1}} \right\}\\
& + C_8\sqrt{\frac{(c_1+1)\log p}{n}}\left\{ \sqrt{\frac{\log p}{n}}+\|F_{j}(x(t))\|_{L_2} \right\} + C_8n^{-1/2}e^{-p}\\
& + C_8\left(\|F_j(x(t))\|_{L_2}\sqrt{\frac{c_1}{n}}+\frac{c_1}{n}\right), 
\end{aligned}
\end{equation*}
for some constant $C_8>0$. Using the change of variable, the following result also holds. That is, with probability at least $1-2p^{-c_1}$, it holds that, 
\begin{equation*}
\begin{aligned}
&\|F_j(x(t))\|_{L_2}^2-\|F_j(x(t))\|_n^2\leq \underset{\begin{subarray}{c}
  g\in\HH,\|g\|_\HH^0\leq \|F_j\|_\HH^0 \\
  \|g\|_{L_2}\leq \|F_j\|_{L_2}
  \end{subarray}}{\sup}\left(\|g\|_{L_2}^2-\|g\|_n^2\right)\\
\leq \; & C_9 \left\{ c_2^{-\frac{4\beta_2}{2\beta_2-1}}\|F_{j}(x(t))\|^2_{L_2}\log^{-2}\|F_{j}(x(t))\|_{L_2} +c_2^{\frac{4\beta_2}{4\beta_2+1}}\left(\frac{n}{\log n}\right)^{-\frac{2\beta_2}{2\beta2+1}} \right\} \\
& + C_9\sqrt{\frac{(c_1+1)\log p}{n}}\left\{ \sqrt{\frac{\log p}{n}}+\|F_{j}(x(t))\|_{L_2} \right\} + C_8n^{-1/2}e^{-p}\\
& + C_9\left\{ \|F_j(x(t))\|_{L_2}\sqrt{\frac{(c_1+1)\log p}{n}}+\frac{(c_1+1)\log p}{n} \right\},
\end{aligned}
\end{equation*}
for some constant $C_9>0$. This completes the proof of Lemma \ref{lem:bdonradamacher}. 
\eop

\subsection{Proof of Theorem \ref{thm:optimalrecovery}}
\label{sec:pfoptimalrecovery}

\noindent
We divide the proof of this theorem to three parts. We first present the main proof in Section \ref{eqn:mainproof}. We then summarize some additional technical assumptions used during the proof in Section \ref{sec:condadd}. We give an auxiliary lemma in Section \ref{sec:auxlemnew}.

\subsubsection{Main proof}
\label{eqn:mainproof}

\noindent
We use the primal-dual witness method to prove that KODE selects all significant variables but includes no insignificant ones. The analysis here extends the techniques in \citet{Ravikumar2010} for the Ising model, where the pairwise interactions have a simple product form. Meanwhile, we also deal with measurement errors in variables.

Consider the optimization problem \eqref{eqn:kodeequiv} that is equivalent to \eqref{eqn:kode}. Recall that, by the representer theorem \citep{wahba1990}, the selection problem becomes \eqref{eqn:regfortheta}; i.e., 
\begin{equation}
\label{eqn:zjlasso}
\begin{aligned}
\min_{\thetabf_j}\left\{(\zbf_j - G\thetabf_j)^\top (\zbf_j - G\thetabf_j) + n \kappa_{nj} \left( \sum_{k=1}^p\theta_{jk}+\sum_{k\neq l, k=1}^p\sum_{l=1}^{p}\theta_{jkl} \right)\right\},
\end{aligned}
\end{equation}
subject to $\theta_k\geq 0,\theta_{kl} \geq 0, k,l = 1, \ldots, p, k\neq l$, where the ``response" is $\zbf_j = (\ybf_j - \bar{\ybf}_j) - (1/2)n\eta_{nj} \cbf_j - Bb_j$, and the ``predictor" is $G \in \R^{n \times p^2}$.  The vector $\theta$ solves (\ref{eqn:zjlasso}) if it satisfies the Karush-Kuhn-Tucker (KKT) condition:
\begin{equation}
\label{eqn:kkttheta}
\frac{2}{n}G^\top(G\theta_j-z_j) + \kappa_{nj}g_j = 0, \quad j=1,\ldots,p,
\end{equation}
where  $G$ contains errors in variable due to the estimated $\widehat{x}(t)$, and
\begin{equation}
\label{eqn:kkttg}
g_j  = \text{sign}(\theta_j) \;\; \text{if }\theta_j\neq 0, \quad\textrm{ and }\quad |g_j| \leq 1 \;\; \text{otherwise}.
\end{equation}

To apply the primal-dual witness method, we next construct an oracle primal-dual pair $(\widehat{\theta}_j,\widehat{g}_j)$ satisfying the KKT conditions (\ref{eqn:kkttheta}) and (\ref{eqn:kkttg}). Specifically,
\begin{itemize}
\item[(a)] We set $\widehat{\theta}_{jkl}=0$ for $(k,l)\not\in S_j$, where $S_j$ is defined as, 
\begin{equation*}
 S_j \equiv \big\{1\leq k\leq l\leq p: \text{ if } F_{jk}\neq 0, \text{ let }l=k; \textrm{ or if } F_{jkl}\neq 0 \textrm{ with }  l \neq k \le p \}.
\end{equation*} 
The definition of  $S_j$ is similar to $M_j$ defined in Section \ref{sec:bayesianci}. However, $S_j$ explicitly includes $\{(k,l):k=l\}$. Let $s_j = \text{card}(S_j)$.

\item[(b)] Let $\widehat{\theta}_{S_j}$ be the minimizer of the partial penalized likelihood, 
\begin{equation}
\label{eqn:consttheta}
(\zbf_j - G_{S_j}\thetabf_{S_j})^\top (\zbf_j - G_{S_j}\thetabf_{S_j}) + n \kappa_{nj} \left( \sum_{k=1}^p\theta_{jk}+\sum_{k\neq l, k=1}^p\sum_{l=1}^{p}\theta_{jkl} \right).
\end{equation}

\item[(c)] Let $S_j^c$ be the complement of $S_j$ in $\{(k,l):1\le k\leq l\leq p\}$.
We obtain $\widehat{g}_{S_j^c}$ from (\ref{eqn:kkttheta}) by substituting in the values of $\widehat{\theta}_j$ and $\widehat{g}_{S_j}$.
\end{itemize}

Next, we verify the support recovery consistency; i.e., 
\begin{equation*}
\label{eqn:supportconsis}
\max_{(k,l)\in S_j}\|\widehat{\theta}_{jkl}-\theta_{jkl}\|_{\ell_2}\leq \frac{2}{3}\theta_{\min},
\end{equation*}
which in turn implies that the oracle estimator $\widehat{\theta}_j$ recovers the support of $\theta_j$ exactly.  

Note that the subgradient condition for the partial penalized likelihood (\ref{eqn:consttheta}) is 
\begin{equation*}
\begin{aligned}
2G_{S_j}^\top(G_{S_j}\widehat{\theta}_{S_j}-z_j)+n\kappa_{nj}\widehat{g}_{S_j} = 0,
\end{aligned}
\end{equation*}  
which implies that
\begin{equation*}
\begin{aligned}
2G_{S_j}^\top(G_{S_j}\widehat{\theta}_{S_j}-G_{S_j}\theta_{S_j})+2G_{S_j}^\top(G_{S_j}\theta_{S_j}-z_j)+n\kappa_{nj}\widehat{g}_{S_j} = 0.
\end{aligned}
\end{equation*}
Define $\mathcal R_{S_j} \equiv 2G_{S_j}^\top G_{S_j}\theta_{S_j} - 2G_{S_j}^\top z_j$.  Then,
\begin{equation}
\label{eqn:diftheta}
\widehat{\theta}_{S_j}-\theta_{S_j} = -\left(2G_{S_j}^\top G_{S_j}\right)^{-1}(\mathcal R_{S_j}+n\kappa_{nj}\widehat{g}_{S_j}).
\end{equation}
For each $(k,l)$, denote the corresponding column of $G$ by $G_{kl}$. Then for $(k,l) \in S_j$,
\begin{equation}
\label{eqn:defofrk}
\mathcal R_{kl} = 2G_{kl}^\top G_{S_j}\theta_{S_j} - 2G_{kl}^\top z_j.
\end{equation}
By Lemma \ref{lem:conds3}, we have $\|\mathcal R_{kl}\|_{\ell_2}\leq\eta_{\mathcal R}$ for any $(k,l)\in S_j$. Then,
\begin{equation}
\label{eqn:rs0}
\|\mathcal R_{S_j}\|_{\ell_2}\leq \eta_{\mathcal R}\sqrt{s_j}.
\end{equation}
By Assumption \ref{network:dependency} given in Section \ref{sec:condadd}, we have $\Lambda_{\min}\left(G_{S_j}^\top G_{S_j}\right)\ge C_{\min}/2$, for some constant $C_{\min}>0$. Henceforth,
\begin{equation*}
 \Lambda_{\max}\left\{\left(2G_{S_j}^\top G_{S_j}\right)^{-1}\right\}\leq \frac{1}{C_{\min}}.
\end{equation*}
Note that for any $(k,l)\in S_j$, $\|\widehat{g}_{jkl}\|_{\ell_2}\leq 1$, which implies that, 
\begin{equation}
\label{eqn:gs0hat}
\|\widehat{g}_{S_j}\|_{\ell_2} \leq \sqrt{s_j}.
\end{equation}
Therefore,
\begin{equation*}
\max_{(k,l)\in S_j}\|\widehat{\theta}_{jkl}-\theta_{jkl}\|_{\ell_2}\leq \|\widehat{\theta}_{S_j}-\theta_{S_j}\|_{\ell_2}\leq \frac{\eta_{\mathcal R}\sqrt{s_j}}{C_{\min}}+n\kappa_{nj}\frac{\sqrt{s_j}}{C_{\min}}\leq \frac{2}{3}\theta_{\min}.
\end{equation*}
where the last inequality is due to Assumption \ref{network:minregeffect} in Section \ref{sec:condadd}.

Next, we verify the strict dual feasibility; i.e., 
\begin{equation*}
\max_{(k,l)\not\in S_j}|\widehat{g}_{jkl}|<1,
\end{equation*}
which in turn implies that the oracle estimator $\widehat{\theta}_j$ satisfies the KKT condition of the KODE optimization problem.  

For any $(k,l)\not\in S_j$, by (\ref{eqn:kkttheta}), we have, 
\begin{equation*}
2G_{kl}^\top(G_{S_j}\widehat{\theta}_{S_j}-z_j)+n\kappa_{nj}\widehat{g}_{jkl} = 0,
\end{equation*}
which implies that
\begin{equation*}
\begin{aligned}
&2G_{kl}^\top(G_{S_j}\widehat{\theta}_{S_j}-G_{S_j}\theta_{S_j})+2G_{kl}^\top(G_{S_j}\theta_{S_j}-z_j)+n\kappa_{nj}\widehat{g}_{jkl} = 0.
\end{aligned}
\end{equation*}
By (\ref{eqn:diftheta}) and (\ref{eqn:defofrk}), we have, 
\begin{equation*}
n\kappa_{nj}\widehat{g}_{jkl} = G_{kl}^\top G_{S_j}(G_{S_j}^\top G_{S_j})^{-1}(\mathcal R_{S_{j}}+n\kappa_{nj}\widehat{g}_{S_j})-\mathcal R_{kl}.
\end{equation*}
By Assumption \ref{network:incoherence} in Section \ref{sec:condadd}, we have that, 
\begin{equation*}
\max_{(k,l)\not\in S_j}\left\|G_{kl}^\top G_{S_j}(G_{S_j}^\top G_{S_j})^{-1}\right\|_{\ell_2}\leq \xi_G.
\end{equation*}
Then by (\ref{eqn:rs0}) and (\ref{eqn:gs0hat}), we have that
\begin{equation*}
|\widehat{g}_{jkl}|\leq \frac{(\xi_G+1)\sqrt{s_j}}{n\kappa_{nj}}\eta_{\mathcal R}+\xi_G\sqrt{s_j},\quad (k,l)\not\in S_j.
\end{equation*}
By Assumption \ref{network:minregeffect} in Section \ref{sec:condadd} that
\begin{equation*}
\frac{(\xi_G+1)\sqrt{s_j}}{n\kappa_{nj}}\eta_{\mathcal R}+\xi_G\sqrt{s_j}<1,
\end{equation*}
we obtain that, 
\begin{equation*}
|\widehat{g}_{jkl}|<1, \quad\text{for any } (k,l)\not\in S_j.
\end{equation*} 

Finally, the selection consistency for $S_j$ implies the selection consistency for $S^0_j$. This completes the proof of Theorem \ref{thm:optimalrecovery}.
\eop

\setcounter{assumption}{2}

\subsubsection{Additional technical assumptions}
\label{sec:condadd}

\noindent
We summarize the additional assumptions used during the proof of Theorem \ref{thm:optimalrecovery}. 
 
\begin{assumption}
\label{network:dependency}
Suppose there exists a constant $C_{\min}>0$ such that the minimal eigenvalue of matrix $G_{S_j}^\top G_{S_j}$ satisfies, 
\begin{equation*}
\Lambda_{\min}\left(G_{S_j}^\top G_{S_j}\right)\ge \frac{1}{2}C_{\min}.
\end{equation*}
\end{assumption}

\begin{assumption}
\label{network:incoherence}
Suppose there exists a constant $0 \leq \xi_G < 1$ such that,
\begin{equation*} 
\max_{(k,l)\not\in S_j}\left\|G_{kl}^\top G_{S_j}(G_{S_j}^\top G_{S_j})^{-1}\right\|_{\ell_2}\leq \xi_G.
\end{equation*}
\end{assumption}

\begin{assumption}
\label{network:minregeffect}
Suppose the following inequalities hold: 
\begin{equation*}
\begin{aligned}
\frac{\eta_{\mathcal R}\sqrt{s_j}}{C_{\min}}+n\kappa_{nj}\frac{\sqrt{s_j}}{C_{\min}}\leq \frac{2}{3}\theta_{\min}, \quad \textrm{ and } \quad 
\frac{(\xi_G+1)\sqrt{s_j}}{n\kappa_{nj}}\eta_{\mathcal R}+\xi_G\sqrt{s_j}<1.
\end{aligned}
\end{equation*}
where $\theta_{\min} = \min_{(k,l)\in S_j}\|\theta_{jkl}\|_{\ell_2}$. 
\end{assumption}

\noindent 
Assumption \ref{network:dependency} ensures the identifiability among the $s_j$ elements in the column set of $G_{S_j}$. The same  condition has been used in \citet{Zhao2006, Ravikumar2010, Chen2017}. Assumption \ref{network:incoherence} reflects the intuition that the large number of irrelevant variables cannot exert an overly strong effect on the subset of relevant variables. This condition is standard in the literature of Lasso regression \citep{meinshausen2006high, Zhao2006, Ravikumar2010}. Assumption \ref{network:minregeffect} imposes some regularity on the minimum regulatory effect. The second inequality characterizes the relationship between the quantities $\xi_G$, the sparse tuning parameter $\kappa_{nj}$, and the sparsity level $s_j$. Similar assumptions have been used in Lasso regression \citep{meinshausen2006high,  Zhao2006, Ravikumar2010, Chen2017}.

We detail Assumptions \ref{network:incoherence} and \ref{network:minregeffect} in three examples, which would lead to a more straightforward  interpretation for the hypotheses. 
 \citet{meinshausen2006high,  Zhao2006} provide examples and results on matrix
families that satisfy a similar type of conditions such as Assumptions \ref{network:incoherence} and \ref{network:minregeffect}, and we show these examples also holds for KODE in dynamic systems.
Recall the definition of the ``predictor" $G \in \R^{n \times p^2}$ in  (\ref{eqn:regfortheta}), where the first $p$ columns of $G$ are $\Sigmabf^k \cbf_j$ with $k=1,\ldots,p$, and the last $p(p-1)$ columns of $G$ are $\Sigma^{kl} \cbf_j$ with $k,l=1,\ldots,p, k\neq l$.
For notational simplicity, denote $\Sigmabf^k\equiv \Sigmabf^{kk}$.
All diagonal elements of $G^\top G$ are assumed to be $1$, which is equivalent to normalizing $\Sigma^{kl}c_j$ to the same scale for any $1\leq k,l\leq p$ since Assumption  \ref{network:incoherence} is invariant under common scaling of   $G^\top G$. 

The first example considers bounded correlations of functional component estimates $\Sigma^{kl}c_j$ for any $1\leq k,l\leq p$. This example has favorable implications for KODE applications; in particular, it implies that Assumption \ref{network:incoherence} holds even for $p$ growing with $n$ as long as $s_j$ remains fixed and consequently ensures that KODE selects the true model asymptotically.
\begin{example}
\label{eg:bdcorrelation}
Suppose that the correlation of $\Sigma^{kl}c_j$ and $\Sigma^{k'l'}c_j$ is bounded by $\frac{\xi_G}{2s_j-1}$ where $1\leq j,k,l,k',l'\leq p$, then Assumption \ref{network:incoherence} holds. 
\end{example}
\begin{proof}
Recall that $\frac{1}{2}C_{\min}$  is a lower bound of the minimal eigenvalue of $G_{S_j}^\top G_{S_j}$ defined in Assumption \ref{network:dependency}. 
We can bound $C_{\min}$ as follows. 
Since  the correlation of $\Sigma^{kl}c_j$ and $\Sigma^{k'l'}c_j$ is bounded by $\frac{\xi_G}{2s_j-1}$ for any $1\leq k,l,k',l'\leq p$, any off-diagonal element of $G^\top G$ is bounded by $\frac{\xi_G}{2s_j-1}$.
Let the vector $u=(u_1,\ldots,u_{s_j})^\top\in\R^{s_j}$. Then
\begin{equation*}
\begin{aligned}
u^\top (G_{S_j}^\top G_{S_j})u & = 1+\sum_{1\leq i_1\neq i_2\leq s_j}u_{i_1}(G_{S_j}^\top G_{S_j})_{(i_1,i_2)}u_{i_2}\geq 1-\frac{\xi_G}{2s_j-1}\sum_{1\leq i_1\neq i_2\leq s_j}|u_{i_1}||u_{i_2}|\\
& \geq 1-\frac{1}{2s_j-1}\sum_{1\leq i_1\neq i_2\leq s_j}|u_{i_1}||u_{i_2}|\geq\frac{s_j}{2s_j-1}.
\end{aligned}
\end{equation*}
Thus, $C_{\min}\geq \frac{2s_j}{2s_j-1}$ and 
\begin{equation*}
\left\|G_{kl}^\top G_{S_j}(G_{S_j}^\top G_{S_j})^{-1}\right\|_{\ell_2}\leq \|G_{kl}^\top G_{S_j}\|_{\ell_2}\frac{2}{C_{\min}}\sqrt{s_j}\leq\frac{\xi_G\sqrt{s_j}}{2s_j-1}\cdot\frac{2s_j-1}{s_j}\cdot\sqrt{s_j}=\xi_G.
\end{equation*}
\end{proof}
The second example shows two instances of Example \ref{eg:bdcorrelation}, where Assumption \ref{network:incoherence} holds under some simplified structures. 
\begin{example}
Assumption \ref{network:incoherence} holds if (i) $s_j=1$, or (ii) $G^\top G$ is orthogonal.
\end{example}
\begin{proof}
(i) If  $s_j=1$, we let $\xi_G \equiv \max_{(k,l)\neq (k',l')}(\Sigma^{kl}c_j)^\top\Sigma^{k'l'}c_j<1$.
Then the condition in Example  \ref{eg:bdcorrelation} holds. Therefore, Assumption \ref{network:incoherence} holds.

(ii) If  the matrix $G^\top G$ is orthogonal, the correlation of $\Sigma^{kl}c_j$ and $\Sigma^{k'l'}c_j$ is zero for any $(k,l)\neq (k',l')$ and hence the condition in Example  \ref{eg:bdcorrelation} holds for any $s_j$ and $\xi_G$. Therefore, Assumption \ref{network:incoherence} holds.
\end{proof}

The third example illustrates Assumption \ref{network:minregeffect} with a natural condition that  the minimal signal term $\theta_{\min}$ does not decay too fast. In particular, it is necessary to have a gap between the decay rate of the minimal signal and $n^{-1/2}$. Since the noise terms aggregates at a rate of $n^{-1/2}$, this condition prevents the estimation and selection from being  dominated by the noise terms.
\begin{example}
Suppose that $\xi_G<s_j^{-1/2}$ and $\theta_{\min}=O\{(\log p)^{\epsilon_p}[(\frac{n}{\log n})^{-\frac{\beta_2}{2\beta_2+1}} + (\frac{\log p}{n})^{\frac{1}{2}} + n^{-\frac{\beta_1}{2\beta_1+1}}]\}$ with  $\epsilon_p>0$. Here,  $\theta_{\min}$ decays at the rate slower than $n^{-1/2}$ as $n$ and $p$ grow. Then there exists tuning parameter $\kappa_{nj}>0$ such that Assumption \ref{network:minregeffect} holds.
\end{example}
\begin{proof}
Recall that $\eta_{\mathcal R} = O_p((\frac{n}{\log n})^{-\frac{2\beta_2}{2\beta_2+1}} + \frac{\log p}{n} + n^{-\frac{2\beta_1}{2\beta_1+1}})$. Then $\P(\theta_{\min} \geq \frac{3}{2}c_\theta\eta_{\mathcal R})\to 1$ for any constant $c_\theta>0$ as $n$ and $p$ grow. Letting $\kappa_{nj} = c_\kappa n^{-1}\eta_{\mathcal R}$ with
\begin{equation}
\label{eqn:upperbdckappa}
0<c_{\kappa}\leq \frac{c_\theta C_{\min}}{\sqrt{s_j}}-1,
\end{equation}
then the first inequality of Assumption \ref{network:minregeffect} holds. Moreover,  letting $\kappa_{nj} = c_\kappa n^{-1}\eta_{\mathcal R}>0$ with
\begin{equation}
\label{eqn:lowerbdckappa}
c_\kappa>\frac{(\xi_G+1)\sqrt{s_j}}{1-\xi_G\sqrt{s_j}},
\end{equation}  
then the second inequality of Assumption \ref{network:minregeffect} also holds. 
By setting $ c_\theta>\frac{(1+\sqrt{s_j})\sqrt{s_j}}{(1-\xi_G\sqrt{s_j})C_{\min}}$, 
 there exists $c_k$ satisfying both (\ref{eqn:upperbdckappa}) and (\ref{eqn:lowerbdckappa}) as long as 
\begin{equation*}
\xi_G<\frac{1}{\sqrt{s_j}}.
\end{equation*}
Therefore, there exists tuning parameter  $\kappa_{nj}>0$ given by $c_\kappa n^{-1}\eta_{\mathcal R}$  such that Assumption \ref{network:minregeffect}  holds. 
\end{proof}

\subsubsection{Auxiliary lemma for Theorem \ref{thm:optimalrecovery}}
\label{sec:auxlemnew}

\noindent 
We present a lemma that is useful for the proof of Theorem \ref{thm:optimalrecovery}. It gives a bound similar to the deviation condition proposed by \citet{loh2012}. The difference is that, the noise in variable $\widehat{x}(t)$ in our setting involves a nonlinear transformation through the kernel $K(\widehat{x}(t),\widehat{x}(s))$.

\begin{lemma}
\label{lem:conds3}
For $j=1,\ldots,p$, we have, 
\begin{equation*}
\|G^\top_{kl} G_{S_j}\theta_{S_j} - G_{kl}z_j\|_{\ell_2}\leq \eta_{\mathcal R}, \;\; \textrm{ where } \; 
\eta_{\mathcal R} = O_p\left( \left(\frac{n}{\log n}\right)^{-\frac{\beta_2}{2\beta_2+1}} + \left(\frac{\log p}{n}\right)^{1/2} + n^{-\frac{\beta_1}{2\beta_1+1}} \right).
\end{equation*}
\end{lemma}
\noindent
\textbf{Proof of Lemma \ref{lem:conds3}}:
Similar to the ``predictor" $G$ defined in \eqref{eqn:regfortheta} in Section \ref{sec:functionalestimation}, we first construct a noiseless version of the predictor $\widetilde{G} \in \R^{n \times p^2}$, whose first $p$ columns are $\widetilde{\Sigmabf}^k \cbf_j$, and the last $p(p-1)$ columns are $\widetilde{\Sigma}^{kl} \cbf_j$, and $\widetilde{\Sigma}^k = (\widetilde{\Sigma}^k_{ii'}), \widetilde{\Sigma}^{kl} = (\widetilde{\Sigma}^{kl}_{ii'})$ are both $n\times n$ matrices whose $(i,i')$th entries are,  
\begin{equation*}
\begin{aligned}
\widetilde{\Sigma}^{k}_{ii'} & = \int_\TT\int_\TT\{ T_i(s)-\bar{T}(s) \} K_{k}(\xbf(t),\xbf(s)) \{ T_{i'}(t)-\bar{T}(t) \} dsdt,   \;\; 1 \leq k \leq p, 1 \leq i, i' \leq n, \\
\widetilde{\Sigma}^{kl}_{ii'} & = \int_\TT\int_\TT\{ T_i(s)-\bar{T}(s) \} K_{kl}(\xbf(t),\xbf(s)) \{ T_{i'}(t)-\bar{T}(t) \} dsdt, \;\; 1 \leq k < l \leq p, 1 \leq i, i' \leq n.
\end{aligned}
\end{equation*}

Next, we consider the term $\left\|G^\top_{kl}z_j - G_{kl}^\top G_{S_j}\theta_{S_j} \right\|_{\ell_2}$, which can be bounded as, 
\begin{equation}
\label{eqn:Gdecomp}
\begin{aligned}
\left\|G^\top_{kl}z_j - G_{kl}^\top G_{S_j}\theta_{S_j} \right\|_{\ell_2} 
\leq \; & \left\|G^\top_{kl}\E[z_j] - G^\top_{kl}\widetilde{G}_{S_j}\theta_{S_j}\right\|_{\ell_2} + \left\|G^\top_{kl}(\widetilde{G}_{S_j}-G_{S_j}) \theta_{S_j} \right\|_{\ell_2} \\ 
& + \left\|G^\top_{kl}(z_j-\E[z_j])\right\|_{\ell_2} \equiv \Delta_7 + \Delta_8 + \Delta_9.
\end{aligned}
\end{equation}
We next bound the three terms $\Delta_7, \Delta_8, \Delta_9$ on the right-hand-side of \refeq{eqn:Gdecomp}, respectively.

For $\Delta_7$, by the Cauchy-Schwarz inequality, we have, 
\begin{equation*}
\begin{aligned}
\Delta^2_7 
\leq \left\| G^\top_{kl} \right\|_{\ell_2}^2 \left\| \E[z_j]-\widetilde{G}_{S_j}\theta_{S_j} \right\|_{\ell_2}^2 \leq C_1\left\| \E[z_j]-\widetilde{G}_{S_j}\theta_{S_j} \right\|_{\ell_2}^2 
= O_p\left( \left(\frac{n}{\log n}\right)^{-\frac{2\beta_2}{2\beta_2+1}} + \frac{\log p}{n} \right),
\end{aligned}
\end{equation*}
for some constant $C_1>0$, where the last step is by (\ref{eqn:step2term1}).

For $\Delta_8$, again by the Cauchy-Schwarz inequality, we have,
\begin{equation*}
\begin{aligned}
\Delta^2_8 \leq \left\| G^\top_{kl} \right\|_{\ell_2}^2 \left\| \left( \widetilde{G}_{S_j}-G_{S_j} \right)\theta_{S_j} \right\|_{\ell_2}^2 \leq C_2 \left\|\left( \widetilde{G}_{S_j}-G_{S_j} \right) \right\|^2_\infty \left\| \theta_{S_j} \right\|^2_{\ell_1} = O_p\left( n^{-\frac{2\beta_1}{2\beta_1+1}} \right),
\end{aligned}
\end{equation*}
for some constants $C_2>0$, where the last step is by (\ref{eqn:bdonerrorinx}) and the fact that $\|\theta_{S_j}\|_{\ell_1}$ is bounded. 

For $\Delta_9$, by Lemma \ref{lem:bdonradamacher}, we have, 
\begin{equation*}
\begin{aligned}
\Delta^2_9 = O_p\left( \left( \frac{n}{\log n} \right)^{-\frac{2\beta_2}{2\beta_2+1}} + \frac{\log p}{n} \right).
\end{aligned}
\end{equation*}

Combining the above three bounds, we obtain that, 
\begin{equation*}
\left\| G^\top_{kl}z_j - G_{kl}^\top G_{S_j}\theta_{S_j} \right\|_{\ell_2} = O_p\left( \left(\frac{n}{\log n}\right)^{-\frac{\beta_2}{2\beta_2+1}} + \left(\frac{\log p}{n}\right)^{1/2} + n^{-\frac{\beta_1}{2\beta_1+1}} \right),
\end{equation*}
which completes the proof of Lemma \ref{lem:conds3}.
\eop

\section{Additional numerical results}
\label{asec:addnumerical}

\noindent
We report some additional numerical results. We begin with a comparison with a family of ODE solutions assuming known functionals $F$. We then carry out a sensitivity analysis to study the robustness of the choice of kernel function and initial parameters. Finally, we report the sparse recovery of the enzymatic regulatory network example studied in Section \ref{sec:enzymatic}, and the gene regulatory network example studied in Section \ref{sec:application} of the paper.

\subsection{Comparison with alternative methods}
\label{asec:morecomparison}

\noindent
We compare the proposed KODE method with a family of alternative ODE solutions, including \citet{gonzalez2014reproducing, zhang2015selection, mikkelsen2017learning}. Particularly, \citet{gonzalez2014reproducing} proposed a penalized log-likelihood approach in RKHS, where the ODE system is used as a penalty. \citet{zhang2015selection} studied a full predator-prey ODE model that takes a special form of two-dimensional rational ODE. \citet{mikkelsen2017learning} learned a class of polynomial or rational ODE systems.  However, the main difference is that, all those solutions assumed the forms of the functionals $F$ are completely known, while KODE does not require so, but instead estimates the functionals adaptively given the data. In addition, both \citet{gonzalez2014reproducing} and \citet{zhang2015selection} focused on the low-dimensional ODE, while our method works for both low-dimensional and high-dimensional ODE. Moreover, none of those solutions tackled post-selection inference, while we do. These differences clearly distinguish our proposal from those existing ones. 

Next, we numerically compare KODE with the least squares approximation (LSA) method of \citet{zhang2015selection}, and the adaptive integral matching (AIM) method of \citet{mikkelsen2017learning}. We did not include \citet{gonzalez2014reproducing} in our numerical comparison, since their code is not available. Moreover, they also assumed known $F$ similarly as the other two works. We implement LSA using the code provided by the Wiley Online Library, and AIM using the R package \texttt{episode} with the Lasso penalty and automatic adaptation of parameter scales. 

\begin{figure}[t!]
\centering
\includegraphics[width=\textwidth, height=2.25in]{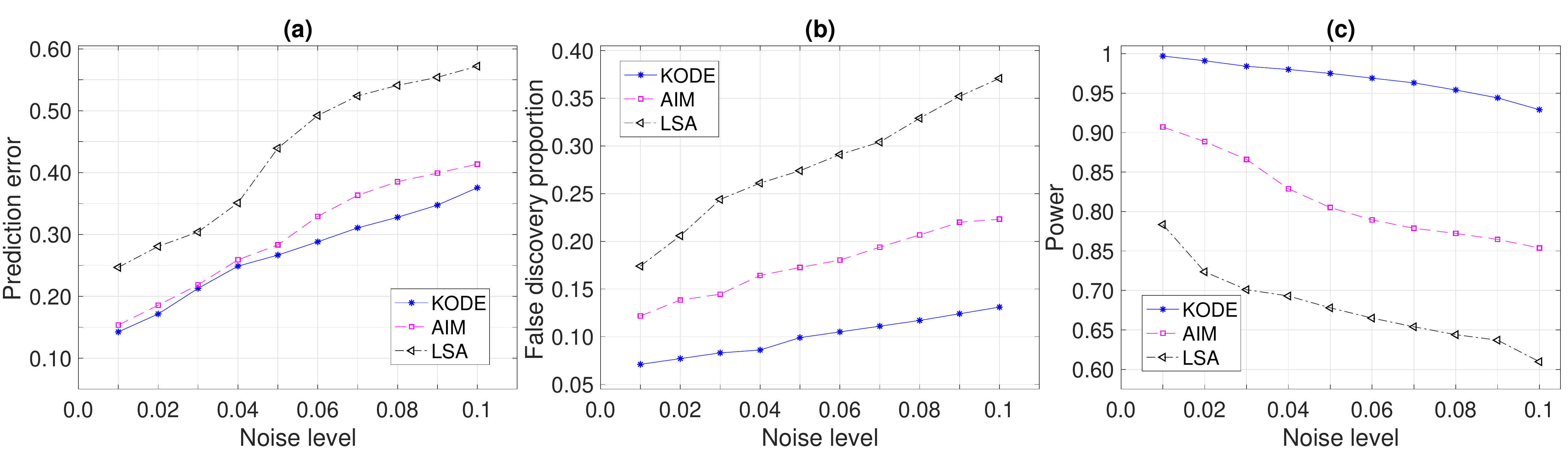}
\caption{The prediction and selection performance of KODE, LSA and AIM with varying noise level. The results are averaged over 500 data replications. (a) Prediction error; (b) False discovery proportion; (c) Empirical power. }
\label{fig:para_estimation}
\end{figure}

Figure \ref{fig:para_estimation} reports the performance of KODE, LSA and AIM for the enzymatic regulatory network example in Section \ref{sec:enzymatic}. It is seen that KODE clearly outperforms both LSA and AIM in terms of both prediction and selection accuracy. This suggests that the  polynomial or rational forms of $F$ imposed by LSA and AIM may not hold in this example. 

\begin{table}[b!]
\centering
\caption{The area under the ROC curve and the $95\%$ confidence interval of KODE, LSA and AIM, for 10 combinations of network structures from GNW. The results are averaged over $100$ data replications.}
\resizebox{\textwidth}{!}{
\begin{tabular}{ccccccc}
\toprule
& \multicolumn{3}{c}{$p=10$} & \multicolumn{3}{c}{$p=100$} \\ \cline{2-4} \cline{5-7}
& KODE & AIM & LSA & KODE & AIM & LSA \\ [0.2ex]
 \midrule
\emph{E.coli1} & $\textbf{0.582}$ & $0.558$ &   $0.437$ & $\textbf{0.711}$ & $0.698$ & $0.614$ \\
 & $(0.577, 0.587)$ & $(0.554, 0.562)$ & $(0.428, 0.446)$ & $(0.708, 0.714)$ & $(0.694, 0.702)$ & $(0.606, 0.622)$ \\ [0.2ex]
\emph{E.coli2} &  $\textbf{0.662}$ & $0.646$ &   $0.541$ & $\textbf{0.685}$ & $0.678$ &   $0.501$ \\
 & $(0.658, 0.666)$ & $(0.641, 0.651)$ & $(0.533, 0.549)$ & $(0.681, 0.689)$ & $(0.673, 0.683)$ & $(0.488, 0.514)$ \\ [0.2ex]
Yeast1 & $\textbf{0.603}$ & $0.539$ &   $0.413$ & $\textbf{0.619}$ & $0.599$ & $0.527$ \\
 & $(0.599, 0.607)$ & $(0.534, 0.544)$ & $(0.401, 0.425)$ & $(0.616, 0.622)$ & $(0.594, 0.604)$ & $(0.511, 0.543)$ \\ [0.2ex]
Yeast2 & $\textbf{0.599}$ & $0.559$ &   $0.497$ & $\textbf{0.606}$ & $0.577$ & $0.518$ \\
 & $(0.595, 0.603)$ & $(0.554, 0.564)$ & $(0.488, 0.506)$ & $(0.603, 0.609)$ & $(0.573, 0.581)$ & $(0.505, 0.531)$ \\ [0.2ex]
Yeast3 & $\textbf{0.612}$ & $0.563$ &   $0.451$ & $\textbf{0.621}$ & $0.609$ & $0.577$ \\
 & $(0.608, 0.616)$ & $(0.558, 0.567)$ & $(0.440, 0.462)$ & $(0.617, 0.625)$ & $(0.604, 0.614)$ & $(0.562, 0.592)$ \\ [0.2ex]
\bottomrule
\end{tabular}
}
\label{table:para_estimation}
\end{table}

Table \ref{table:para_estimation} reports the performance of the three methods for the gene regulatory network example in Section \ref{sec:application}. The results are averaged over $100$ data realizations for all ten combinations of network structures. Again, it is clearly seen that KODE outperforms LSA and AIM in all cases. 

Together with the numerical results that compare KODE with linear ODE \citep{Zhang2015} and additive ODE \citep{Chen2017} reported in the paper, it demonstrates that the proposed KODE is a competitive and useful tool for modeling complex dynamic systems.

\subsection{Sensitivity analysis}
\label{asec:sensitivity}

\noindent
We carry out a sensitivity analysis to investigate the robustness of the choice of kernel function and initial parameters in KODE. 

First, we consider three commonly used kernels and study their performances using the enzymatic regulatory network example in Section \ref{sec:enzymatic}. Recall the first-order Mat\'{e}rn kernel used in our analysis in Section \ref{sec:enzymatic},
\begin{equation*}
\label{eqn:matern1}
K^{(1)}_\FF(x,x') = (1+\sqrt{3}\|x-x'\|/\nu)\exp(-\sqrt{3}\|x-x'\|/\nu).
\end{equation*}
In addition, we also consider the second-order Mat\'{e}rn kernel,
\begin{equation*}
\label{eqn:matern2}
K^{(2)}_\FF(x,x') = (1+\sqrt{5}\|x-x'\|/\nu+5\|x-x'\|^2/3\nu^2)\exp(-\sqrt{5}\|x-x'\|/\nu),
\end{equation*}
and the Gaussian kernel, 
\begin{equation*}
\label{eqn:gaussian}
K^{(3)}_\FF(x,x') =\exp(-\|x-x'\|^2/2\nu^2).
\end{equation*}
It is known that the RKHS generated by $K^{(1)}_\FF$ and $K^{(2)}_\FF$ contains once differentiable and twice differentiable functions, respectively \citep{gneiting2010matern}, while the RKHS generated by $K^{(3)}_\FF$ contains infinitely differentiable functions \citep{lin2004statistical}. We couple the proposed KODE method with these three kernels: KODE-1 with the first-order Mat\'{e}rn kernel, KODE-2 with the second-order Mat\'{e}rn kernel, and KODE-3 with the Gaussian kernel. We continue to choose the bandwidth $\nu$ using tenfold cross-validation.

\begin{figure}[t!]
\centering
\includegraphics[width=\textwidth, height=2.25in]{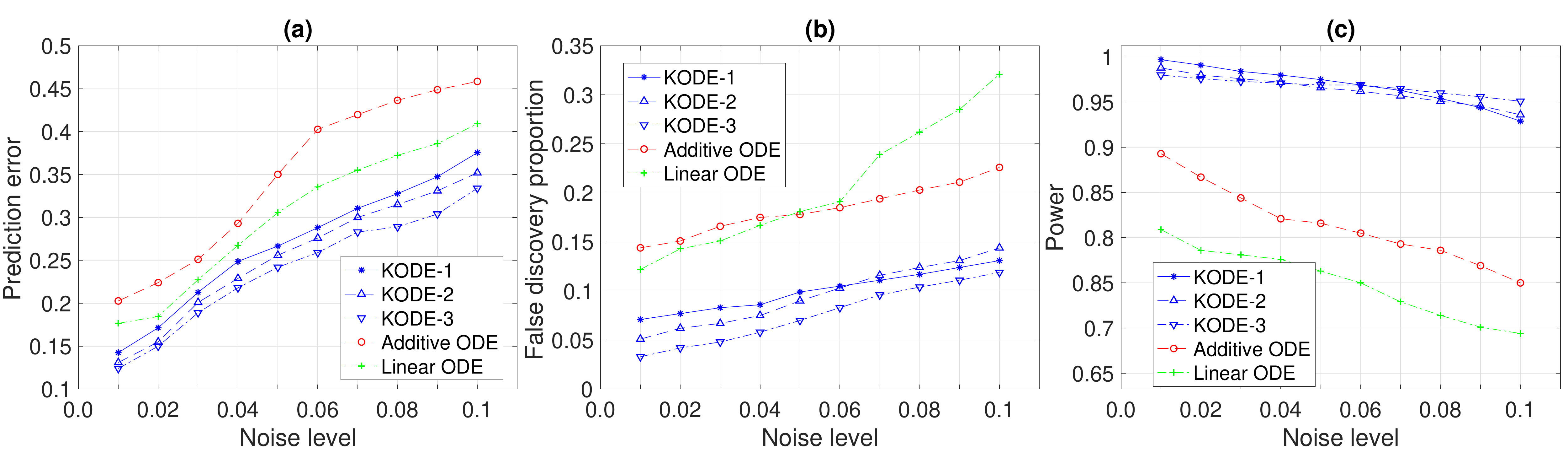}
\caption{The prediction and selection performance with varying noise level, for KODE with three different kernels, plus linear ODE and additive ODE. The results are averaged over $500$ data replications. (a) Prediction error; (b) False discovery proportion; (c) Empirical power. }
\label{fig:differentkernels}
\end{figure}

Figure \ref{fig:differentkernels} reports the prediction and selection performance of the KODE method with the three kernels, plus the linear and additive ODE methods. It is seen that the performances of the three KODE methods are fairly close. The relative prediction errors differ at most $4.6\%$, the false discovery proportions differ at most $0.8\%$, and the empirical powers differ at most $1.3\%$, across different noise levels. Besides, they all outperform the linear and additive ODE considerably. These results demonstrate that the proposed KODE is relatively robust to the choice of the kernel function.

We also make remarks about some general principles of choosing kernel functions. In practice, if there is prior knowledge about the smoothness of the trajectory $x$ and the ODE system $F$, we may choose kernels such that the corresponding RKHS has the same order of smoothness as $x$ and $F$, respectively. In this case, Theorems \ref{thm:optimalestofpredictor} and \ref{thm:optimalestoffunctional} ensure that KODE is minimax optimal for the estimation of $x$ and $F$. On the other hand, if there is no such prior knowledge, we may then choose kernels with a higher-order smoothness. This recommendation is supported by the observation that the performance of KODE-3 based on the Gaussian kernel is slightly better than those of KODE-1 and KODE-2 based on the Mat\'{e}rn kernels when the noise level is large. This recommendation also agrees with the usual recommendation in classical kernel learning, e.g., density estimation \citep{hall1988choice}, and nonparametric function estimation \citep{lin2004statistical}. Lastly, we comment that one can use cross-validation to choose the best performing kernel function as well \citep{duan2003evaluation, meyer2003support}.

Next, we study the sensitivity of KODE with respect to the choice of initial parameters. Toward that end, we divide the parameters in Algorithm \ref{alg:trainofkode} that require initialization into four subsets: (i) the parameters: $\{\theta_{jk},j,k=1,\ldots,p\}$; (ii) the parameters: $\{\theta_{jkl},j,k,l=1,\ldots,p, k\neq l\}$; (iii) the tuning parameters: $\{\eta_{nj}, j=1,\ldots,p\}$; and (iv) the tuning parameters: $\{\kappa_{nj}, j=1,\ldots,p\}$. We then consider the following initialization schemes: First, we uniformly draw $200$ i.i.d.\ vectors from $[0.5,1.5]^{p^2}$ as the initial values for $\{\theta_{jk},j,k=1,\ldots,p\}$ in (i), and fix the initial values of the parameters in (ii) to (iv) to $1$. Second, we uniformly draw $200$ i.i.d.\ vectors from $[0.5,1.5]^{p^2(p-1)}$ as the initial values for $\{\theta_{jkl},j,k,l=1,\ldots,p, k\neq l\}$ in (ii), and fix the initial values of the rest of the parameters in (i), (iii) and (iv) to $1$. Third, we uniformly draw $200$ i.i.d.\ values from $[10^{-5},1]$ as the initial for $\{\eta_{nj}, j=1,\ldots,p\}$ in (iii), and fix the initial values of the rest of parameters to $1$. Fourth, we uniformly draw $200$ i.i.d.\ values from $[10^{-5},1]$ as the initial for $\{\kappa_{nj}, j=1,\ldots,p\}$ in (iv),  and fix the initial values of the rest of parameters to $1$. Finally, we initialize all the parameters in (i) to (iv) to 1, and take this as a reference. For each setting of parameter initialization, we apply Algorithm  \ref{alg:trainofkode} to the  enzymatic regulatory network example in Section \ref{sec:enzymatic} with the noise level $\sigma_j=0.05, j=1,2,3$.  

\begin{figure}[t!]
\centering
\includegraphics[width=\textwidth, height=2.25in]{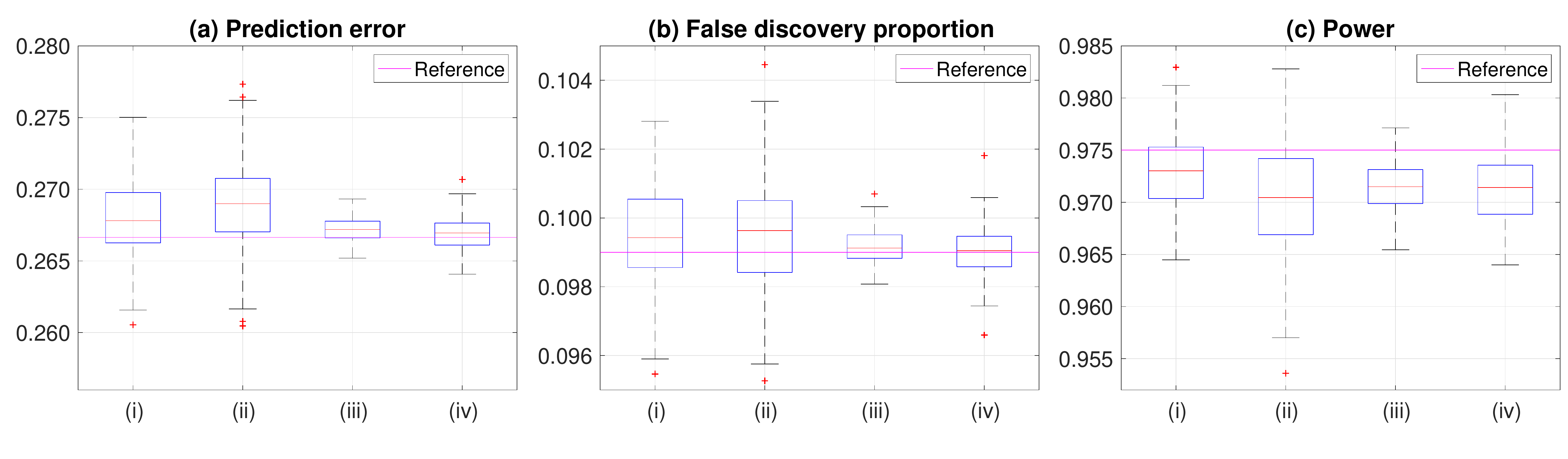}
\caption{The prediction and selection performance of KODE with different initialization schemes. The boxes range from the lower to the upper quartile, and the whiskers extend to the most extreme data point that is no more than $1.5$ times the interquartile range from the box. The solid horizontal lines denote the reference initialization scheme. The results are averaged over $500$ data replications. (a) Prediction error; (b) False discovery proportion; (c) Empirical power.}
\label{fig:initialization}
\end{figure}

Figure \ref{fig:initialization} reports the prediction and selection performance of KODE with different schemes of parameter initialization, and the results are averaged over 500 data replications. It is seen that the performances under different initialization schemes are close. For the majority of cases, the relative prediction errors to the reference differ at most $1.9\%$, the false discovery proportions differ at most $0.2\%$ compared to the reference, and the empirical powers differ at most $0.8\%$ compared to the reference. This example shows that the proposed KODE is fairly robust to the choice of the initial values. In the paper, we simply employ the reference initialization scheme, i.e., setting all the initial values to 1.

\subsection{Enzymatic regulatory network recovery by KODE}
\label{asec:enzymatic}

\begin{figure}[t!]
\centering
\includegraphics[width=\textwidth, height=2.25in]{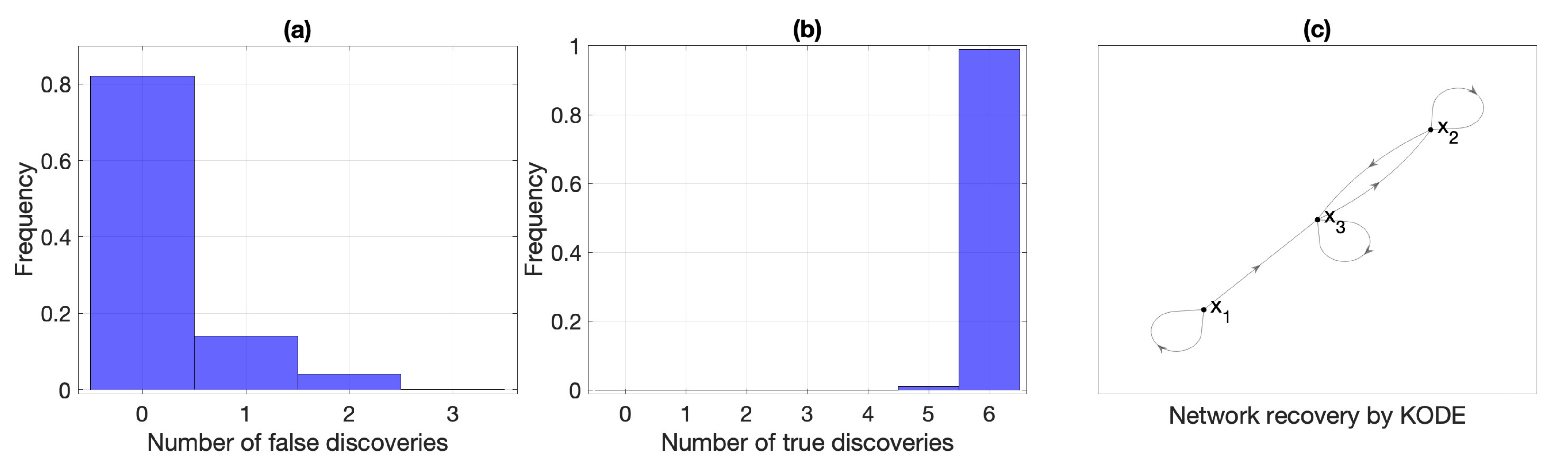}
\caption{The selection performance of  KODE for the enzymatic regulatory network, based on $500$ data replications. (a) Number of false discoveries in the estimated model. (b) Number of true discoveries in the estimated model. (c) Network recovery thresholded at the $90\%$ frequency.}
\label{fig:recoveryplot}
\end{figure}

\noindent
Figure \ref{fig:eg2_2}(b)-(c) in Section \ref{sec:enzymatic} of the paper reported the selection performance of KODE in terms of false discovery proportion and power when recovering the enzymatic regulatory network under different noise levels. Here we present additional results about the recovery of this network at a given noise level. Figure \ref{fig:recoveryplot} reports the results based on 500 data replications, where the noise level is set at $\sigma_j=0.01, j=1,2,3$. Figure \ref{fig:recoveryplot}(a) shows that, in more than $80\%$ of the cases, KODE is able to recover the network without making any false discovery, whereas Figure \ref{fig:recoveryplot}(b) shows that, in  more than $99\%$ of the cases, KODE recovers all the true edges. Figure \ref{fig:recoveryplot}(c) reports a sparse recovery of the network, where an arrowed edge is drawn if it appears in more than $90\%$ of the estimated networks out of 500 data replications. It is seen that KODE successfully recovers the true regulatory network.

\subsection{Gene regulatory network recovery by KODE}
\label{asec:genenet}

\begin{figure}[b!]
\centering
\includegraphics[width=\textwidth, height=2.25in]{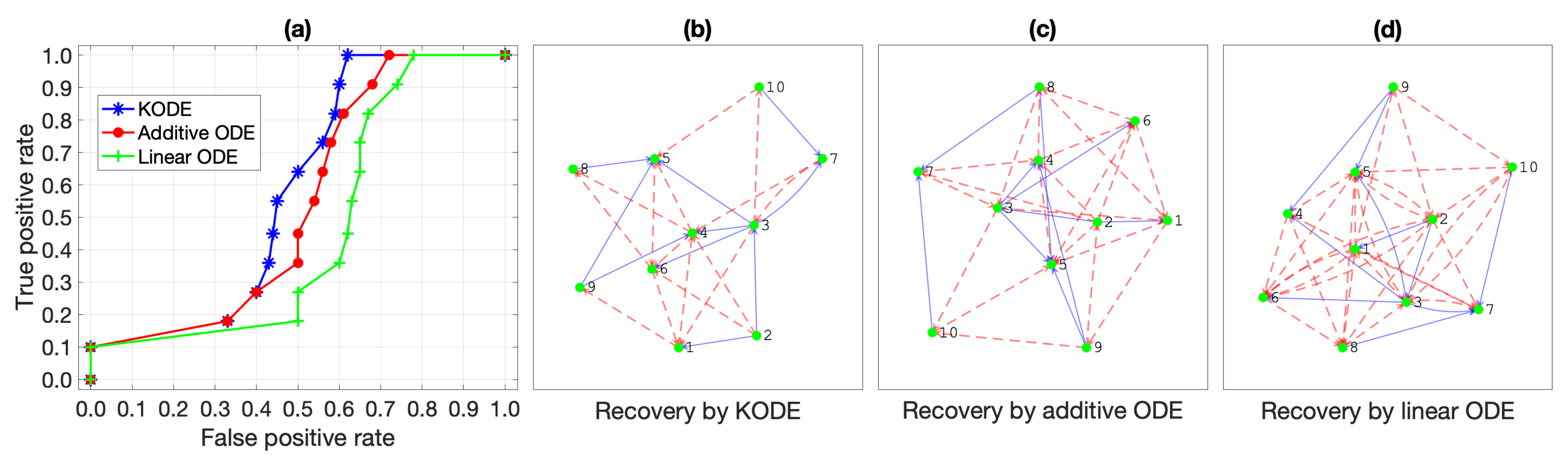}
\caption{(a) Median ROC for recovering the 10-node gene regulatory network of \emph{E.coli1}, based on 100 data replications. (b)-(d) The network recovery by KODE, additive ODE, and linear ODE. The solid and dashed arrowed lines denote the true and false discoveries, respectively.}
\label{fig:roccurve}
\end{figure}

\noindent
Table \ref{table:gnw} in Section \ref{sec:application} of the paper reported the selection performance of KODE in terms of the area under the ROC curve for all 10 combinations of gene regulatory network structures from GeneNetWeaver. Here we present some additional results about the recovery of one such structure, the 10-node \emph{E.coli1} network. Figure \ref{fig:roccurve} reports the results based on 100 data replications. Figure \ref{fig:roccurve}(a) shows the median ROCs for KODE, additive ODE, and linear ODE. It is seen that KODE achieves the fastest recovery rate, as well as the largest AUC of 0.582, compared to 0.541 for additive ODE and 0.460 for linear ODE. Figure \ref{fig:roccurve}(b)-(d) report the sparse recovery of the network, based on KODE, additive ODE, and linear ODE, respectively, where an arrowed edge is drawn if it appears in more than $90\%$ of the estimated networks out of 100 data replications. It is seen that KODE achieves a better selection accuracy compared to linear ODE and additive ODE. 

\end{document}